\documentclass[a4paper,12pt]{article}
\usepackage{longtable}

\usepackage[utf8]{inputenc}

\usepackage{url}
\usepackage[dvipdfmx]{graphicx}
\usepackage{tikz}
\usetikzlibrary{decorations.pathmorphing}

\usepackage{braket}

\usepackage{amsmath}

\usepackage{mathrsfs}

\usepackage[dvipdfmx]{graphicx}
\usepackage{tikz}
\usepackage{bm}
\usepackage{arydshln}
\usepackage{color}
\usepackage{ulem}
\usepackage{amssymb}

\usepackage{comment}

\newcommand{\be}{\begin{equation}}
\newcommand{\ee}{\end{equation}}
\newcommand{\ba}{\begin{eqnarray}}
\newcommand{\ea}{\end{eqnarray}}
\newcommand{\f}{\frac}
\newcommand{\s}{\sqrt}

\DeclareMathOperator{\Pf}{Pf}

\DeclareMathOperator{\Tr}{Tr}

\usepackage{geometry}
\numberwithin{equation}{section}
\allowdisplaybreaks

\begin{document}

\newcommand{\hiduke}[1]{\hspace{\fill}{\small [{#1}]}}
\newcommand{\aff}[1]{${}^{#1}$}
\renewcommand{\thefootnote}{\fnsymbol{footnote}}

\begin{titlepage}
\begin{flushright}
{\footnotesize preprint SISSA 30/2019/FISI}
\end{flushright}
\begin{center}
{\Large\bf
Quantum Chaos,  Thermodynamics and Black Hole Microstates in the mass deformed SYK model
}\\
\bigskip\bigskip
\bigskip\bigskip
{\large Tomoki Nosaka,\footnote{\tt nosaka@yukawa.kyoto-u.ac.jp}}\aff{1,2}
{\large and Tokiro Numasawa,\footnote{\tt tokiro.numasawa@mail.mcgill.ca}}\aff{3}\\
\bigskip\bigskip
\aff{1}: {\small
\it INFN Sezione di Trieste, Via Valerio 2, 34127 Trieste, Italy
}\\
\aff{2}: {\small
\it International School for Advanced Studies (SISSA), Via Bonomea 265, 34136 Trieste, Italy
}\\
\bigskip
\aff{3}: {\small
\it Department of Physics, McGill University, 3600 Rue University ,Montreal, Quebec H3A 2T8, Canada 
}\\
\end{center}
\bigskip
\bigskip
\begin{abstract}
We study various aspects of the mass deformation of the SYK model which makes the black hole microstates escapable.
SYK boundary states are given by a simple local boundary condition on the Majorana fermions and then evolved in Euclidean time in the SYK Hamiltonian.
We study the ground state of this mass deformed SYK model in detail. 
We also use SYK boundary states as a variational approximation to the ground state of the mass deformed SYK model.
We compare variational approximation with the exact ground state results and they showed a good agreement.
We also study the time evolution of the mass deformed ground state under the SYK Hamiltonian.
We give a gravity interpretation of the mass deformed ground state and its time evolutions.
In gravity side, mass deformation gives a way to prepare black hole microstates that are similar to pure boundary state black holes.
Escaping protocol on these ground states simply gives a global AdS$_2$ with an IR end of the world brane.
We also study the thermodynamics and quantum chaotic properties of this mass deformed SYK model.
Interestingly, we do not observe the Hawking Page like phase transition in this model in spite of similarity of the Hamiltonian  with eternal traversable wormhole model where we have the phase transition.

\end{abstract}

\bigskip\bigskip\bigskip

\end{titlepage}

\renewcommand{\thefootnote}{\arabic{footnote}}
\setcounter{footnote}{0}

\tableofcontents

\newpage
\section{Introduction and Summary}
Recently there are many progresses in understanding the spacetime dynamics like the chaotic behavior in black holes \cite{Shenker:2013pqa, Maldacena:2015waa} or making traversable wormholes \cite{Gao:2016bin,Maldacena:2017axo,Maldacena:2018lmt}.
The Sachdev-Ye-Kitaev (SYK) model \cite{PhysRevLett.70.3339,KitaevTalk} plays an important role to study these behaviors.
The SYK model is a UV complete model that shares the same low energy dynamics with the Jackiw-Teitelboim gravity \cite{JACKIW1985343,Teitelboim:1983ux,Maldacena:2016upp}.
We can solve the SYK model analytically in the large $N$, low energy limit \cite{Maldacena:2016hyu}.
We can also study the large $N$ Schwinger-Dyson equation directly numerically \cite{Maldacena:2016hyu} and further we can study the model numerically at finite $N$ \cite{Cotler:2016fpe,Garcia-Garcia:2017bkg,Garcia-Garcia:2019poj} by exact diagonalization, which serves a complemental way to analyze the system.

Traversable wormholes are made by introducing a direct coupling between two asymptotic boundaries \cite{Gao:2016bin,Maldacena:2018lmt}.
After turning off the  coupling between two sides, we can make a two sided black holes \cite{Maldacena:2018lmt,Chen:2019qqe}.
Therefore, traversable wormhole protocol also gives a way to prepare the thermofield double states.
We can consider the similar question for a single sided system whether we can prepare a ``thermofield double" state in a single copy of CFT from some massive deformation. 
An analog of the thermofield double states for single sided case is the boundary state \cite{Cardy:1989ir,Onogi:1988qk}, which have a simple entanglement structure.
In gravity side,  the dual spacetimes are  black hole microstates \cite{Hartman:2013qma,Cooper:2018cmb,Numasawa:2018grg}, and by state dependent mass deformation we can reveal the interior of these single sided black holes \cite{Almheiri:2018ijj}.
Employing the proposal for holographic duals of boundary CFT (BCFT) \cite{Takayanagi:2011zk,Fujita:2011fp}, these microstates are modeled by geometry with end of the world (EOW) branes. 
Now the problem is whether we can prepare the black hole microstates from massive deformations.
In the field theory side, the relation between boundary states and gapped ground states are discussed in the literature.
Boundary states and the time evolution of them are used to model so called quantum quench \cite{Calabrese:2009qy, Calabrese:2005in}.
The quantum quench\footnote{Precisely speaking, here we consider the sudden quantum quench with a time dependent mass term, which is a special case of quantum quenches.} is the time dependent process where we suddenly turn off the mass term and evolve the state by the Hamiltonian at critical points.
In this context, boundary states are used to approximate the initial gapped ground states.
This variational approximation is also used to study the massive deformations of conformal field theory \cite{Cardy:2017ufe} and gives a qualitative picture of the phase diagram.

In the SYK model, an analog of boundary states and their gravity interpretation  are proposed in \cite{Kourkoulou:2017zaj}.
We can also reveal the interior by the mass deformation \cite{Kourkoulou:2017zaj,Brustein:2018fkr}.
Motivated by the above observations on the boundary states, gapped deformation of CFT and their connection to black hole microstates,
in this paper we study the ground state of a mass deformed SYK model that also first appeared in \cite{Kourkoulou:2017zaj} and its relation to the SYK boundary state in detail.
A merit to consider the SYK model is that we can analyze directly the mass deformed theory itself.
This enables us to compare the exact ground state and the variational approximation and check the validity of the variational approximation, which is usually difficult.

Another motivation to study this model is to gain an insight to the gravity counterpart of the chaotic/integrable transition.
The relevance of the quantum chaotic properties to the black hole physics was pointed out in \cite{Hawking:2014tga,Hayden:2007cs,Sekino:2008he}.
One way to characterize the quantum chaos is the exponentially growing behavior of the out-of-time-ordered correlation functions (OTOC) \cite{1969JETP...28.1200L} in time which is quantified by the quantum Lyapunov exponent.
The OTOC was also studied in the SYK model \cite{Maldacena:2016hyu} in the large $N$ limit, where the quantum Lyapunov exponent was found to saturate the bound proposed in \cite{Maldacena:2015waa} in the strong coupling limit.
This also supports the relation between black holes and the quantum chaos.

Recently the quantum chaotic property was also studied in a variety of deformations of the SYK model \cite{Garcia-Garcia:2017bkg,Garcia-Garcia:2019poj,Azeyanagi:2017drg,Ferrari:2019ogc,Sorokhaibam:2019qho,Nosaka:2018iat,Garcia-Garcia:2018pwt,Garcia-Garcia:2018ruf,Qi:2018bje,Banerjee:2016ncu}.
Among these developments in \cite{Garcia-Garcia:2017bkg} the authors considered a deformation of the SYK Hamiltonian by a random mass term and found that the quantum Lyapunov exponent decreases as the mass parameter is increased.
More interestingly, the authors also found that the Lyapunov exponent vanishes at some finite value of mass parameter, where the system exhibits the ``chaotic/integrable transition'' \cite{Garcia-Garcia:2017bkg}.
The chaotic/integrable transition was also captured by another characterization of the quantum chaos through the level statistics \cite{1977RSPSA.356..375B,Bohigas:1983er,Guhr:1997ve,1984LNP...209....1B} as a sharp transition.

With the relation between quantum chaos and the black holes in one's mind, it may be natural to speculate the gravitational phenomenon dual to the chaotic/integrable transition to be the Hawking-Page transition \cite{Hawking:1982dh}.
Note, however, that these two phenomena do not necessarily happen at the same time.
The Hawking-Page like transition can be captured as a first order phase transition through the entropy, or equivalently, the free energy $F=-\frac{1}{\beta}\log Z(\beta)=-\frac{1}{\beta}\int dE\langle \rho(E)\rangle e^{-\beta E}$ where $\rho(E)$ is the number density of the states.
On the other hand, it is known that the level statistics diagnoses the chaoticity of a quantum system correctly only after the energy spectrum of the system is unfolded \cite{1984LNP...209....1B,Guhr:1997ve}, which essentially subtract the information of $\langle \rho(E)\rangle$ from the spectrum.
Also, the connection between chaos and the thermalization property of the system implies that one cannot see whether the system is chaotic or not just by looking thermodinamic quantities such as the free energy \cite{Nandkishore:2014kca}.
Nevertheless, there are some examples \cite{Garcia-Garcia:2019poj,Ferrari:2019ogc} where we have multiple evidences that they are indeed correlated.
It would be interesting to ask what kind of additional properties of a model can relate the chaotic/integrable transition and the Hawking-Page like transition {\it indirectly}, which will define an interesting class of theories.

\subsection{Summary of the paper}

We have studied various properties of the mass deformed Hamiltonian which is first proposed in \cite{Kourkoulou:2017zaj}.
The model consists from $N$ Majorana fermions $\psi_i$ and the Hamiltonian is given by 
\begin{align}
H_{\text{def}} &= H _{SYK} + H_M, \notag \\ 
H_{SYK} &= i^{\f{q}{2}} \sum_{i_1 < \cdots <i_q } J_{i_1\cdots i_q} \psi_{i_1}\cdots \psi_{i_q}, \qquad
H_M = i \mu \sum _{k = 1}^{\f{N}{2}} s_k \psi_{2k-1}\psi_{2k},  \label{eq:IntroHdef}
\end{align}
with mean $\braket{J_{i_1\cdots i_q}} = 0$ and variance $\braket{J_{i_1\cdots i_q}^2} = \f{J^2}{N^{q-1}}(q-1)!  = \f{1}{q} \f{\mathcal{J}^2 (q-1)!}{(2N)^{q-1}}$.
We give an effective action in terms of the collective variables and derive the large $N$ Schwinger-Dyson equation for the mass deformed Hamiltonian. 
We study this Schwinger-Dyson equation numerically.
In the zero temperature case, we found the analytical solution for this Schwinger-Dyson equation in the small mass parameter limit.
The diagonal correlation function is related to the SYK correlation function by the conformal transformation.
The off diagonal correlation function is also determined in the conformal limit.
Using these Euclidean correlation functions in the ground state, we study the several physical observables, which show the non trivial scaling with respects to the mass parameter $\mu$ in (\ref{eq:IntroHdef})  for small $\mu$ limit.

We have also used the SYK boundary state, which is also first proposed in  \cite{Kourkoulou:2017zaj} and interpreted as a black hole microstate, as a variational approximation for the ground state of the mass deformed Hamiltonian (\ref{eq:IntroHdef}).
We studied this variational approximation both numerically and analytically.
We compare the numerical results in exact ground state and in variational approximation  and two results show good agreement in the entire mass parameter regime.
In the small $\mu$ limit, we compare the analytical results in exact ground state and in variational approximation.
We found that the scaling with respect to the mass parameter coincide but the proportional constants are different.
However the coefficients themselves are also very close.
Therefore, the variational approximation is a good approximation but is not a perfect approximation and has an order $N$ difference even in the conformal limit.

In section \ref{section_thermodynamic} we have also computed the large $N$ free energy by solving the Schwinger-Dyson equations at finite temperature numerically.
In contrast to the results in the models with a similar Hamiltonian \cite{Maldacena:2018lmt,Azeyanagi:2017drg,Ferrari:2019ogc}, in our model we have not found a phase transition.

We have also studied our model for finite $N$ in section \ref{sec_finiteN}.
We have computed the overlap between the boundary state and the true ground state, and have found these two states are close to each other.\footnote{
While we treat the average over the random couplings by the annealed average in the large $N$ analysis, assuming that the annealed average is a good approximation to the original quenched average in the large $N$ limit, in section \ref{sec_finiteN} we directly adopt the quenched average.
}
This result supports the validity of the variational approximation for the ground state in section \ref{sec_variationalansatz}.
We have also diagnosed the quantum chaoticity of the system by computing the adjacent gap ratio 
\cite{2013PhRvL.110h4101A,2007PhRvB..75o5111O,2016PhRvB..94n4201B,2015PhRvB..91h1103L}
(we explain more detail in section \ref{sec_chaoticproperty}).
As a result we found that the system is chaotic for any value of the deformation parameter and at all energy scale.

In section \ref{sec_largeq} we have studied the mass deformed model in the large $q$ limit where we can analyze the mass deformed theory analytically beyond the conformal limit.
The large $q$ results are consistent with the conformal limit results when the mass parameter $\mu$ is small.\footnote{
This is a non-trivial statement as the two results agrees not only in the strict SYK limit $\mu=0$ but also in the $\mu$-dependent sub-leading corrections (see \eqref{eq:GDEnConf} and \eqref{eq:gdenergylargeq1}).
}
In the large $q$ limit, the variational approximation actually coincides with the exact ground state in all mass parameter regime.
We checked this agreement from the calculation of the ground state energy and the other observables also perfectly coincide.
Furthermore, in the large $q$ limit we compute the overlap between exact ground state and the SYK boundary state for the variational approximation.
We found the saddle point solution that gives the maximal overlap $1$.
Even at finite temperature, we can solve the system analytically.
We checked analytically that there are no phase transition in the model (\ref{eq:IntroHdef}).
At the order of $\beta \sim q$, the chaos exponents grows from $0$ to $\f{2\pi}{\beta}$ that is maximal \cite{Maldacena:2015waa}.

Finally, we give a gravity interpretation of the mass deformed ground state.
We studied the time evolution of the mass deformed ground state under the SYK Hamiltonian.
This is a setup of quantum quench where we suddenly turn off the mass term.
We solve this quench problem analytically in the conformal limit, determined the time evolution of the reparametrization mode and found that the system thermalizes.
We interpret this dynamics of the reparametrization mode in gravity.
The original geometry is interpreted as the global AdS$_2$ with EOW brane which is static under the global time translation.
The quench corresponds to the black hole generation.
Therefore, we interpret the mass deformation as a protocol to obtain atypical black hole microstates that are similar to pure boundary state black holes.
This is a single sided analog of the preparation of the thermofield double from the two coupling mass deformation \cite{Maldacena:2018lmt,Cottrell:2018ash}.
We have also applied our gravity interpretation to the escaping interior protocol starting from the mass deformed ground state.
The insertion of the mass deformed Hamiltonian before the quench just delays the generation of the black hole and shifts the position of the black hole horizon.
As a special case, we can apply the deformed Hamiltonian eternally and we can make the interior escapable eternally.
Actually, this is exactly the global AdS$_2$ with EOW brane.
This is a single sided counterpart of the identification of eternal traversable wormholes and global AdS$_2$.
We also point out that the mismatch of the sign of spins in the ground state and the mass term leads to the excitation.
Too much excitation leads to the huge excitation and we expect that they finally lead to the black hole generation because the system shows the chaotic behavior at high energy as we studied in this paper. 
Therefore, the escaping protocol is successful only when we choose the mass term in a correct state dependent way.

The organization of this paper is as follows.
{\bf  In section 2}, we review the mass deformed SYK model  and boundary states those are introduced in \cite{Kourkoulou:2017zaj}.
We also review their gravity interpretation as black hole microstates and their time evolution.
{\bf  In section 3}, we study the mass deformed SYK model in the large $N$ limit.
We derive the large $N$ Schwinger-Dyson equation for the mass deformed SYK model and solve them both analytically and numerically.
{\bf  In section 4}, we study the overlap and the level statistics at finite $N$ and compare with the large $N$ results.
{\bf  In section 5}, we study the large $q$ limit of the mass deformed SYK model.
We calculate various quantities analytically and checked the consistency with the results in large $N$, finite $q$ analysis.
{\bf  In section 6}, we describe the gravity interpretation of the mass deformed SYK model.
{\bf  In section 7}, we discuss some implications of our results and possible future works.
In  {\bf  appendix A}, we present the derivation of the large $N$ effective action for collective variables.
In  {\bf appendix B}, we discuss the numerical method to solve the large $N$ equation of the mass deformed SYK model.
In  {\bf appendix C}, we show the detail of large $q$ analysis.
In  {\bf appendix D}, we discuss the detail of the large $N$, $q=4$ case.

\section{The model}

The main purpose of this paper is to study the following Hamiltonian 
\begin{align}
H_{\text{def}} &= H _{SYK} + H_M, \notag \\ 
H_{SYK} &= i^{\f{q}{2}} \sum_{i_1 < \cdots <i_q }^N J_{i_1\cdots i_q} \psi_{i_1}\cdots \psi_{i_q}, \notag \\
H_M &= i \mu \sum _{k = 1}^{\f{N}{2}} s_k \psi_{2k-1}\psi_{2k} \equiv -\f{\mu}{2} \sum_{k=1}^{\f{N}{2}}s_k S_{k}, \label{Hdef}
\end{align}
with mean $\braket{J_{i_1\cdots i_q}} = 0$ and variance
$\braket{J_{i_1\cdots i_q}^2} = \f{J^2}{N^{q-1}}(q-1)!  = \f{1}{q} \f{2^{q-1}\mathcal{J}^2 (q-1)!}{N^{q-1}}$.
We also defined $S_k = - 2i \psi_{2k-1}\psi_{2k}$.
This Hamiltonian was first introduced in \cite{Kourkoulou:2017zaj}.
First we review some properties of this Hamiltonian with its connection to particular pure states in the SYK model.
We also review their gravity interpretation of pure states and the evolution under mass deformed Hamiltonian.

\subsection{A review of the SYK model}

The Sachdev-Ye-Kitaev (SYK) model \cite{PhysRevLett.70.3339,KitaevTalk} is the system of $N$ Majorana fermions $\psi_i$, which obey the anti commutation relation $\{\psi_i, \psi_j \} = \delta _{ij}$, with the Hamiltonian
\be
H_{SYK} = i^{\f{q}{2}} \sum_{i_1 < \cdots <i_q }^N J_{i_1\cdots i_q} \psi_{i_1}\cdots \psi_{i_q},
\ee
with mean $\braket{J_{i_1\cdots i_q}} = 0$ and variance
$\braket{J_{i_1\cdots i_q}^2} = \f{J^2}{N^{q-1}}(q-1)!  = \f{1}{q} \f{2^{q-1}\mathcal{J}^2 (q-1)!}{N^{q-1}}$.
At large $N$ , the correlation function $G(\tau_1,\tau_2) = \f{1}{N}\sum_{i=1}^N \braket{\psi_i(\tau_1)\psi_i(\tau_2)}$ satisfies the following Schwinger-Dyson equation
\be
\partial _{\tau_1} G(\tau_1,\tau_2) - \int d\tau_3 \Sigma(\tau_1,\tau_3) G(\tau_3,\tau_2) = \delta(\tau_1-\tau_2),\qquad \Sigma(\tau_1,\tau_2) = \f{\mathcal{J}^2}{q} (2G(\tau_1,\tau_2))^{q-1}. \label{eq:SDSYK}
\ee
This Schwinger-Dyson equation comes from the Euclidean action
\be
-\f{S_E}{N} =  \log \text{Pf} (\partial _{\tau} - \Sigma) - \f{1}{2} \int d\tau_1 \int d\tau_2 \Bigg[ \Sigma(\tau_1,\tau_2) G(\tau_1,\tau_2) - \f{\mathcal{J}^2}{2q} (2 G(\tau_1,\tau_2))^q \Bigg].
\ee
We can solve the equation (\ref{eq:SDSYK}) numerically.
In the large $q$ limit, we can solve the equation (\ref{eq:SDSYK}) analytically.
It is also possible to solve at long time ($1 \ll \mathcal{J}\tau_{12}  \ll N $) regime by ignoring the derivative term $\partial_\tau$ in (\ref{eq:SDSYK}) as
\be
G(\tau_1,\tau_2) = \f{c_{\Delta}}{|\mathcal{J}(\tau_1 - \tau_2)|^{2\Delta}} \text{sgn}(\tau_1 -\tau_2), \label{eq:SYKGDlow}
\ee
where the scaling dimension $\Delta$ and the coefficient $c_{\Delta}$ is given by 
\be
\Delta = \f{1}{q}, \qquad 
 c_{\Delta} = \f{1}{2} \Big[ \Big(1 -2\Delta\Big) \f{\tan \pi \Delta}{\pi \Delta} \Big]^{\Delta}.
\ee
When we ignore the term $\partial_\tau$, the equation of motion have a reparametriztion symmetry $G(\tau_1,\tau_2) \to [f'(\tau_1)f'(\tau_2)]^{\Delta}G(f(\tau_1),f(\tau_2))$ and $\Sigma(\tau_1,\tau_2) \to [f'(\tau_1)f'(\tau_2)]^{1-\Delta}\Sigma(f(\tau_1),f(\tau_2))$.
In the low temperature ($1 \ll \mathcal{J}\beta  \ll N $) and long time $(1 \ll \mathcal{J}\tau_{12} )$, the thermal correlation function is obtained from the ground state answer (\ref{eq:SYKGDlow}) with the reparametrization $f(\tau) = \tan \f{\pi}{\beta}\tau$.

The low energy reparametriztion symmetry is actually broken by the UV effect.
This leading breaking term is given by the Schwarzian action \cite{Maldacena:2016hyu}
\be
S = - \f{N \alpha _S}{\mathcal{J}} \int d\tau \{ f(\tau), \tau \}, \qquad  \{ f(\tau), \tau \} = \f{f'''(\tau)}{f'(\tau)} -  \f{3}{2}\Big( \f{f''(\tau)}{f'(\tau)}\Big)^2.
\ee
The constant $\alpha_S$ can be determined numerically. 
For example, $\alpha _S \approx 0.00709$ for $q=4$, $\alpha _S \approx 0.00403$ for $q=6$ and $\alpha _S \approx 0.00257$ for $q=8$ \cite{Maldacena:2016hyu}.
At large $q$, $\alpha_S$ goes as $\alpha_S \sim \f{1}{4q^2}$.  

\subsection{A review of pure states and mass deformation of the SYK model}
In the SYK model we can also study the real time evolution of particular pure states.
To define such states, first we define a set of spin operators from Majorana fermion operators.
They are defined as 
\be
S_{k} = -2 i\psi_{2k-1}\psi_{2k}.
\ee
They satisfy $S_{k}^2 = 1$, which means that the eigenvalues of $S_k$'s  are $\pm 1$.
Moreover, they are mutually commuting with each others $[S_k , S_{k'}] = 0$.
Therefore, we can consider the simultaneous eigenstates of the spin operators $S_k$'s:
\be
S_k \ket{B_{\bm{s}}} = s_k \ket{B_{\bm{s}}} \label{Bs},
\ee
where $\bm{s} = (s_1 ,\cdots , s_{\f{N}{2}})$ is a set of eigenvalues.
These $2^{\f{N}{2}}$ states span the SYK Hilbert spaces.
This condition can also be written as 
\be
\psi_{2k-1} \ket{B_{\bm{s}}} = -i s_k \psi_{2k}\ket{B_{\bm{s}}} \label{eq:fermioncondition}.
\ee
We can produce lower energy states by including the Euclidean evolution $\ket{B_{\bm{s}}(\beta)} = e^{-\f{\beta}{2}H_{SYK}}\ket{B_{\bm{s}}}$.
We can interpret the (\ref{eq:fermioncondition}) as a transparent boundary condition between fermion field $\psi_{2k-1}$ and $\psi_{2k}$
in the path integral language.
Because the states $\ket{B_{\bm{s}}}$ form a basis, the average of the correlators over all choices of $s_k$ reproduces the thermal ensemble exactly:
\be
\sum_{s_k = \pm 1} \bra{B_{\bm{s}}(\beta)}  \mathcal{O}\ket{B_{\bm{s}}(\beta)}  = \Tr[ e^{-\beta H_{SYK}} \mathcal{O}]. \label{eq:averages} 
\ee
This is a true statement for any operator $\mathcal{O}$.

In the large $N$ limit, the model possesses an emergent $O(N)$ symmetry.
This $O(N)$ symmetry includes an element $f_1$ that flips the sign of $\psi_2$. Similarly, there are elements $f_k$ that flips the sign of $\psi_{2k}$.
Each of $f_k$'s also flips the sign of the spin operator $S_k$.
This flip element $f_k$ maps the $\ket{B_{\bm{s}}}$ to other state $\ket{B_{\bm{s}'}}$ where $\bm{s}'$ is given by the flip of $s_k$ from $\bm{s}$.
Therefore, the norms $\braket{B_{\bm{s}}(\beta)|B_{\bm{s}}(\beta)}$ have the same value in all $\ket{B_{\bm{s}}}$ states because of this emergent symmetry.
On the other hand, we saw in (\ref{eq:averages}) that the average over all the $\ket{B_{\bm{s}}}$ is equivalent to the thermal one.
Therefore, the norms of $|B_{\bm{s}}(\beta)\rangle$ is equal to the thermal partition function in the leading of the $1/N$ expansion:
\begin{align}
\braket{B_{\bm{s}}(\beta)|B_{\bm{s}}(\beta)} &= 2^{-\f{N}{2}}  \sum_{a_1 = \pm 1}\sum_{a_2 = \pm 1}\cdots\sum_{a_{\f{N}{2}} = \pm 1} \braket{B_{\bm{s}}(\beta)|(f_1^{a_1}f_2 ^{a_2} \cdots f_{\f{N}{2}} ^{a_\f{N}{2}}) (f_1^{a_1}f_2 ^{a_2} \cdots f_{\f{N}{2}} ^{a_\f{N}{2}})|B_{\bm{s}}(\beta)}   \notag \\
&= 2^{-\f{N}{2}} \Tr(e^{-\beta H_{SYK}}).
\end{align}
Similarly, the two point functions such as $\psi_1(\tau)\psi_1(\tau')$ or $\psi_2(\tau)\psi_2(\tau')$ are also individually invariant under the flip groups.
They are called diagonal corerlators \cite{Kourkoulou:2017zaj}.
They also become the same with the thermal correaltors in the large $N$ limit:
\be
\braket{B_{\bm{s}}(\beta)|\psi_i(\tau)\psi_i(\tau') |B_{\bm{s}}(\beta)} = 2^{-\f{N}{2}} \Tr(e^{-\beta H_{SYK}} \psi_i(\tau)\psi_i(\tau'))
\ee

For off diagonal correlators like $\braket{B_{\bm{s}}(\beta)|\psi_1(\tau)\psi_2(\tau') |B_{\bm{s}}(\beta)}$ we can do similar argument by inserting $S_k (\tau) = -2i\psi_{2k-1} (\tau)\psi_{2k}(\tau)$ at $\tau = -\f{\beta}{2}$.
Because
\be
S_k \Bigl(-\f{\beta}{2}\Bigr)\ket{B_{\bm{s}}(\beta)} = e^{-\f{\beta}{2}H_{SYK}}S_k  \ket{B_{\bm{s}}} = s_k  e^{-\f{\beta}{2}H_{SYK}}\ket{B_{\bm{s}}} = s_k \ket{B_{\bm{s}} (\beta)},
\ee
we obtain 
\ba
 \braket{B_{\bm{s}}(\beta)| \psi_1(\tau)\psi_2(\tau') s_1  |B_{\bm{s}}(\beta)}
= -2i \braket{B_{\bm{s}}(\beta)| \psi_1(\tau)\psi_2(\tau') \psi_1 (-\beta/2)\psi_2(-\beta/2) |B_{\bm{s}}(\beta)} .
\ea
Now, the off diagonal correlator times the boundary condition $s_k$ becomes a $4$ point function with 2 $ \psi_1$'s and 2 $\psi_2$'s.
Therefore, in the large $N$ limit these are flip group invariant correlation function.
Then,
\ba
&& s_k \braket{B_{\bm{s}}(\beta)| \psi_{2k-1}(\tau)\psi_{2k}(\tau')  |B_{\bm{s}}(\beta)} \notag \\
&=& 2^{-\f{N}{2}}\sum_{\bm{s}} -2i \braket{B_{\bm{s}}(\beta)| \psi_{2k-1}(\tau)\psi_{2k}(\tau') \psi_{2k-1} (-\beta/2)\psi_{2k}(-\beta/2) |B_{\bm{s}}(\beta)} \notag \\
&=& -2i\times 2^{-\f{N}{2}} \Tr [e^{-\beta H_{SYK}} \psi_{2k-1}(\tau)\psi_{2k}(\tau') \psi_{2k-1} (-\beta/2)\psi_{2k}(-\beta/2)] .
\ea
We also know that in the large $N$ limit the normalization factor becomes $\braket{B_{\bm{s}}(\beta)|B_{\bm{s}}(\beta)} = 2^{-\f{N}{2}} \Tr(e^{-\beta H_{SYK}})$, 
and therefore the off diagonal correlator becomes
\be
s_k \f{\braket{B_{\bm{s}}(\beta)| \psi_{2k-1}(\tau)\psi_{2k}(\tau')  |B_{\bm{s}}(\beta)} }{  \braket{B_{\bm{s}}(\beta)|B_{\bm{s}}(\beta)} } = -2 i \f{\Tr [e^{-\beta H_{SYK}} \psi_{2k-1}(\tau)\psi_{2k}(\tau') \psi_{2k-1} (-\beta/2)\psi_{2k}(-\beta/2)]}{ \Tr(e^{-\beta H_{SYK}})}.\label{eq:BdiaglargeN}
\ee
Further because we are taking the large $N$ limit, four point function factorizes to the product of $2$ point functions and the off diagonal correlator becomes 
\be
G_{\text{off}}(\tau,\tau') \equiv s_k \f{\braket{B_{\bm{s}}(\beta)| \psi_{2k-1}(\tau)\psi_{2k}(\tau')  |B_{\bm{s}}(\beta)} }{  \braket{B_{\bm{s}}(\beta)|B_{\bm{s}}(\beta)}} = 2i G_{\beta}(\tau + \beta/2)G_{\beta}(\tau ' + \beta/2) + \mathcal{O}(1/N).\label{eq:BofflargeN}
\ee
Here we obtain another minus sign because we need to contract $\psi_{2k}$'s and to do that we need to exchange the order of $\psi_{2k-1}$ and $\psi_{2k}$.

We can think of the state $\ket{B_{\bm{s}}(\beta)}$ as a state after projection measurement of thermofield double state \cite{Numasawa:2016emc,Almheiri:2018xdw}:
\be
_L\braket{B_{\bm{s}}|TFD(\beta)}_{LR} = \ket{B_{\bm{s}}(\beta)}_R.
\ee

\subsubsection{conformal limit and the symmetry of correlation functions}

In the conformal limit ($\beta{\cal J}\gg 1$, ${\cal J}|\tau-\tau'|\gg 1$, $\tau,\tau'>-\frac{\beta}{2}$), the correlation function becomes 
\be
G(\tau,\tau') = \f{\braket{B_{\bm{s}}(\beta)|\psi_i(\tau)\psi_i(\tau') |B_{\bm{s}}(\beta)}}{ \braket{B_{\bm{s}}(\beta)|B_{\bm{s}}(\beta)}}= 
c_{\Delta} \Bigg[ \f{\pi}{\mathcal{J}\beta \sin \f{\pi |\tau-\tau'|}{\beta}} \Bigg]^{2\Delta} \text{sgn}(\tau-\tau').
\ee
\be
G_{\text{off}}(\tau,\tau') =  s_k\f{\braket{B_{\bm{s}}(\beta)| \psi_{2k-1}(\tau)\psi_{2k}(\tau')  |B_{\bm{s}}(\beta)} }{  \braket{B_{\bm{s}}(\beta)|B_{\bm{s}}(\beta)} } = 2 i (c_{\Delta})^2  \Bigg[  \f{\pi ^2 }{(\mathcal{J}\beta)^2 \cos \f{\pi \tau}{\beta}  \cos \f{\pi \tau'}{\beta}}   \Bigg] ^{2\Delta}
\ee
Especially, by analytically continuing to real time $\tau \to i t $, we obtain 
\be
G_{\text{off}}(t,t') =  2 i c_{\Delta}^2  \Bigg[  \f{\pi ^2 }{(\mathcal{J}\beta)^2 \cosh \f{\pi t}{\beta}  \cosh \f{\pi t'}{\beta}}   \Bigg] ^{2\Delta}.
\ee
Therefore, for example the spin operator expectation value $\braket{S_k(t)} = -2 i G_{\text{off}}(t,t)$ is 
\be
\braket{S_k(t)} = 4 s_k (c_{\Delta})^2  \Bigg[  \f{\pi }{\mathcal{J}\beta \cosh \f{\pi t}{\beta}  }   \Bigg] ^{4\Delta} \label{eq:expodecayS1},
\ee
which decays exponentially in time $t$.
Under the reparametrization 
\be
\tau_{P} = \f{\pi}{\beta \mathcal{J}^2} \tan \f{\pi \tau}{\beta}, \label{eq:repara1}
\ee
the correlators become 
\be
G(\tau_{P},\tau_{P}') = \f{c_{\Delta}}{|\mathcal{J}(\tau_{P} -\tau_{P}')|^{2\Delta}} \text{sgn}(\tau_{P} -\tau_{P}'),
\ee
\be
G_{\text{off}}(\tau_{P},\tau_{P}') = 2 i (c_{\Delta})^2.
\ee
In this coordinate, it is manifest that the translation $\tau_P \to \tau_P + c$ is a symmetry of both of the diagonal correlator and the off diagonal correlator.
This is the same symmetry of the Poincare patch in AdS$_2$.
Later we consider the gravity setup with similar symmetry.

\subsubsection{evolution under the mass deformed Hamiltonian}
Now we consider the evolution under the mass deformed Hamiltonian:
\be
H_{\text{def}} = H _{SYK} + H_M = i^{\f{q}{2}} \sum_{i_1 < \cdots <i_q }^N J_{i_1\cdots i_q} \psi_{i_1}\cdots \psi_{i_q} + i \mu \sum_{k=1}^{\f{N}{2}} s_k \psi_{2k-1}\psi_{2k}.
\ee
In the low energy limit and small $\mu$ limit, we can treat this deformation as 
\be
\braket{e^{-i  \int dt H_M(t)}} \sim \int \mathcal{D}f e^{i S[f] -i  \int dt \braket{H_M(f(t))}}.
\ee
The Schwartzian action is
\be
S[f] = -\f{N\alpha_S}{\mathcal{J}} \int \{ f, t\} =- \f{N\alpha_S}{\mathcal{J}} \int \Bigl\{ \f{\pi}{\mathcal{J}^2 \beta} \tanh \f{\pi \varphi(t)}{\beta}, t\Bigr\}. 
\ee
The term $\braket{H_M(t)}$ is evaluated as
\ba
\frac{1}{\mu}\f{\braket{B_{\bm{s}}(\beta)|H_M(t) |B_{\bm{s}}(\beta)}}{\braket{B_{\bm{s}}(\beta) |B_{\bm{s}}(\beta)}} &=& i \sum_{k=1}^{\f{N}{2}} s_k\f{\braket{B_{\bm{s}}(\beta)| \psi_{2k-1}(t)\psi_{2k} (t)|B_{\bm{s}}(\beta)}}{\braket{B_{\bm{s}}(\beta) |B_{\bm{s}}(\beta)}}\notag \\
&=& i\f{N}{2 }G_{\text{off}}(t,t)
= -\f{N (c_{\Delta})^2}{[\f{\mathcal{J}\beta}{\pi} \cosh \f{\pi t}{\beta}]^{4\Delta}}.
\ea 
Therefore, the coupling to the reparametrization mode becomes
\be
\frac{H_M(f(t))}{\mu} = - \f{N (c_{\Delta})^2 \varphi'(t)^{2\Delta}}{[\f{\mathcal{J}\beta}{\pi} \cosh \f{\pi \varphi(t)}{\beta}]^{4\Delta}} = -N (c_{\Delta})^2  (f')^{2\Delta}.
\ee
Then, we obtain a Lagrangian for $f(t)$:
\be
S = -\f{N\alpha_S}{\mathcal{J}} \int dt \{f ,t\} + N \mu (c_{\Delta})^2  \int dt (f')^{2\Delta}.
\ee
We can write the Schwartzian term using a Lagrange multiplier $\lambda (t)$ as 
\ba
-\f{N\alpha_S}{\mathcal{J}} \int dt \{f ,t\}
&=& \f{N\alpha_S}{2\mathcal{J}} \int dt \Big[\phi'(t)^2 + \mathcal{J} \lambda(t) (e^{\phi(t)} - f'(t)) \Big].
\ea
When we integrate over $\lambda(t)$, this impose the condition $\phi(t) = \log f'$ and the action reduces to the original one.
By introducing $\hat{\eta} = \f{ \mu (c_{\Delta})^2 }{\mathcal{J}\alpha_S}$, the low energy action becomes 
\be
S = \f{N\alpha_S}{2} \int dt \Bigg[\f{1}{\mathcal{J}} \Big( \f{d \phi}{dt} \Big)^2 + \lambda(t) (e^\phi - f') + 2\mathcal{J}\hat{\eta} e^{2\Delta \phi} \Bigg].
\ee
On the other hand, the initial condition we consider is set by the Euclidean evolution with the Euclidean action 
\be
S = \f{N \alpha_S}{2} \int d\tau \Bigg[ \f{1}{\mathcal{J}} \Big( \f{d \phi}{d\tau} \Big)^2 -\lambda(\tau) (e^{\phi(\tau)} - f'(\tau)) \Bigg].
\ee
Here we put $\hat{\eta} = 0$ because the Euclidean evolution is given by the Hamiltonian without mass deformation.
The solution we are interested in is 
\be
f (\tau) = \f{\pi}{\mathcal{J}^2 \beta} \tan \f{\pi \tau}{\beta}.
\ee
The equation of motion for $\lambda(t)$ gives
\be
e^{\phi(\tau)} = f '(\tau)  = \f{\pi^2}{\mathcal{J}^2\beta^2} \f{1}{\cos ^2 \f{\pi \tau}{\beta}}.
\ee
The EOM for $f$ gives
\be
\lambda'(\tau) = 0.
\ee
Therefore, $\lambda(\tau)$ should be constant.
The equation of motion for $\phi(\tau)$ gives
\be
\f{2}{\mathcal{J}} \f{d^2 \phi }{d \tau^2} + \lambda e^{\phi(\tau)} = 0.
\ee 
This determines $\lambda = - 4 \mathcal{J}$.

\begin{figure}[ht]
\begin{center}
\includegraphics[width=8cm]{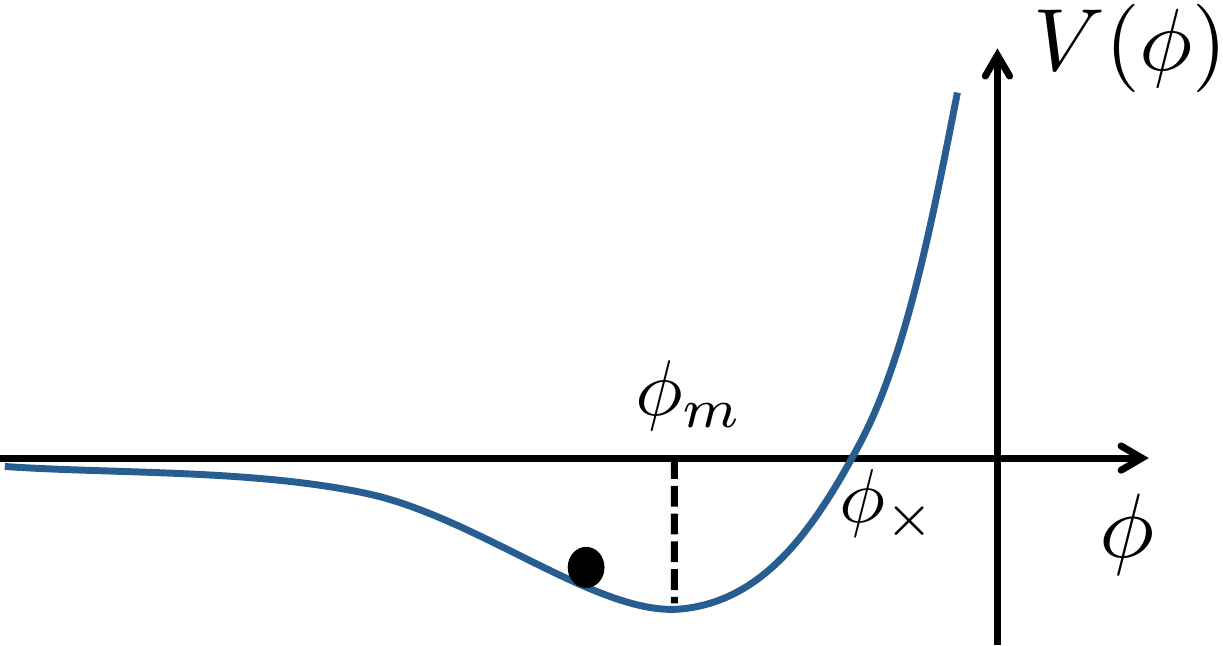}
\caption{A schematic form of the potential for a particle $\phi(t)$.}  
\label{fig:Potential1}
\end{center}
\end{figure}

Now we consider the time evolution with the initial condition $\phi'(0) = 0, e^{\phi(t=0)} =  \f{\pi^2}{\mathcal{J}^2\beta^2}$.
Because the equation of motion for $f$ implies that $\lambda(t)$ is constant, we can set $\lambda = -4\mathcal{J}$.
The Lagrangian is now
\be
S = \f{N\alpha_S}{2} \int dt \Bigg[\f{1}{\mathcal{J}} \Big( \f{d \phi}{dt} \Big)^2 -4\mathcal{J} (e^\phi - f') + 2 \mathcal{J}\hat{\eta} e^{2\Delta \phi} \Bigg].
\ee
Therefore, the evolution is simply given by the motion of a particle with a potential 
\be
V(\phi) = 4\mathcal{J}e^{\phi} - 2\mathcal{J}\hat{\eta} e^{2\Delta \phi}. \label{eq:potentialKM}
\ee
This potential crosses $0$ at $\phi = \phi_{\times}$ that is given by
\be
e^{(1-2\Delta) \phi_{\times} } = \f{\hat{\eta}}{2},
\ee
for $0 < \Delta < \f{1}{2}$.
The bottom of the potential is 
\be
V'(\phi_m) = 0 \qquad \leftrightarrow \qquad 4\mathcal{J} (e^{\phi_m} - \hat{\eta} \Delta e^{2\Delta \phi_m}) = 0,
\ee 
which gives
\be
e^{\phi_m} = (\hat{\eta}\Delta)^{\f{1}{1-2\Delta}}.
\ee
The Lorentzian dynamics is simply described by the particle motion under this potential with the initial condition 
\be
e^{\phi(0)} = e^{\phi_0} = \f{\pi^2}{(\beta \mathcal{J})^2}, \qquad \phi'(0) = 0.
\ee
A schematic form of the potential is described in Fig.~\ref{fig:Potential1}.
When the deformation parameter $\hat{\eta}$ satisfies $\phi_0 < \phi_\times$, the motion of the particle is confined in a finite region.
In the SYK model, this especially means that the expectation value $\braket{S_k(t)}$ does not decay and the system does not thermalize. 
Especially, when $\phi_0 = \phi_m$, the particle sits on the bottom of the potential and does not oscillate.
In this case, using $f' = e^{\phi}$ the expectation value of the spin operator becomes
\be
\braket{S_k(t)} = 4 s_k (c_{\Delta})^2  (f')^{2\Delta} = 4 s_k(c_{\Delta})^2 \Big( \f{\pi}{\beta \mathcal{J}} \Big)^{4\Delta}.
\ee
Therefore, the time evolution keeps the expectation value of the spin operator and prevents thermalization.
$\phi_0 = \phi_m$ gives a relation between $\beta$ and $\mu$, which becomes 
\be
\Big(\f{\pi}{\beta(\mu) \mathcal{J}}\Big)^2 = (\hat{\eta}\Delta)^{\f{1}{1-2\Delta}}\qquad \rightarrow \qquad \f{1}{\beta(\mu)\mathcal{J}} = \f{1}{\pi}\Big( \f{ \mu (c_{\Delta})^2 \Delta}{\mathcal{J}\alpha_S}\Big)^{\f{1}{2(1-2\Delta)}}.
\ee

\subsubsection{the large $q$ limit}
For later purpose, we also consider the large $q$ limit of the pure states $\ket{B_{\bm{s}}(\beta)}$.
The correlators are approximated as 
\ba
G(\tau,\tau') &=& \f{1}{2} \text{sgn}(\tau-\tau') \Big(1 + \f{1}{q} g(\tau,\tau') + \cdots \Big), \notag \\
G_{\text{off}}(\tau,\tau') &=& \f{i}{2}\Big( 1 + \f{1}{q}g_{\text{off}}(\tau,\tau') + \cdots \Big).
\ea
The correlation functions in the large $q$ limit become
\be
e^{g(\tau_1,\tau_2)} = \frac{\check{\alpha}^2}{\mathcal{J}^2\sin^2 (\check{\alpha}|\tau_1 - \tau_2| + \check{\gamma})},
\ee 
\be
e^{g_{\text{off}}(\tau_1,\tau_2)} = \frac{\check{\alpha}^2}{\mathcal{J}^2\cos ^2(\check{\alpha}\tau_1  )}\frac{\check{\alpha}^2}{\mathcal{J}^2\cos^2 (\check{\alpha}\tau_2 )},
\ee 
where $\check{\alpha} = \mathcal{J} \sin \check{\gamma}$, and  $\check{\alpha} \f{\beta}{2} + \check{\gamma} = \f{\pi}{2}$\footnote{In the notation of \cite{Maldacena:2016hyu}, we can write the correlation functions as 
\be
e^{g(\tau_1,\tau_2)} = \Bigg[\frac{\cos \f{\pi v}{2}}{\cos (\pi v(\frac{1}{2} - \frac{|\tau_1 - \tau_2|}{\beta}))} \Bigg]^2,  \qquad e^{g_{\text{off}}(\tau_1,\tau_2)} = \Bigg[\frac{\cos^2 \f{\pi v}{2}}{\cos (\frac{\pi v}{\beta} \tau_1 )\cos (\frac{\pi v}{\beta} \tau_2)} \Bigg]^2,
\ee
where $v \in [0,1]$ and $v$ satisfies $\f{\pi v}{\cos \f{\pi v}{2}} = \mathcal{J}\beta$.
The relation with that in our paper is given by $\check{\alpha} = \f{\pi v }{\beta}$ and $\check{\gamma} = \f{\pi}{2} - \f{\pi v}{2}$.}.
This solution can also be written as 
\be
e^{g(\tau_1,\tau_2)} = \f{\check{h}_1'(\tau_1)\check{h}_2'(\tau_2)}{\mathcal{J}^2 (\check{h}_1(\tau_1) - \check{h}_2(\tau_2))^2 }, \qquad e^{g_{\text{off}}(\tau_1,\tau_2)} = \check{f}_1(\tau_1)\check{f}(\tau_2),
\ee
where 
\ba
\check{h}_1(\tau) &=& \tan \Big(\check{\alpha}\tau + \f{\check{\gamma}}{2} \Big) , \qquad \check{h}_2(\tau) = \tan \Big(\check{\alpha}\tau - \f{\check{\gamma}}{2} \Big), \notag \\
\check{f}_1(\tau) &=& \check{f}_2(\tau) = \f{\check{\alpha}^2}{\mathcal{J}^2 \cos^2(\check{\alpha} \tau) }.
\ea

\subsection{Gravity interpretation of pure states}
According to \cite{Kourkoulou:2017zaj} here we consider the gravity configuration that have features in common with the SYK setup.
Currently we do not know the precise dual gravity theory of the SYK model.
However, the Nearly-AdS$_2$ gravity has some features in common with the low energy limit of the SYK model.
Especially, they share the same low energy theory that is described by the Schwarzian action \cite{Maldacena:2016upp,Maldacena:2016hyu}.
Therefore, we consider the  gravity setup that is similar to the SYK pure states.

\begin{figure}[ht]
\begin{center}
\includegraphics[width=10cm]{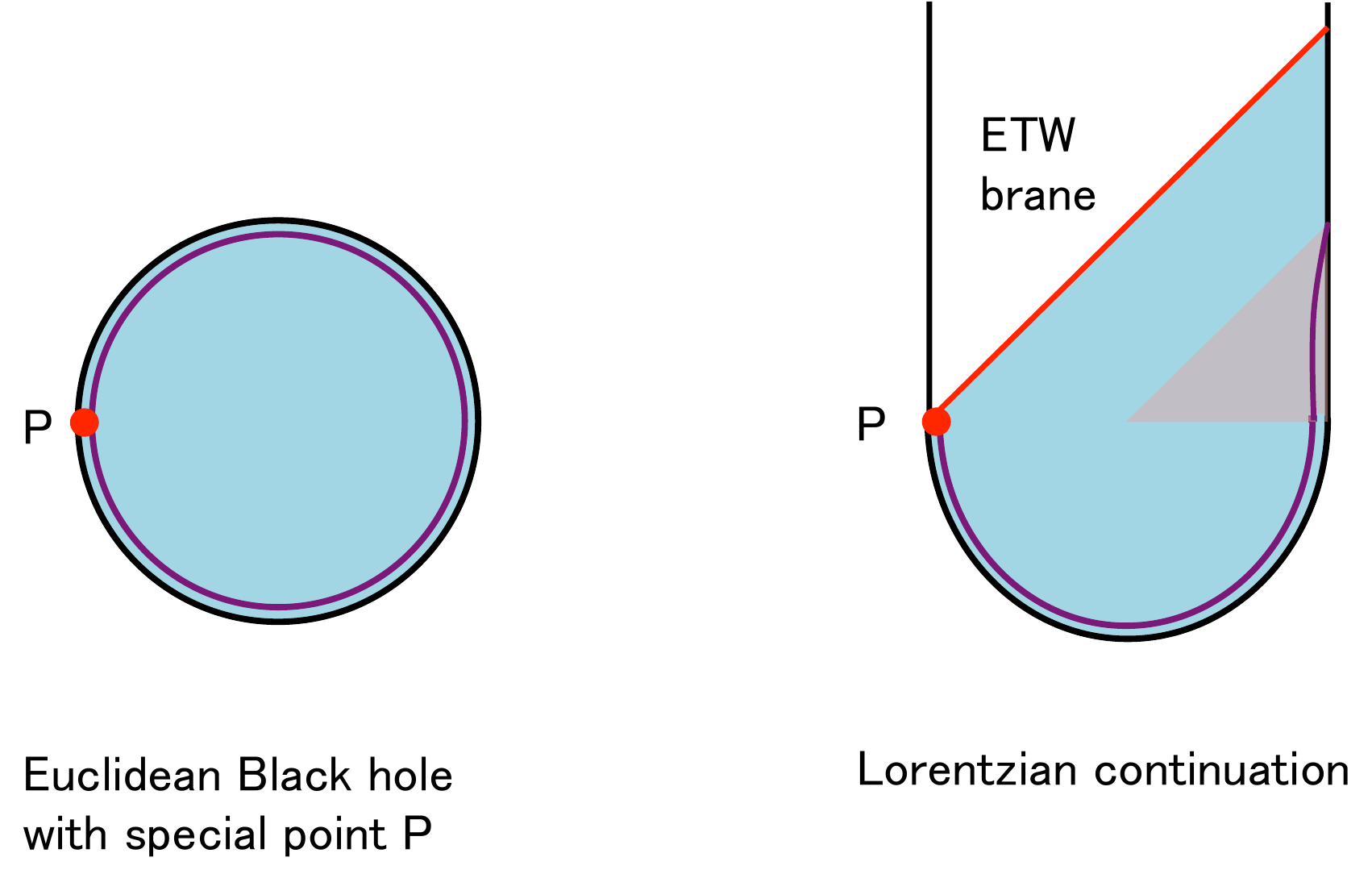}
\caption{The gravity interpretation of the SYK pure states.
The left picture describes the gravity interpretation of pure states in Euclidean signature.
The right picture describes the gravity interpretation in Lorentzian signature.
The purple line is the UV cutoff surface in Nearly AdS$_2$ gravity \cite{Maldacena:2016upp}. }  
\label{fig:KMgeometry1}
\end{center}
\end{figure}

In Euclidean signature, the diagonal correlator is the same with the thermal correlator.
This is interpreted as the Euclidean black hole or hyperbolic disc $H_2$ and we imagine that there is a boundary at some finite but very large circle \cite{Maldacena:2016upp}.
The difference is the existence of the special point $P$ that corresponds to the insertion of projection operator $\ket{B_{\bm{s}}}\bra{B_{\bm{s}}}$.
Imagining the existence of $N$ bulk fields, this is interpreted as the boundary condition that relates the bulk fields in pairs like $\psi_{2k-1} = is_k \psi_{2k}$ at the point $P$.
Except $P$ we impose the same, standard boundary conditions with the thermal case.
Other property is the symmetry of the correlation function.
We saw that both of diagonal and off diagonal correlation function have the symmetry of Poincare patch in AdS$_2$, where the metric is 
\be
ds_E^2 = \f{d\tau_P^2+dz^2 }{z^2}.
\ee
In this coordinate, the special point is sent to infinity $\tau_P =\pm\infty$ and $z= \infty$.
In summary, the Euclidean gravity configuration is the Euclidean black hole with a special point $P$ with boundary conditions on the bulk fields on this point, see Fig.~\ref{fig:KMgeometry1}. 
In nearly AdS$_2$ setup, we interpret this as the special point at large $z$.

\begin{figure}[ht]
\begin{center}
\includegraphics[width=12cm]{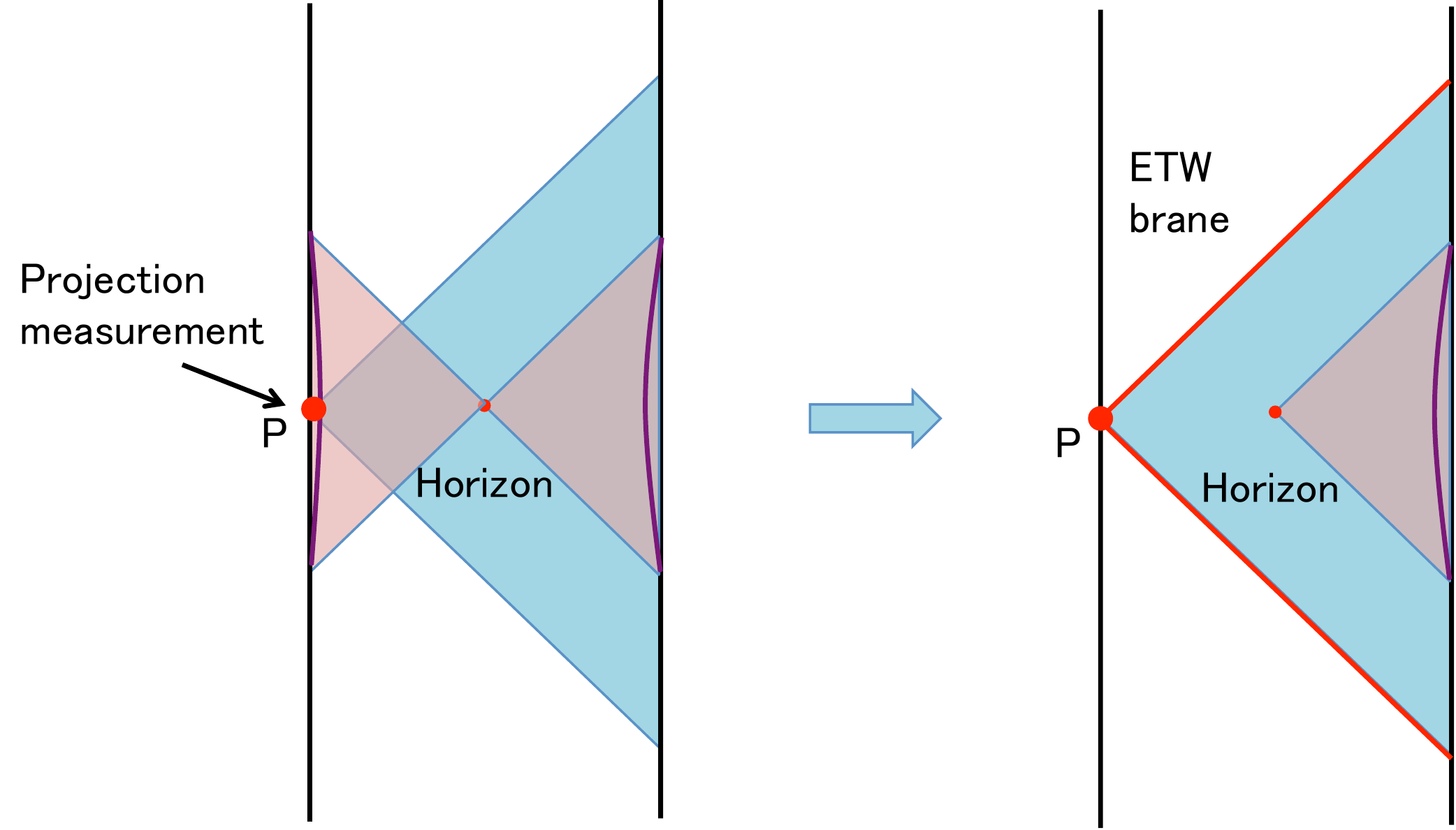}
\caption{The gravity interpretation of the thermofield double states and the projection on them.
Measurements create a shock wave which  propagates along the red line.
}  
\label{fig:projectionG1}
\end{center}
\end{figure}

Next, we consider the Lorentzian continuation.
The AdS$_2$ metric in Poincare coordinate is 
\be
ds_L^2 = \f{-dt_P^2 + dz^2}{z^2} = \f{-dx^+ dx^-}{4(x^+-x^-)^2},
\ee
where we defined $x^{\pm} = z \pm t_P$.
Because of the Poincare time translation symmetry of the SYK correlation function, we are interested in the Lorentzian geometry with this symmetry.
Especially, the boundary condition at special point should be invariant under the Poincare time translation.
This is interpreted as the end of the world line at large $z$ with the same boundary condition with that on the special point $P$.
We can think of this end of the world (EOW) brane as a shock wave that is created by the projection measurement on the left of the thermofield double state and falling to the bulk of AdS$_2$ spacetime \cite{Nozaki:2013wia}, see Fig.~\ref{fig:projectionG1}.

Though Poincare time translation is the symmetry of the diagonal and off diagonal correlation function, the physical time $t$ is related to the Poincare time by the reparametrization (\ref{eq:repara1}).
This corresponds to the Rindler Patch.
The coordinate transformation $t_P = f(t)$ is extended to the bulk by $x_{\pm} = f(y_{\pm})$ where $x_\pm =z \pm t_P $ and $y_{\pm} = X\pm t_R$ with the radial direction $X$ in Rindler patch.

In summary, the Lorentzian configuration consists from the AdS$_2$ geometry  with the end of spacetime at large $z$ with the boundary conditions for bulk fields.
The cutoff boundary is located on the constant $X$.
The Lorentzian configuration are drawn in Fig.~\ref{fig:KMgeometry1}.

\begin{figure}[ht]
\begin{center}
\includegraphics[width=10cm]{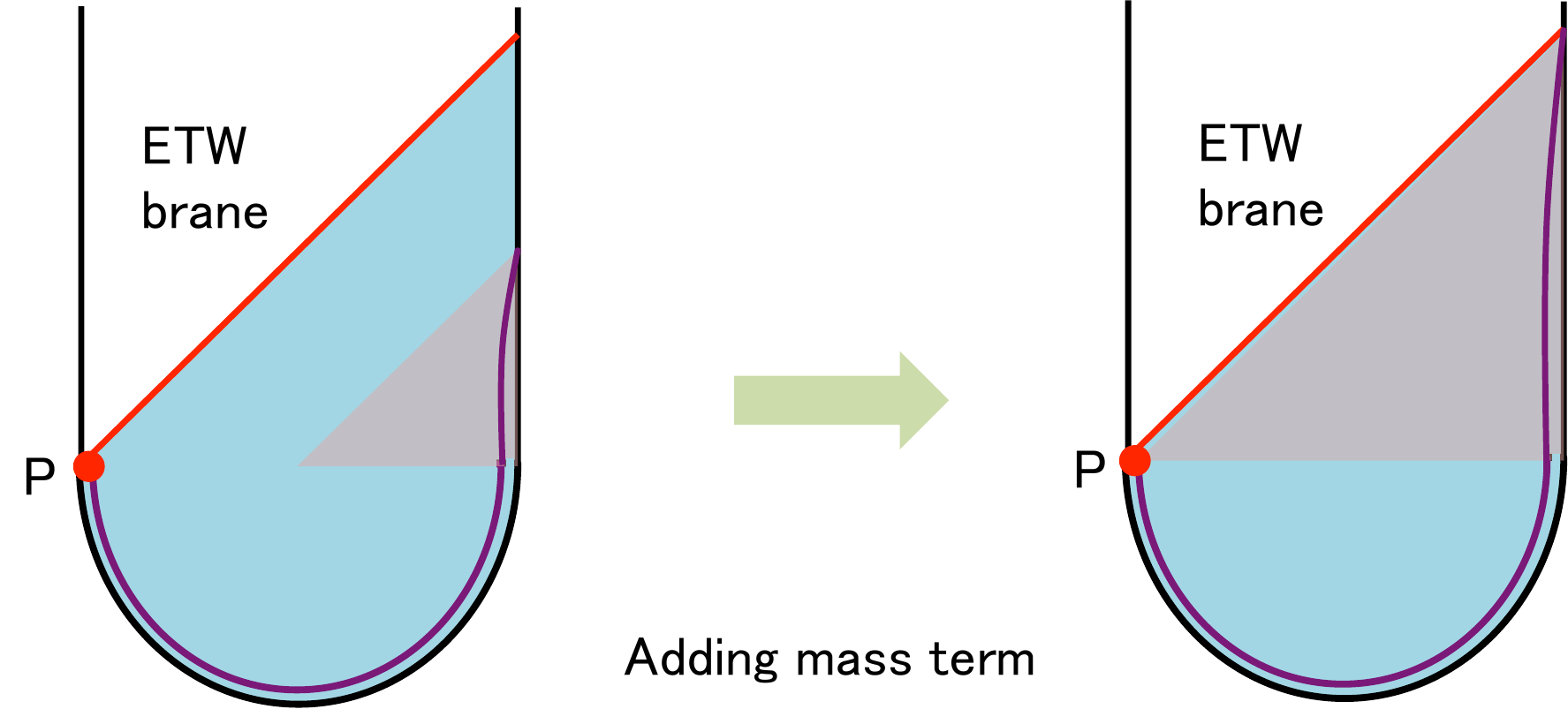}
\caption{The gravity interpretation of evolution in different Hamiltonian.
The left figure is the case where we evolve the state by the SYK Hamiltonian.
The motion of the UV cutoff particle terminates at the finite Poincare time and correspondingly only the inside of the Rindler patch is visible from the boundary.
The right figure is the case where we evolve the state by the mass deformed Hamiltonian.
The motion of the UV cutoff particle extends to whole the Poincare time and whole the spacetime within the EOW brane is visible.}  
\label{fig:KMmass1}
\end{center}
\end{figure}

We can also evolve the SYK model with the mass deformed Hamiltonian (\ref{Hdef}).
In this case, the location of physical boundary is oscillating around the constant $z$ and the coordinate covers whole the Poincare patch.
Therefore, we can see behind the original horizon in the evolution with deformed Hamiltonian  as depicted in Fig.~\ref{fig:KMmass1}.
In gravity side, this interaction is interpreted as a change of boundary conditions on the bulk field on AdS boundary.
These are interpreted as quantum teleportation\cite{Gao:2016bin,Maldacena:2017axo,Numasawa:2016emc}, where we measure the left side of TFD state and then apply the measurement dependent time evolution.

The underlying physics of this teleportation protocol is that we try to put each black hole microstate on a ground state of the deformed Hamiltonian to prevent the black hole generation.
This is the gravity interpretation of preventing thermalization in the SYK.
This essentially depends on how the ground state is close to the ground state of the deformed Hamiltonian and its gap.
This motivate us to study the property of the mass deformed Hamiltonian.
From next section, we study this Hamiltonian in various methods.

\section{Large $N$, finite $q$ analysis}
\label{sec_largeNfiniteqanalysis}
In this section, we analyze the Hamiltonian 
\ba
H_{\text{def}} &=& H _{SYK} + H_M, \notag \\ 
H_{SYK} &=& i^{\f{q}{2}} \sum_{i_1 < \cdots <i_q } J_{i_1\cdots i_q} \psi_{i_1}\cdots \psi_{i_q}, \notag \\
H_M &=& i \mu \sum _{k = 1}^{\f{N}{2}}s_k \psi_{2k-1}\psi_{2k} \equiv -\f{\mu}{2}\sum_{k=1}^{\f{N}{2}} s_kS_{k},
\ea
in the large $N$ limit.
 
Our starting point of the analysis is the Schwinger-Dyson equation for this model with collective degrees of freedom $G, \Sigma$ \cite{PhysRevLett.70.3339,Maldacena:2016hyu}.
According to \cite{Maldacena:2018lmt}, we also introduce these collective variables for off diagonal component.
We study them in Euclidean time.
The effective action in the large $N$ limit is 
\begin{align}
&-S_E \notag   \\ 
= &
\f{N}{2} \log \text{Pf} \Big(
\begin{pmatrix}
1 & 0 \\
0 & 1
\end{pmatrix}
\partial_{\tau}  - 
\begin{pmatrix}
\Sigma & \Sigma_{\text{off}} \\
-\Sigma_{\text{off}}^T & \Sigma
\end{pmatrix}
\Big) \notag \\
& -\f{N}{2} \int d\tau \int d\tau'
\Bigg \{ \f{1}{2}\Tr \Big[
\begin{pmatrix}
\Sigma(\tau,\tau') & \Sigma_{\text{off}}(\tau,\tau') \\
-\Sigma_{\text{off}}(\tau',\tau) & \Sigma(\tau,\tau')
\end{pmatrix}
\begin{pmatrix}
G(\tau,\tau') & -G_{\text{off}}(\tau',\tau) \\
G_{\text{off}}(\tau,\tau') & G(\tau,\tau')
\end{pmatrix}
\Big] - \f{J^2}{q} G(\tau,\tau')^q  \Bigg \} \notag \\
&  - \f{N}{2} i\mu \int d \tau G_{\text{off}}(\tau,\tau).
 \label{eq:effectiveS}
\end{align}
The derivation is shown in the appendix \ref{appendix_ZinGandSigma}.
The Schwinger-Dyson equation arises as the equation of motion for this effective action.
They become \footnote{Using the convolution, we can also write the equation more symbolically as 
\ba
&&\partial_{\tau} G - \Sigma* G +   \Sigma_{\text{off}} * G_{\text{off}}= \delta,  \notag  \\ 
&&\partial_{\tau} G_{\text{off}} -  \Sigma * G_{\text{off}}  -  \Sigma_{\text{off}}*G = 0. \notag 
\ea}
\ba
&&\partial_{\tau} G(\tau,\tau') - \int d\tau'' \Sigma(\tau,\tau'') G(\tau'',\tau') + \int d\tau'' \Sigma_{\text{off}}(\tau,\tau'') G_{\text{off}}(\tau'',\tau') = \delta (\tau - \tau'), \label{eq:SD1} \\ 
&&\partial_{\tau} G_{\text{off}}(\tau,\tau') - \int d\tau'' \Sigma(\tau,\tau'') G_{\text{off}}(\tau'',\tau')  - \int d\tau'' \Sigma_{\text{off}}(\tau,\tau'') G(\tau'',\tau') = 0, \label{eq:SD2}
\ea
and 
\ba
\Sigma(\tau,\tau') &=& J^2 G(\tau,\tau')^{q-1} = \f{\mathcal{J}^2}{q} (2G(\tau,\tau'))^{q-1}, \label{eq:selfe1}
 \\ 
\Sigma_{\text{off}}(\tau,\tau') &=& - i\mu \delta(\tau-\tau'). \label{eq:selfe2}
\ea
By substituting $\Sigma_{\text{off}}(\tau,\tau') =  - i\mu \delta(\tau-\tau')$ into (\ref{eq:SD1}) and (\ref{eq:SD2}), we obtain 
\ba
&&\partial_{\tau} G(\tau,\tau') - \int d\tau'' \Sigma(\tau,\tau'') G(\tau'',\tau') -  i\mu   G_{\text{off}}(\tau,\tau') = \delta (\tau - \tau'),  \\ \label{eq:SD3}
&&\partial_{\tau} G_{\text{off}}(\tau,\tau') - \int d\tau'' \Sigma(\tau,\tau'') G_{\text{off}}(\tau'',\tau')  + i \mu G(\tau,\tau') = 0, \label{eq:SD4}
\ea
with $\Sigma(\tau,\tau') = J^2 G(\tau,\tau')^{q-1}$.
It is also useful to rewrite the Schwinger-Dyson equation in the frequency space.
In this representation, we can decouple $G_{\text{off}}$ from the diagonal part $G, \Sigma$.
The equation becomes 
\ba
G(\omega) &=& -\f{i\omega + \Sigma(\omega)}{(i\omega + \Sigma(\omega))^2 - \mu^2}, \notag \\
G_{\text{off}}(\omega) &=& \f{-i\mu}{(i\omega + \Sigma(\omega))^2 - \mu^2}.\label{eq:SDqrFreq}
\ea
We can determine $G_{\text{off}}(\omega)$ after solving the equation for diagonal part $G,\Sigma$.
In the finite temperature, the frequency  $\omega$ is quantized to the Matsubara frequency $\omega _n = \f{2\pi}{\beta}(n + \f{1}{2})$.

Once we solve the Schwinger-Dyson equation, the energy can be calculated from the Green functions in the following way:
\be
\f{E}{N} = \frac{\langle H_\text{SYK}+H_M\rangle}{N} = \f{1}{q} \partial_{\tau} G(\tau,0)\Big |_{\tau = 0^+}  + i\f{\mu}{2}\Big( 1 -\f{2}{q} \Big)G_{\text{off}}(0,0) . \label{eq:gsenergy}
\ee
This is derived from 
\ba
N\partial_{\tau}G(\tau,0)\Big | _{\tau = 0^+} &=& \sum_i \braket{\partial_{\tau}\psi_i\psi_i} = \sum_i\braket{[H,\psi_i]\psi_i} = \braket{q H_{SYK} + 2H_M}, \notag \\
\f{N}{2} G_{\text{off}}(0,0) &=& \sum_{k =1}^{\f{N}{2}}s_k \braket{\psi_{2k-1}\psi_{2k}} = \f{1}{i\mu} \braket{H_M}.
\ea
This is the exact relation between the energy and correlation function for the deformed SYK model even before the large $N$ limit or the disorder average.

We can also rewrite the energy using the Schwinger-Dyson equation as  
\ba
\f{1}{q} \partial_{\tau} G(\tau,0) |_{\tau \to 0^+} &=& \f{1}{q}\int d\tau'' \Sigma(0,\tau'') G(\tau'' ,0) + i\f{\mu}{q} G_{\text{off}}(0,0) \notag \\
&=& -\f{\mathcal{J}^2}{2q^2} \int d\tau'' (2G(\tau'',0))^q + i\f{\mu}{q} G_{\text{off}}(0,0)
.
\ea
Here we used $\Sigma(\tau_1,\tau_2) = -\Sigma(\tau_2,\tau_1)$.
Therefore, in the large $N$ limit the energy becomes
\ba
\f{E}{N} &=&  \f{1}{q} \partial_{\tau} G(\tau,0)\Big |_{\tau = 0^+}  + i\f{\mu}{2}\Big( 1 -\f{2}{q} \Big)G_{\text{off}}(0,0) 
 \notag\\
 &=&  -\f{\mathcal{J}^2}{2q^2} \int d\tau (2G(\tau,0))^q  + i\f{\mu}{2}G_{\text{off}}(0,0) .
\ea
This expression is useful when we compute the free energy numerically.

\subsection{Solving the model in the conformal limit}
In this section, we study the ground state of the Hamiltonian $H_{\text{def}}$.
We can study the Schwinger-Dyson equation (\ref{eq:SD4}) numerically where the detail of the numerical calculation is shown in the appendix \ref{sec_solveSDnumerical}.
From the numerical analysis for various parameter regions of $\mu$ and $q$, we confirmed that the system has a mass gap above the ground state.
We also find that the numerical solution agrees well with the correlation function that is obtained from the reparametrization $f(\tau) = \tanh (\alpha \tau)$ of the SYK correlation function in the small $\mu$ limit.
This is expected since the long time behavior of the SYK model is controlled by the reparametrization (or conformal symmetry) \cite{Maldacena:2018lmt,Maldacena:2016hyu} and the mass term affects the long time behavior in small $\mu$ limit.
Therefore, in this section we consider to solve the mass deformed theory using the approximate conformal symmetry of the SYK model.

\begin{figure}[ht]
\begin{center}
\includegraphics[width=10cm]{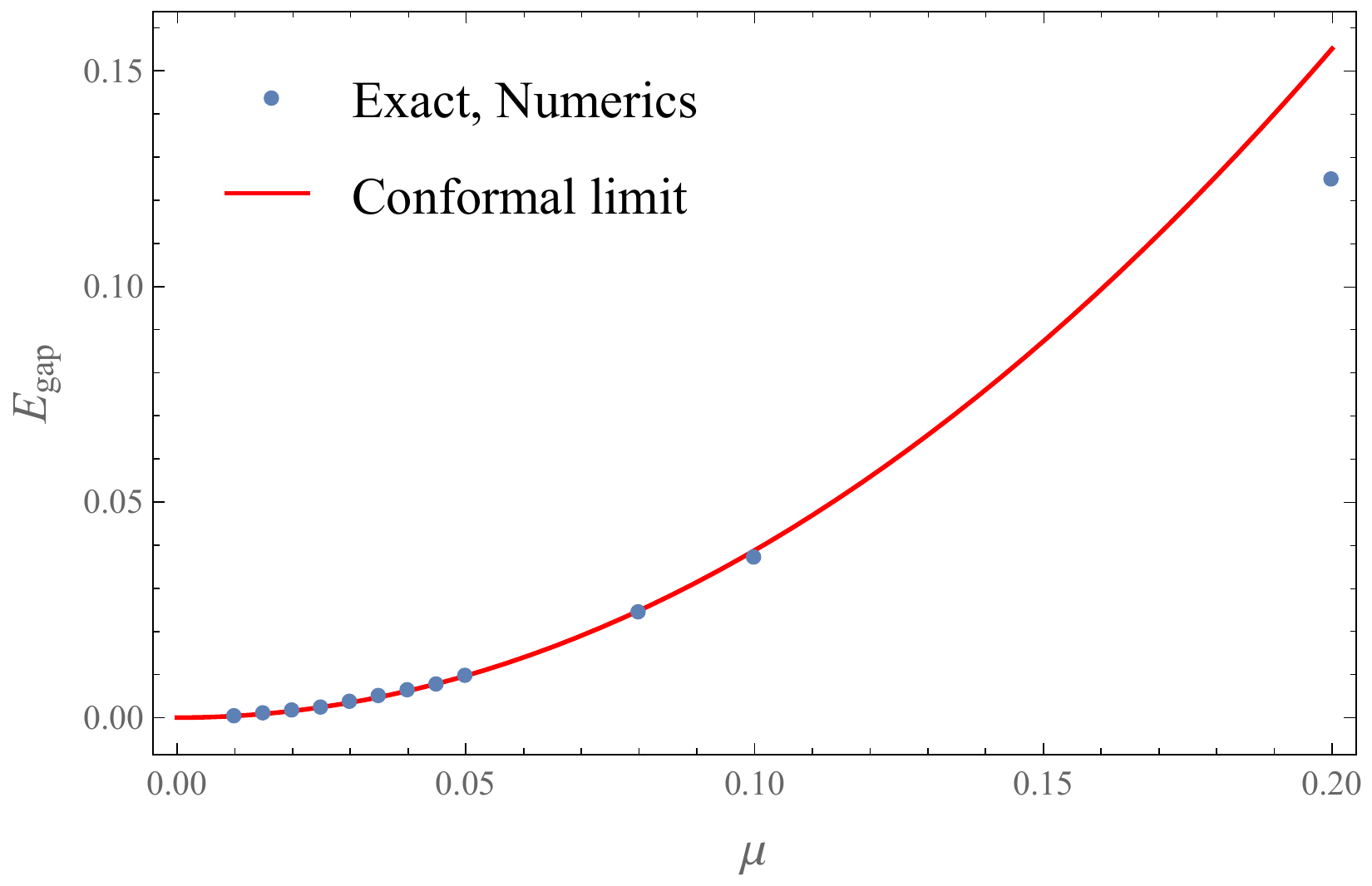}
\caption{The plot of the mass gap $E_{\text{gap}}$, which is defined as the exponential decay rate $G(\tau) \sim e^{-E_{\text{gap}}\tau}$ of the correlation functions, for $q=4, J = 1$ case.
In the conformal limit, the mass gap is given by $E_{\text{gap}} = 2\alpha \Delta$.
For small $\mu$, the result in the conformal limit agrees with the numerics well.
}  
\label{fig:Egapq4plot}
\end{center}
\end{figure}

The diagonal correlation function in the conformal limit is given by
\ba
G_c(\tau-\tau') &=& c_{\Delta}  \Big( \f{\alpha}{\mathcal{J} \sinh \alpha |\tau-\tau'|}\Big)^{2\Delta} \text{sgn}(\tau-\tau') \notag \\
 &=&  c_{\Delta}\Big( \f{f'(\tau)f'(\tau')}{\mathcal{J} |f(\tau) - f(\tau')|^2} \Big)^{\Delta}\text{sgn}(\tau-\tau'), \qquad f(\tau) = \tanh (\alpha \tau). 
\ea
\ba
\Sigma_c(\tau-\tau') 
&=& \f{\mathcal{J}^2}{q}(2c_{\Delta})^{q-1}\Big( \f{f'(\tau)f'(\tau')}{\mathcal{J} (f(\tau) - f(\tau'))^2} \Big)^{(1-\Delta)}\text{sgn}(\tau-\tau').
\ea
$\alpha$ is a function of $\mu$ with $\alpha(\mu = 0) =0$, which we will determine later. 
The Fourier transformation of the conformal limit correlation function becomes 
\ba
G_c(\omega) 
 &=&  c_{\Delta}2^{2\Delta }i \f{\alpha^{2\Delta -1}}{\mathcal{J}^{2\Delta}}\Gamma(1-2\Delta) \f{\cos \pi \Delta}{\pi }\Gamma\Big(\Delta + i\f{\omega}{2\alpha}\Big)\Gamma\Big(\Delta - i\f{\omega}{2\alpha}\Big) \sinh \f{\pi \omega}{2\alpha}, \label{eq:Gfreqconf}
\ea
and 
\begin{align}
&\Sigma_c(\omega) \notag \\ 
=& \f{\mathcal{J}^2}{q}(2c_{\Delta})^{q-1}2^{2(1-\Delta)}  \f{\alpha^{1-2\Delta}}{\mathcal{J}^{2(1-\Delta)}} i \Gamma(2\Delta-1)\f{\cos \pi(1-\Delta)}{\pi} \Gamma\Big(1-\Delta + i\f{ \omega}{2\alpha}\Big) \Gamma\Big(1-\Delta - i\f{ \omega}{2\alpha}\Big) \sinh \f{\pi \omega}{2\alpha}. \notag \\ \label{eq:Sigfreqconf}
\end{align}
We can easily confirm that in the limit $\omega \gg \alpha$ these reduce to the conformal limit of the SYK ground state correlation function
\be
G_c^{SYK}(\omega) = i c_{\Delta}\f{1}{\mathcal{J}^{2\Delta}} 2^{1-2\Delta}\s{\pi} \f{\Gamma(1-\Delta)}{\Gamma(\f{1}{2} + \Delta)} |\omega|^{2\Delta -1 } \text{sgn}(\omega),
\ee
\be
\Sigma_c^{SYK}(\omega) =  i\f{\mathcal{J}^2}{q}(2c_{\Delta})^{q-1}\f{1}{\mathcal{J}^{2(1-\Delta)}} 2^{2\Delta-1}\s{\pi} \f{\Gamma(\Delta)}{\Gamma(\f{3}{2} - \Delta)} |\omega|^{1-2\Delta}\text{sgn}(\omega).
\ee
This says that only the low frequency part $\omega \ll \alpha$ is affected by the mass term, as we expected.
This also implies that the solution satisfies the Schwinger-Dyson equation (\ref{eq:SDqrFreq}) in the regime $\alpha \ll \omega \ll \mathcal{J}$ where we can ignore the mass term $\mu$ and the UV term $\partial _{\tau}$.
Now, we solve the Schwinger-Dyson equation at $\omega \ll \alpha$.
In this regime, the $G_c(\omega) $ and $\Sigma_c(\omega)$ are linear in $\omega$.
However, the slope is very large and we can ignore the first term $\omega$ in $\omega + \Sigma(\omega)$ in (\ref{eq:SDqrFreq}).
Therefore, we can approximate the Schwinger-Dyson equation for diagonal part as 
\be
-\Sigma(\omega)  + \f{\mu^2}{\Sigma(\omega)} = \f{1}{G(\omega)}.
\ee
Because $\Sigma(\omega)$ is small in $\omega \ll \alpha$, we can also ignore the first term $\Sigma(\omega)$.
Then we solve the Schwinger-Dyson equation in the leading of $\omega$ expansion by inserting the expression for $G_c(\omega)$ and $\Sigma_c(\omega)$ (\ref{eq:Gfreqconf}) and (\ref{eq:Sigfreqconf}).
When we expand as $G_c(\omega) = g_c(\alpha)\omega + \cdots$ and $\Sigma_c(\omega) = \sigma_c(\alpha)\omega + \cdots$, the Schwinger-Dyson equation gives
\be
\f{\sigma_c(\alpha)}{g_c(\alpha)} = \mu^2. 
\ee
This determines $\alpha$ as a function of $\mu$ as 
\ba
\Big(\f{2\alpha}{\mathcal{J}} \Big)^{2(1-2\Delta)} =  
 \f{\Gamma(2-2\Delta)\Gamma(\Delta)^2}{\Gamma(2\Delta +1 )\Gamma(1-\Delta)^2} \f{1}{(2c_\Delta)^{(q-2)}}\Big(\f{\mu}{\mathcal{J}} \Big)^2,
\ea
or 
\be
\alpha(\mu) =  \f{1}{2} \mathcal{J}  \Bigg[  \f{\Gamma(2-2\Delta)\Gamma(\Delta)^2}{\Gamma(2\Delta +1 )\Gamma(1-\Delta)^2} \f{1}{(2c_\Delta)^{(q-2)}}\Bigg]^{\f{1}{2(1-2\Delta)}}\Big(  \f{\mu}{\mathcal{J}}\Big)^{\f{1}{1-2\Delta}} . \label{eq:EgapConf}
\ee
The power of $\mu$ is given by $\f{1}{1-2\Delta}$, which is always larger than $1$.
Therefore, in the low energy limit the physical mass gap is much smaller than the naive mass gap $\mu$.
This is in contrast with the two coupled SYK model \cite{Maldacena:2018lmt} where the physical mass gap is much greater than the naive gap $\mu$.
We also compute the mass gap numerically and for small $\mu$ the numerics agrees with the conformal limit result (\ref{eq:EgapConf}).
See Fig.~\ref{fig:Egapq4plot}.

\begin{figure}[ht]
\begin{minipage}{0.49\hsize}
\begin{center}
\includegraphics[width=8cm]{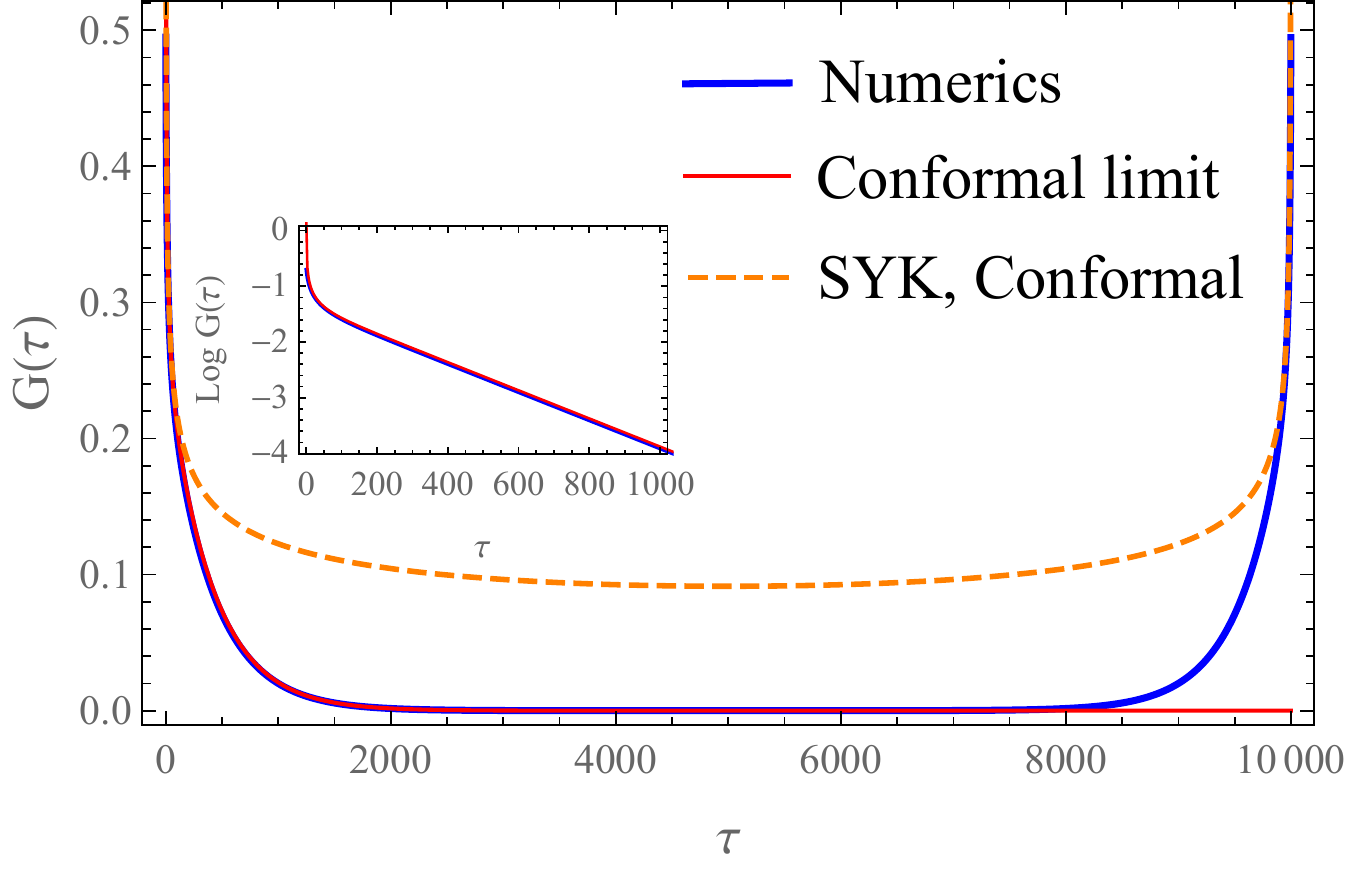} 
\end{center}
\end{minipage}
\begin{minipage}{0.49\hsize}
\begin{center}
\includegraphics[width=8cm]{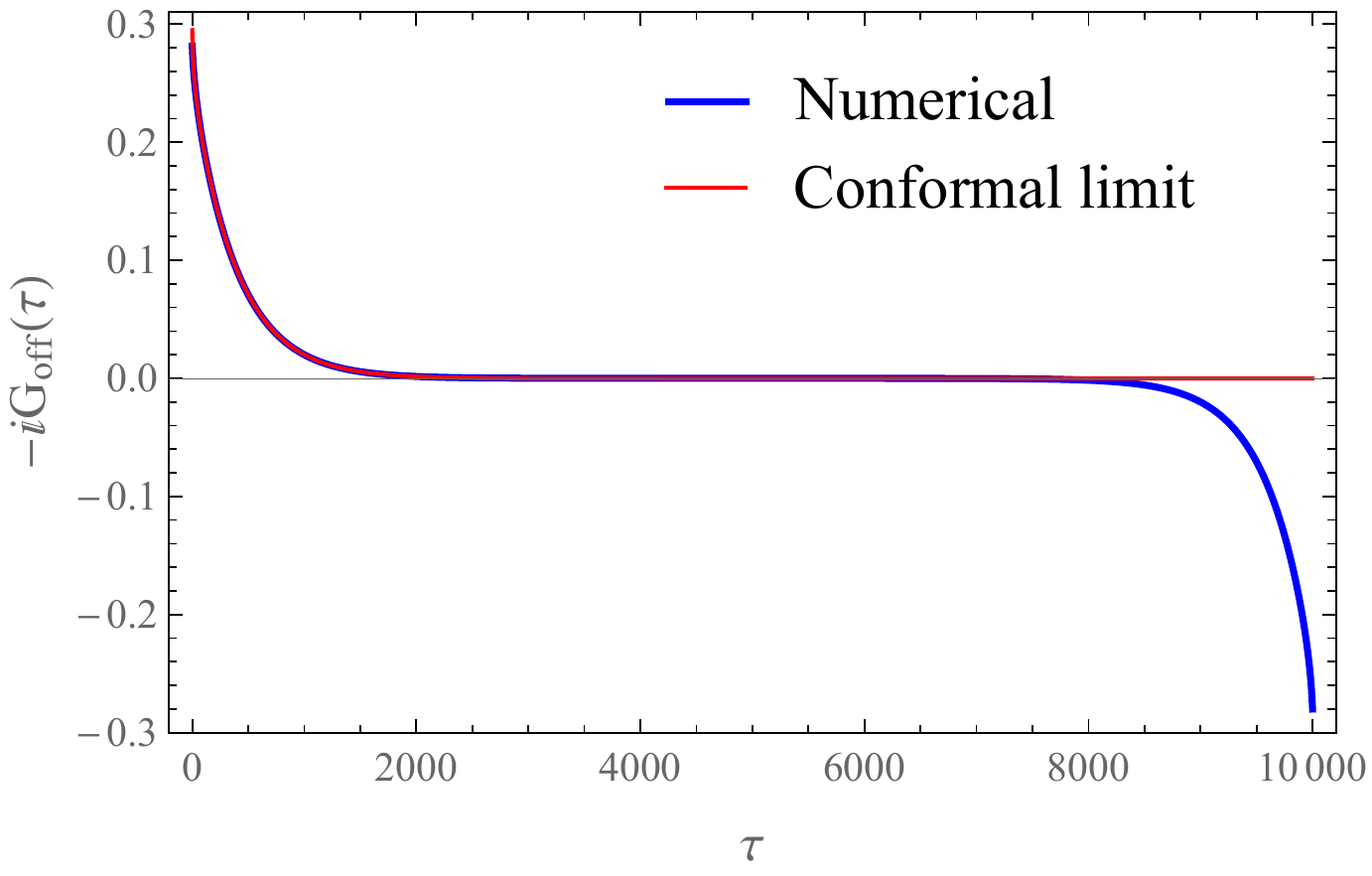}
\end{center}
\end{minipage}
\caption{The plot of the correlation functions for $q = 8$, $\beta = 10000$, $J = 1$ and $\mu = 0.005$  mass deformed SYK model.
{\bf Left}: The plot of the diagonal correlation functions. We plot the numerical solution for the Schwinger-Dyson equation, the conformal limit and the conformal limit of the SYK model. 
{\bf Right:} The plot of the off diagonal correlation functions. We plot the numerical solution and the conformal limit. }  
\label{fig:plotgdcor1}
\end{figure}

Once we determine the conformal limit of the diagonal correlation functions, we can also determine the off diagonal correlation function.
It is convenient to rewrite the Schwinger-Dyson equation as 
\be
G_{\text{off}}(\omega) = \f{i\mu G(\omega)}{i\omega + \Sigma(\omega)}.
\ee
In the conformal limit, we can ignore the $\omega$ in the denominator and approximate $G, \Sigma$ by the conformal limit $G_c(\omega), \Sigma_c(\omega)$.
Therefore, $G_{\text{off}}(\omega)$ becomes
\be
G_{\text{off}}(\omega)  = i \mu \f{G_c(\omega)}{\Sigma_c(\omega)} = i \mu^{-1} \f{\Gamma(1-\Delta)^2}{\Gamma(\Delta)^2}  \f{\Gamma(\Delta + i\f{\omega}{2\alpha})\Gamma(\Delta - i\f{\omega}{2\alpha})}{\Gamma(1-\Delta + i\f{\omega}{2\alpha})\Gamma(1-\Delta - i\f{\omega}{2\alpha})} \label{eq:Goffomega}
\ee
The Euclidean time off diagonal correlator is obtained by the inverse Fourier transformation of $G_{\text{off}}(\omega)$.
This inverse Fourier transformation becomes 
\be
G_{\text{off}}(\tau) = 2i \alpha(\mu) \mu^{-1} \f{\Gamma(1-\Delta)^2}{\Gamma(\Delta)^2} \f{\Gamma(2\Delta)}{\Gamma(1-2\Delta)} e^{- 2\alpha \Delta |\tau|} {}_2F_1(2\Delta,2\Delta;1;e^{-2\alpha|\tau|}).
\ee
We compare the conformal limit and the exact numerical solution for the Schwinger-Dyson equation in Fig.~\ref{fig:plotgdcor1} and they show good agreements.

The $\tau =0$ value of the off diagonal correlator gives the expectation value of the spin operator $S_k = -2i \psi_{2k-1}\psi_{2k}$ .
In the conformal limit, this becomes \footnote{The result (\ref{eq:SpinExConf}) contains $\Gamma(1-4\Delta)$, which is divergent when $q=4$.
This means that the spin operator expectation value is not determined in the conformal limit but is regulated by the UV effect.
As a consequence, the scaling behavior with respect to $\mu$ is violated in $q=4$ case.
We treat this case in the appendix \ref{sec:q4case}. }
\be
\braket{G_{\bm{s}}(\mu)|S_k|G_{\bm{s}}(\mu)} = -2i s_k G_{\text{off}}(0) = 4 s_k \alpha(\mu) \mu^{-1} \f{\Gamma(1-\Delta)^2\Gamma(2\Delta)\Gamma(1-4\Delta)}{\Gamma(\Delta)^2\Gamma(1-2\Delta)^3}. \label{eq:SpinExConf}
\ee
Using $G_{\text{off}}(0)$, we can calculate the ground state energy:
\be
\f{1}{N}\mu\f{\partial E_0(\mu)}{\partial\mu} = \mu\f{i}{2}G_{\text{off}}(0)   = -\alpha(\mu) \f{\Gamma(2\Delta)\Gamma(1-\Delta)^2\Gamma(1-4\Delta)}{\Gamma(\Delta)^2\Gamma(1-2\Delta)^3}. 
\ee
The first relation comes from the relation for the free energy $\f{1}{N}\f{\partial (\beta F)}{ \partial \mu} =  \f{i \beta}{2} G_{\text{off}}(0)$ and by specializing this relation to the ground state $\beta \to \infty$.
By integrating this differential equation, we obtain the ground state energy as 
\be
\f{E_0(\mu)}{N} =  \f{E_0}{N} -\alpha(\mu) (1-2\Delta)\f{\Gamma(2\Delta)\Gamma(1-\Delta)^2\Gamma(1-4\Delta)}{\Gamma(\Delta)^2\Gamma(1-2\Delta)^3}, \label{eq:GDEnConf} 
\ee
where $E_0$ is the ground state energy of the SYK model.
Using the relation $H_{\text{def}} = H_{SYK} + H_M$, we can also compute the expectation value of the SYK Hamiltonian under the ground state of the deformed Hamiltonian as
\begin{align}
\f{1}{N}\braket{G_{\bm{s}}(\mu)|H_{SYK}|G_{\bm{s}}(\mu)} &= \f{E_0(\mu)}{N}  -i \f{\mu}{2}G_{\text{off}}(0)  \notag \\
&=\f{E_0}{N}+\alpha(\mu)\f{\Gamma(2\Delta+1)\Gamma(1-\Delta)^2\Gamma(1-4\Delta)}{\Gamma(\Delta)^2\Gamma(1-2\Delta)^3}. \label{eq:SYKEnConf}
\end{align}
The $\ket{G_{\bm{s}}(\mu)}$ has larger energy than the SYK ground state and the energy expectation value of $\ket{G_{\bm{s}}(\mu)}$ does not depend on $\bm{s}$. 
Therefore we can prepare $2^{\f{N}{2}}$ (=dimension of the SYK Hilbert space) states from the mass deformation with the same energy expectation value.

\subsection{variational approximation for the ground state}
\label{sec_variationalansatz}
To study how the SYK ``black hole microstate" is close to the ground state of the deformed Hamiltonian, we apply the variational method for the deformed Hamiltonian by the microstate $\ket{B_{\bm{s}}(\beta)}$. 
This is an SYK analog of variational approximation by smeared boundary states for mass deformations of $(1+1) d$  CFT \cite{Cardy:2017ufe}.

For variational approximation, we need to evaluate the mass deformed Hamiltonian in the microstate $\ket{B_{\bm{s}}(\beta)}$.
Here we use the same collection of spins $\bm{s} = \{s_1,\cdots , s_\f{N}{2}\}$ with the mass deformation $H_M=\frac{\mu}{2}\sum_ks_kS_k$.
Using the relation
\be
N \partial_{\tau} G(\tau,0)|_{\tau = 0} = \sum_ i \f{\braket{B_{\bm{s}}(\beta)|\partial_\tau \psi _i  \psi_i| B_{\bm{s}}(\beta)} }{\braket{B_{\bm{s}}(\beta)|B_{\bm{s}}(\beta)}} = \sum_ i \f{\braket{B_{\bm{s}}(\beta)|[H_{SYK},\psi _i]  \psi_i| B_{\bm{s}}(\beta)} }{\braket{B_{\bm{s}}(\beta)|B_{\bm{s}}(\beta)}} = \braket{q H_{SYK}}_{B_{s}},
\ee
\be
\f{N}{2} G_{\text{off}}(0,0) = \sum_{ k =1}^{\f{N}{2}} \f{\braket{B_{\bm{s}}(\beta)| \psi _{2k-1}  \psi_{2k}| B_{\bm{s}}(\beta)} }{\braket{B_{\bm{s}}(\beta)|B_{\bm{s}}(\beta)}} = \sum_{k = 1}^{\f{N}{2}} \braket{\psi_{2k-1}\psi_{2k}}_{B_{\bm{s}}} = \f{1}{i\mu} \braket{H_M}_{B_{\bm{s}}},
\ee
we can compute the expectation value of the mass deformed Hamiltonian $\braket{H_{SYK} + H_M}_{B_{\bm{s}}}$ as 
\be
\f{\braket{H_{SYK} + H_M}_{B_{\bm{s}}}}{N} = \f{1}{q} \partial_{\tau}  G(\tau,0)|_{\tau \to 0_+} + i\f{\mu}{2} G_{\text{off}}(0,0) \label{eq:trialenergy1}.
\ee
Here correlation functions are evaluated in the state $\ket{B_{\bm{s}}(\beta)}$.
Using the equation (\ref{eq:BdiaglargeN}) and (\ref{eq:BofflargeN}), we can represent this expectation value completely in terms of the SYK thermal correlation function:
\be
\f{\braket{H_{SYK} + H_M}_{B_{\bm{s}}}}{N}
= \f{1}{q} \partial_{\tau}  G_{\beta}(\tau)|_{\tau \to 0_+} -\mu G_{\beta}(\beta/2)^2. \label{eq:trialenergy}
\ee
The first term is the thermal energy in the SYK model \cite{Maldacena:2016hyu}:
\be
\f{1}{q} \partial_{\tau}  G_{\beta}(\tau)|_{\tau \to 0_+} = -\f{\mathcal{J}^2}{2q^2} \int _0^{\beta} (2G_{\beta}(\tau))^q = -\f{\partial}{\partial \beta}\log Z = E.
\ee
As usual, we minimize the energy evaluated on the trial wavefunction (\ref{eq:trialenergy}), to achieve the best approximation for ground state energy.

\subsubsection{variational approximation in conformal limit}

In the low energy limit, the partition function have the expansion \cite{Maldacena:2016hyu}
\be
\log Z = -\beta E_0 + S_0 + \f{c}{2\beta} + \cdots.
\ee
Here $c = \f{4\pi^2 \alpha_S N}{\mathcal{J}}$ is the specific heat of the SYK model and $E_0, S_0$ are the ground state energy and the zero temperature entropy in the SYK model that is not calculated analytically.
Therefore the energy expectation value becomes 
\be
\f{\braket{H_{SYK}}}{N} = -\f{\partial }{\partial \beta} \log Z = \f{E_0}{N} + \f{c}{2 \beta^2 N} = \f{E_0}{N} + \f{2 \pi^2 \mathcal{J} \alpha _S}{(\beta\mathcal{J})^2 }.
\ee
On the other hand, at low energy limit $G_{\beta}(\beta/2) = c_{\Delta} \big( \f{\pi}{\mathcal{J}\beta}\big)^{2\Delta}$.
Therefore, the expectation value of the deformation term becomes 
\be
\f{\braket{H_M}}{N} = -\mu G_{\beta}(\beta/2)^2 = -\mu (c_{\Delta})^2 \Big( \f{\pi}{\mathcal{J}\beta}\Big)^{4\Delta}.
\ee
Therefore, the total variational energy is 
\ba
\f{\braket{H_{SYK} + H_M}}{N} - \f{E_0}{N}&=& \f{2 \pi^2 \mathcal{J} \alpha _S}{(\beta\mathcal{J})^2 } -\mu (c_{\Delta})^2 \Big( \f{\pi}{\mathcal{J}\beta}\Big)^{4\Delta} \notag \\
&=&  \mathcal{J}\alpha_S ( 2 e^{\phi_0}- \hat{\eta} e^{2\Delta \phi_0}) \equiv V(\phi_0).
\ea
Here we put $e^{\phi_0} = \f{\pi^2}{\mathcal{J}^2 \beta^2 }$ and $\hat{\eta} = \f{\mu(c_{\Delta})^2 }{\mathcal{J}\alpha_S}$.
We should note that this potential is exactly the same with (\ref{eq:potentialKM}).
The derivative becomes $\beta \partial_{\beta} = - 2\partial_{\phi_0}$ and the minimal value of the variational energy is the minimal value of the potential $V$.
This potential has a unique minimal that is given by
\be
V'(\phi_0) =2 \mathcal{J}\alpha_S  (e^{\phi_0} - \hat{\eta}\Delta e^{2\Delta \phi_0}) = 0.
\ee
Therefore, the relation between $\beta$ and $\mu$ becomes 
\be
e^{\phi_0/2} =\Big( \f{\pi}{\mathcal{J}\beta(\mu)}\Big) = \Big(\f{\mu (c_{\Delta})^2 \Delta}{\mathcal{J}\alpha_S}\Big) ^{\f{1}{2(1-2\Delta)}}. \label{eq:valbetamulow}
\ee  

\begin{figure}[ht]
\begin{minipage}{0.49\hsize}
\begin{center}
\includegraphics[width=8cm]{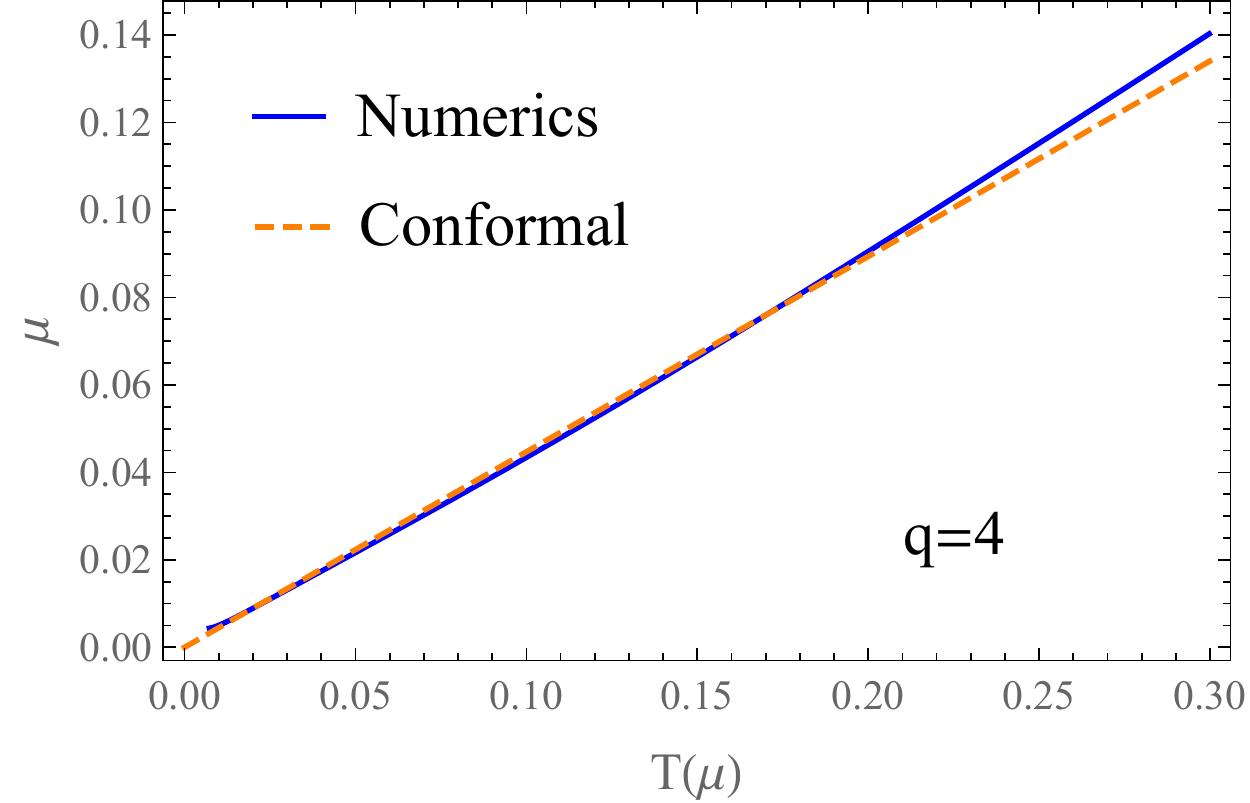} 
\end{center}
\end{minipage}
\begin{minipage}{0.49\hsize}
\begin{center}
\includegraphics[width=8cm]{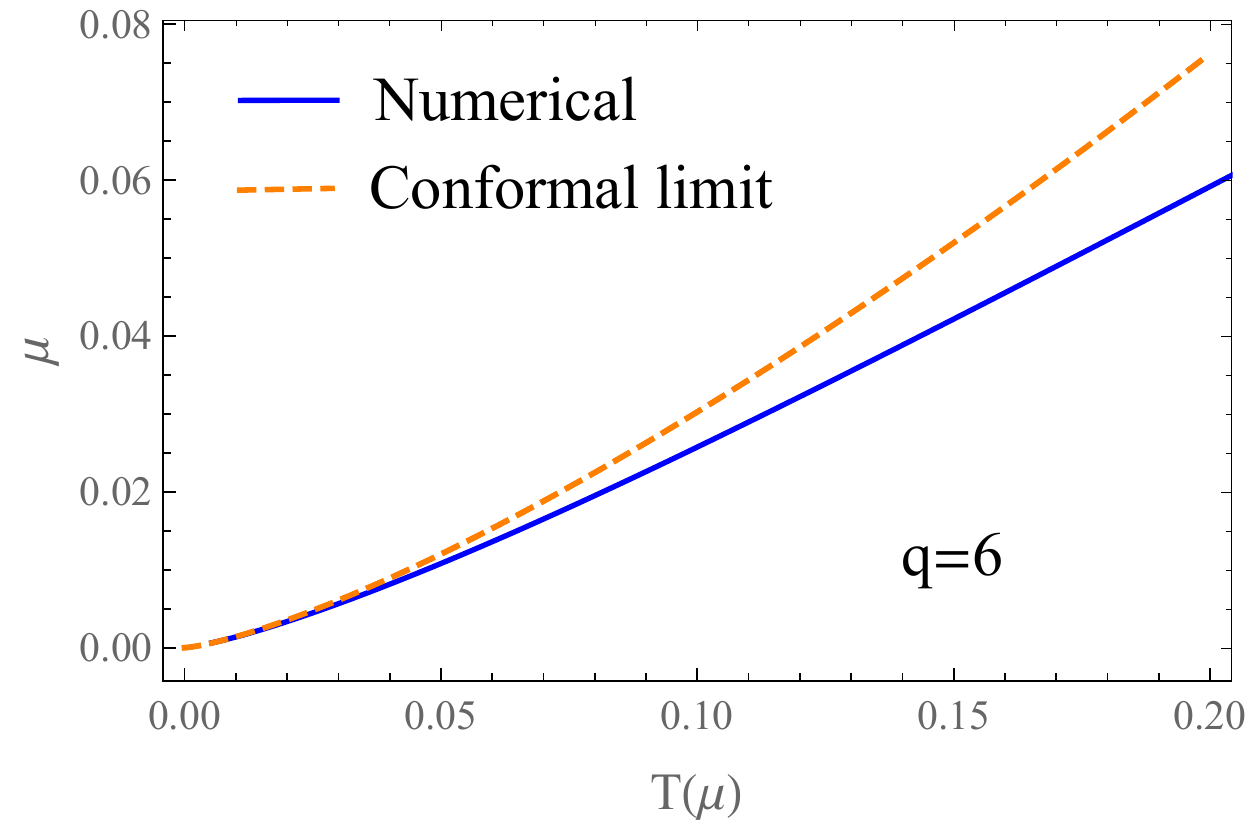}
\end{center}
\end{minipage}
\caption{Comparing the $\mu $ dependence  $T(\mu) = \beta(\mu)^{-1}$ from  variational method for low energy approximation and exact numerical calculation for $q=4$ and $q =6$ case.
The conformal limit is given by (\ref{eq:valbetamulow}).} 
\label{fig:valTvsMuq6}
\end{figure}
The variational energy becomes
\be
\f{\braket{H_{SYK} + H_M}}{N}  = \f{E_0 }{N}+ V(\phi_0)
= \f{ E_0}{N}- \mathcal{J}\alpha_S \f{1 - 2\Delta}{\Delta} \Big(\f{\mu (c_{\Delta})^2 \Delta}{\mathcal{J}\alpha_S}\Big) ^{\f{1}{(1-2\Delta)}}. \label{eq:ValGDEnConf}
\ee

Using the variational wave function, we can compute several physical observables.
For example, we can compute the spin operator expectation value $\braket{S_k} = -2i\braket{\psi_{2k-1}\psi_{2k}}$, which is essentially the off diagonal correlation function at $\tau = 0$.
The half of the spin operator expectation value becomes
\be
\f{1}{2}\braket{S_k} = -iG_{\text{off}}(0) =2 s_k (c_{\Delta})^2 \Big( \f{\pi}{\beta(\mu) \mathcal{J}} \Big)^{4\Delta} = 2 s_k (c_{\Delta})^2 \Big(\f{\mu (c_{\Delta})^2 \Delta}{\mathcal{J}\alpha_S}\Big) ^{\f{2\Delta}{(1-2\Delta)}} .\label{eq:ValSpinExConf}
\ee

Another observable we can compute is the energy of the SYK Hamiltonian $\braket{H_{SYK}}$ that gives the energy of the ground state of the deformed Hamiltonian as an excited state of the SYK Hamiltonian.
This becomes 
\be
\f{\braket{H_{SYK}}}{N} = \f{E_0}{N} + 2 \mathcal{J} \alpha _S\Big(\f{\pi}{\beta(\mu)\mathcal{J} } \Big)^2  =  \f{E_0}{N}  + 2 \mathcal{J} \alpha _S\Big(\f{\mu (c_{\Delta})^2 \Delta}{\mathcal{J}\alpha_S}\Big) ^{\f{1}{(1-2\Delta)}}.\label{eq:ValSYKEnConf}
\ee

As a consistency check, we also solve the minimization condition for the trial energy (\ref{eq:trialenergy}) using the numerical solution for thermal SYK correlation functions.
The comparison of numerics and the analytical results in conformal limit is shown in Fig.~\ref{fig:valTvsMuq6}.

\subsubsection{Comparison of variational approximation and Exact ground state}

\begin{figure}[ht]
\includegraphics[width=5.75cm]{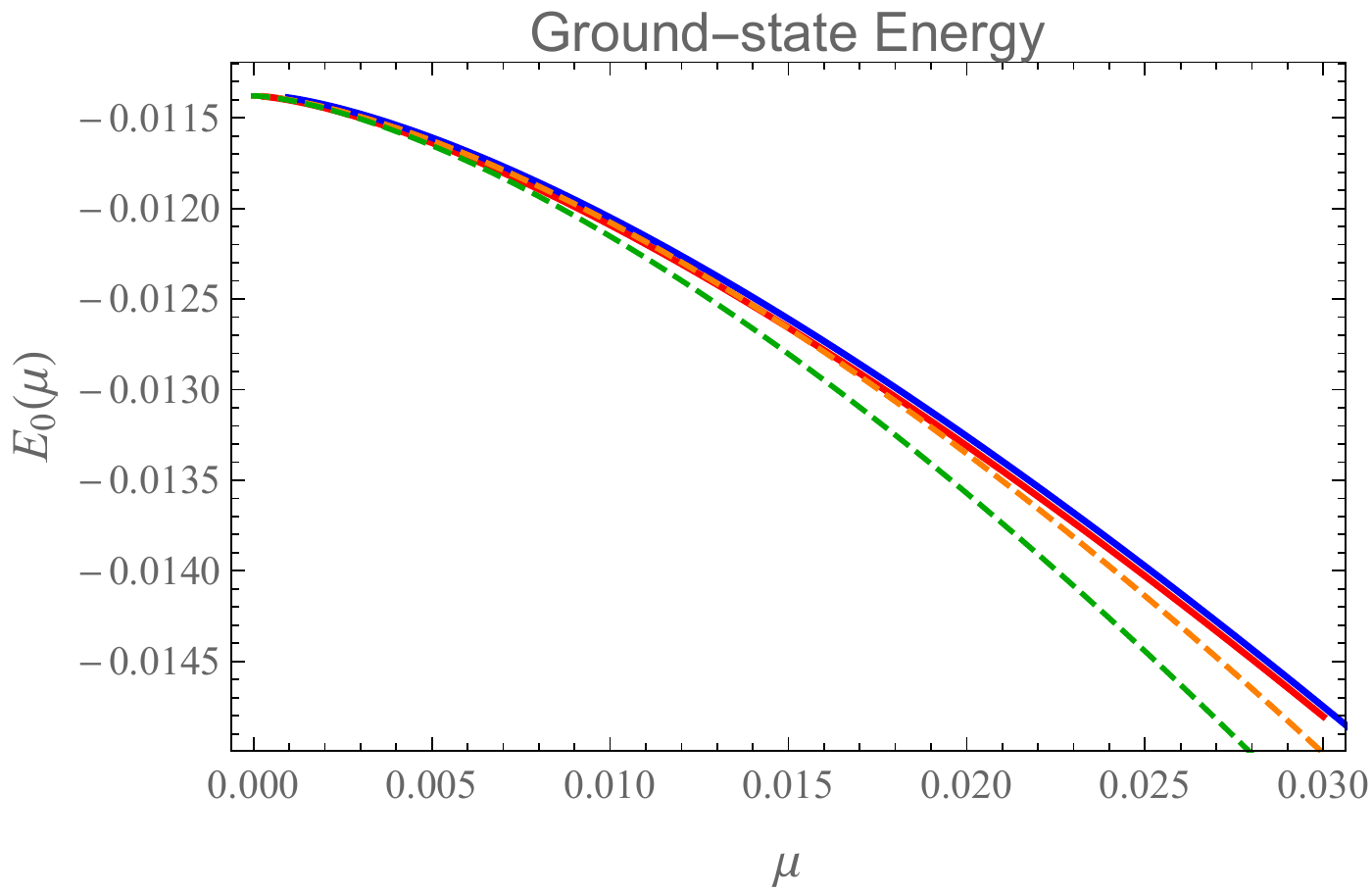} 
\includegraphics[width=5.4cm]{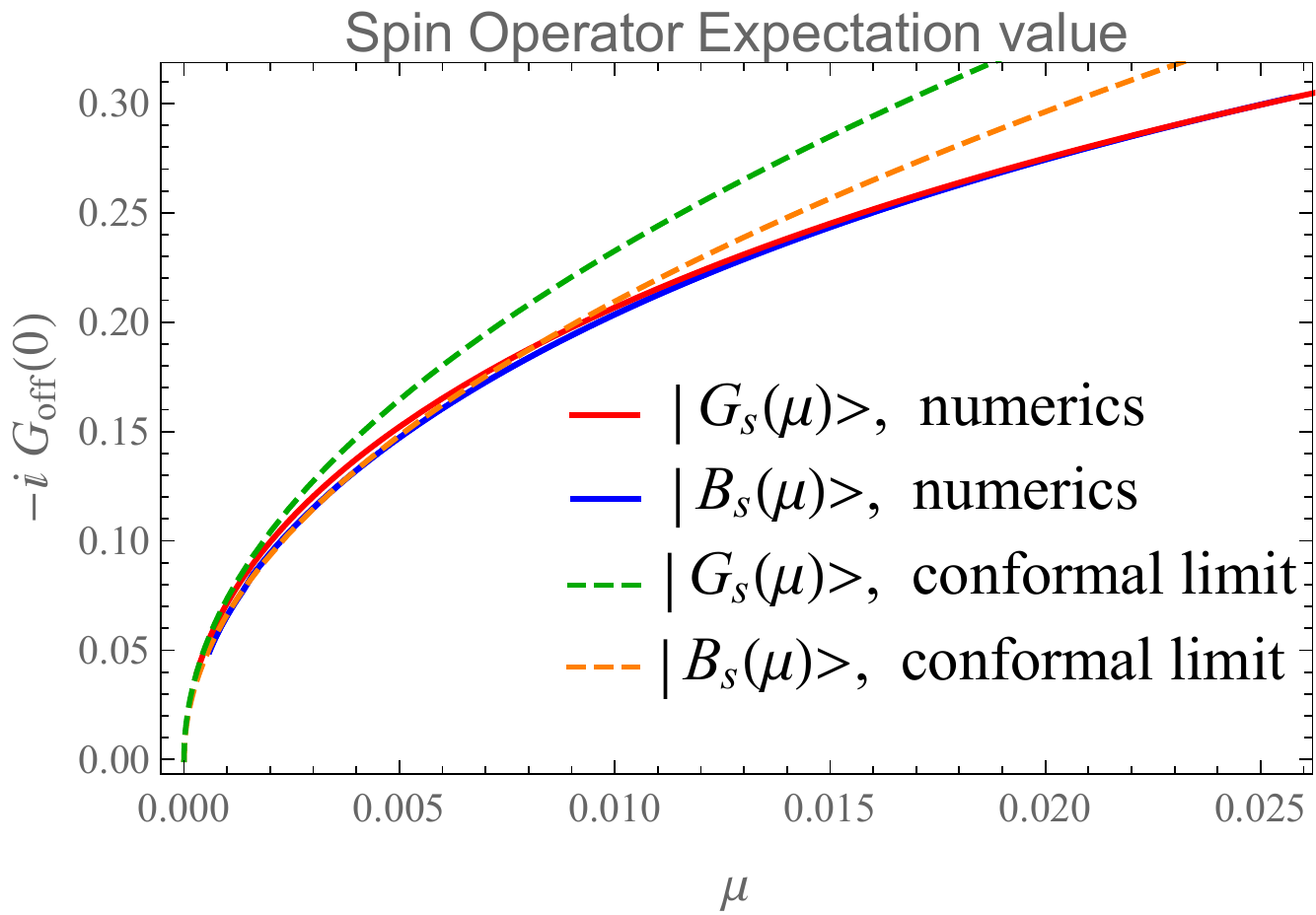}
\includegraphics[width=5.7cm]{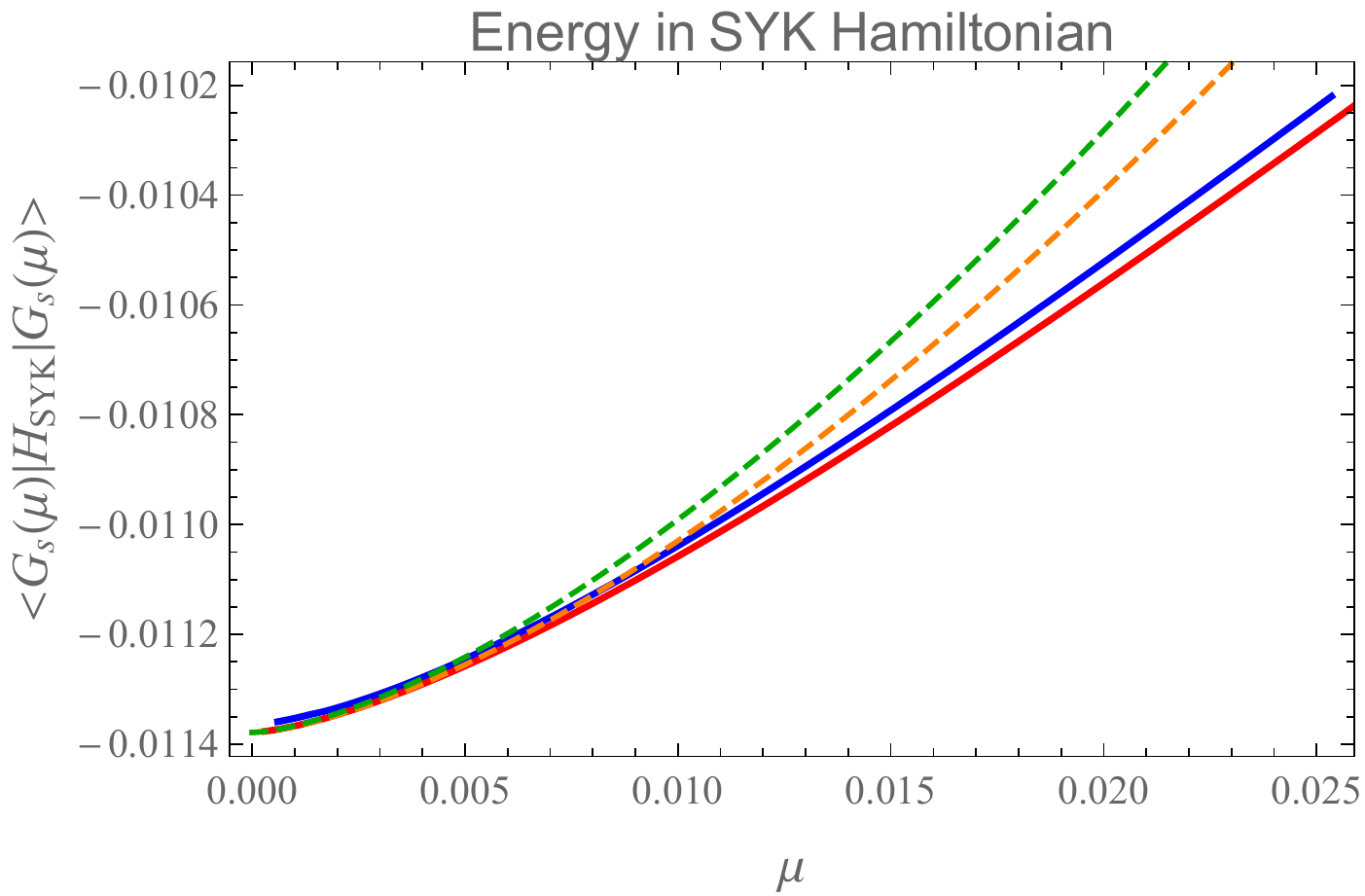}
\caption{The plot of observables both in the exact ground state $\ket{G_{\bm{s}}(\mu)}$ and the variational approximation $\ket{B_{\bm{s}}(\beta(\mu))}$.
Here we choose the parameter to be $q=6$ and $J = 1$.
As written in the central picture, the solid lines represent the numerics and the dashed lines represent the conformal limit answer.
{\bf Left:} The plot of the ground state $E_0$ as a function of $\mu$. Conformal limit results are given in (\ref{eq:GDEnConf}) and (\ref{eq:ValGDEnConf}). 
{\bf Middle:} The plot of the half of the absolute value of the spin operator expectation value $|\braket{S_k}|$, which is equal to the $\tau = 0$ off diagonal correlation function $-iG_{\text{off}}(0)$, as a function of $\mu$.
Conformal limit results are given in (\ref{eq:SpinExConf}) and (\ref{eq:ValSpinExConf}).
{\bf Right:} The plot of the energy in the SYK Hamiltonian $\braket{H_{SYK}}$ as a function of $\mu$. 
Conformal limit results are given in (\ref{eq:SYKEnConf}) and (\ref{eq:ValSYKEnConf}).
 }  
\label{fig:ValVsExact1}
\end{figure}

Even Beyond the conformal limit, we can study both of the variational approximation and the ground state numerically.
Especially, we can compare both results in the whole parameter region.
In Fig.~\ref{fig:ValVsExact1}, we show the numerical results for the spin operator expectation value $\braket{S_k}$, ground state energy $E_0(\mu)$ and energy in the SYK Hamiltonian $\braket{H_{SYK}}$ for both of the exact ground state $\ket{G_{\bm{s}}(\mu)}$ and variational approximation $\ket{B_{\bm{s}}(\beta(\mu))}$.
We found that these observables in $\ket{G_{\bm{s}}(\mu)}$ and $\ket{B_{\bm{s}}(\beta(\mu))}$ are very close and $\ket{B_{\bm{s}}(\beta)}$ is a good approximation for the ground state.
We also checked that the true ground state energy never goes beyond that in the variational approximation, which is expected.

In the conformal limit, we have analytic expression both for the exact ground state and the variational approximation.
By comparing the results, we can find that the variational approximation reproduce the correct scaling with respect to the mass parameter $\mu$.
On the other hand, the coefficients are different.
This means that the variational approximation is not perfect even in the small $\mu$ limit.
This is in contrast with the two coupled SYK model \cite{Maldacena:2018lmt} where the observables in the exact ground state and the thermofield double state perfectly agree in the small mass parameter limit.

However, in the large $q$ limit, the observables in $\ket{G_{\bm{s}}(\mu)}$ perfectly agrees with those in $\ket{B_{\bm{s}}(\beta(\mu))}$.
Actually, we can study the large $q$ limit analytically in the whole parameter regime and we can confirm that the variational approximation is perfect in any $\mu$ as we will see later.

\subsection{Thermodynamics of the deformed SYK model}
\label{section_thermodynamic}

In this section we study the thermodynamic property of the deformed SYK model \eqref{Hdef}.
In the complex SYK model with a similar deformation, an interesting phase structure was found \cite{Azeyanagi:2017drg,Ferrari:2019ogc} through the analysis of the large $N$ free energy $\frac{F}{N}=-\frac{1}{N\beta}\log Z$: the first order phase transition in $\mu$-$T$ plane\footnote{
The parameter in the complex SYK model playing the same role as $\mu$ in the Hamiltonian is the chemical potential dual to the $\text{U}(1)$ charge.
}
 and the disappearance of the phase transition above some critical values of $\mu$ and the temperature $T$.
The similar phase structure was also found in the two coupled real SYK model with equal random couplings \cite{Maldacena:2018lmt}.
It would be natural to expect a similar phase structure also in our setup.

The large $N$ free energy can be evaluated by solving the Schwinger-Dyson equations \eqref{eq:selfe1}, \eqref{eq:SD3} and then evaluating the partition function on that solution.
As we are interested in the phase structure at finite $(\mu,T)$, we solve \eqref{eq:selfe1}, \eqref{eq:SD3} directly without any further approximation and numerically by discretizing $\tau$ direction.
See appendix \ref{sec_solveSDnumerical} for detail.
The Schwinger-Dyson equations are discretized as \eqref{eqsforGandSigma_fornumerics} and the free energy is evaluated through \eqref{free_numerical_final}.
Here we have chosen the discretization parameter as $\tau=\frac{\beta m}{2\Lambda}$ $(m=1,2,\cdots,2\Lambda)$ with $\Lambda=10^6$.
For each $\mu$, we have first solved the Schwinger-Dyson equation for $T=0.3$ numerically by an iterative method \cite[appendix G]{Maldacena:2016hyu} with initial values for $G$ and $\Sigma=J^2G^{q-1}$ chosen as ${\widetilde G}_n=\frac{i}{\omega_n}$ ($\omega_n=\frac{2\pi}{\beta}(n+\frac{1}{2})$).
Then we have decreased the temperature slowly by solving the equation for the temperature $T-\Delta T$ with the initial condition chosen as the solution obtained for the temperature $T$, with $\Delta T=5\times 10^{-5}$.
Once we reach a sufficiently small temperature, we solve the Schwinger-Dyson equation again by slowly increasing the temperature in the similar way.
This recursive technique is similar to the technique employed in \cite{Maldacena:2018lmt,Azeyanagi:2017drg,Ferrari:2019ogc}.
If we find two different free energy for the increasing $T$ and the decreasing $T$, crossing with each other at some temperature $T_c$, we conclude that there is a first-order phase transition as $T=T_c$.

\begin{figure}
\begin{center}
\includegraphics[width=12cm]{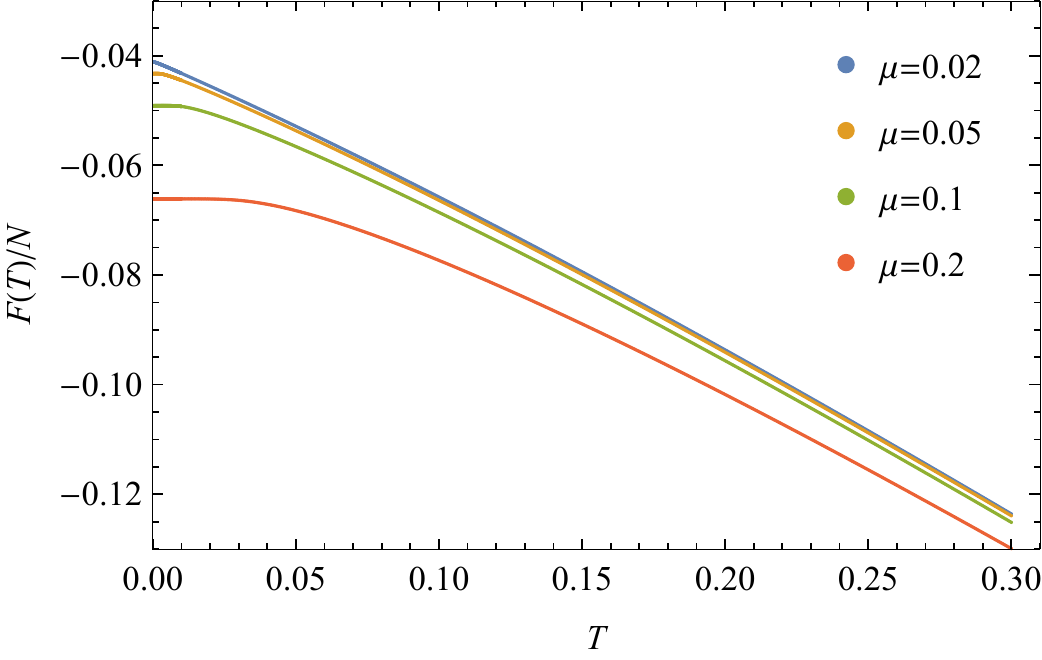}
\end{center}
\caption{
The large $N$ free energy $\frac{F}{N}$ of the deformed SYK model \eqref{free_numerical_final} computed by solving the Schwinger-Dyson equation numerically.
Here the horizontal axis is the temperature $T$.
}
\label{SDeq_freeenergy}
\end{figure}

The results are summarized in Fig.~\ref{SDeq_freeenergy}.
We find that the free energy for each $\mu$ interpolates two extreme behaviors: $F=const.$ (i.e., gapped) for low temperature and $F\approx F_\text{SYK}$ at high temperature, which is consistent with the structure of the deformed Hamiltonian \eqref{Hdef}.
From the observations \cite{Maldacena:2018lmt,Azeyanagi:2017drg,Ferrari:2019ogc} we suspected that the system exhibits a first order phase transition in the intermediate temperature (for example, $T\sim 0.04$ for $\mu=0.2$).
However, we have not observed the aformentioned hysteretic behavior which would indicate the first order phase transition.

\begin{figure}
\begin{center}
\includegraphics[width=12cm]{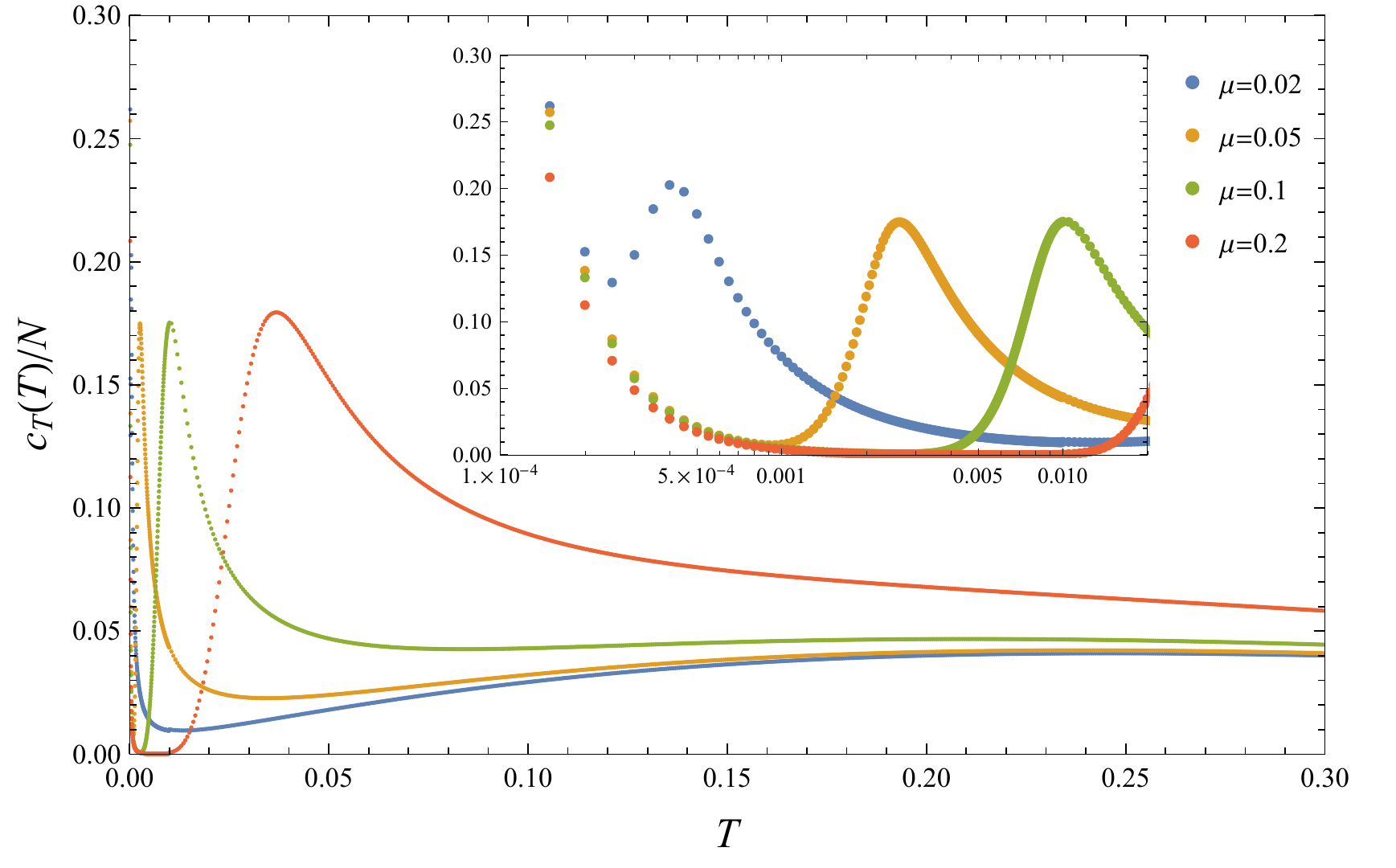}
\end{center}
\caption{
The large $N$ specific heat $\frac{c_T}{N}$ \eqref{cheat} of the deformed SYK model \eqref{Hdef}, here the horizontal axis is the temperature $T$.
Note that the universal increasing behavior at $T\approx 0$ is a numerical artifact due to the fact that the numerical UV cutoff $|\omega_n|<\frac{2\pi\Lambda}{\beta}$ is not large enough.
}
\label{SDeq_cheat}
\end{figure}

We further examine the presence of the second order phase transition by calculating the large $N$ specific heat
\begin{align}
c_T=-T\frac{\partial^2 F}{\partial T^2},
\label{cheat}
\end{align}
which would diverge at the second order phase transition point.
See Fig.~\ref{SDeq_cheat}.
Though the specific heat exhibits a peak at some temperature in the intermediate regime, we find that the peak is finite and smooth.

From these result we conclude that our model exhibits neither the first order phase transition nor the second order phase transition.\footnote{
Strictly speaking, our analysis is not a proof of the absence of the phase transition.
For example, it is not ensured that our algorithm exhausts all the solutions to the Schwinger-Dyson equation which are relevant in the limit of $\Lambda\rightarrow\infty$.
Nevertheless in the large $q$ limit we can explicitly prove that there are no phase transition in this model.
See section \ref{sec_largeqbetaqlogq} for more detail.
}
This result is rather surprising and we discuss possible explanation in section \ref{sec_discussion}.

\section{Finite $N$ analysis of the model}
\label{sec_finiteN}
In this section, we study the mass deformed Hamiltonian \eqref{Hdef} at finite $N$.
We focus on the case with $q=4$ and $J=1$ of this model.

Since the canonical anti-commutation relation of $\psi_i$, $\{\psi_i,\psi_j\}=\delta_{ij}$ can be realized by the Gamma matrices $\Gamma_i$ as $\psi_i=\frac{1}{\sqrt{2}}\Gamma_i$, the Hamiltonian $H_\text{def}$ \eqref{Hdef} for finite $N$ is written as the following $2^{N/2}\times 2^{N/2}$ matrix
\begin{align}
H_\text{def}=H_\text{SYK}+H_M,\quad
H_\text{SYK}=\frac{1}{4}\sum_{i<j<k<\ell}J_{ijk\ell}\Gamma_i\Gamma_j\Gamma_k\Gamma_\ell,\quad
H_M=\frac{i\mu}{2}\sum_{j=1}^{N/2}\Gamma_{2j-1}\Gamma_{2j},
\label{HdefinGamma}
\end{align}
with $J_{ijk\ell}$ random coupling chosen out of Gaussian distribution with the mean $\langle J_{ijk\ell}\rangle=0$ and the variance $\langle J_{ijk\ell}^2\rangle=\frac{6}{N^3}$.

Note that $H_\text{def}$ commutes with the following chirality (i.e. fermion number in $\psi_i$) matrix 
\begin{align}
\Gamma_c=i^{-\frac{N}{2}}\Gamma_1\Gamma_2\cdots\Gamma_N
\label{Gammac}
\end{align}
whose eigenvalues are $\pm 1$.
Hence with an appropriate choice of basis, $H_\text{def}$ takes a block diagonal form
\begin{align}
H_\text{def}=H_\text{def}^{(+)}\oplus H_\text{def}^{(-)}
\label{blockdiagonal}
\end{align}
with $H_\text{def}^{(\pm)}=H_\text{def}\frac{1 \pm \Gamma_c}{2}$, regardless of the choice of $J_{ijk\ell}$.

\begin{figure}
\includegraphics[width=16cm]{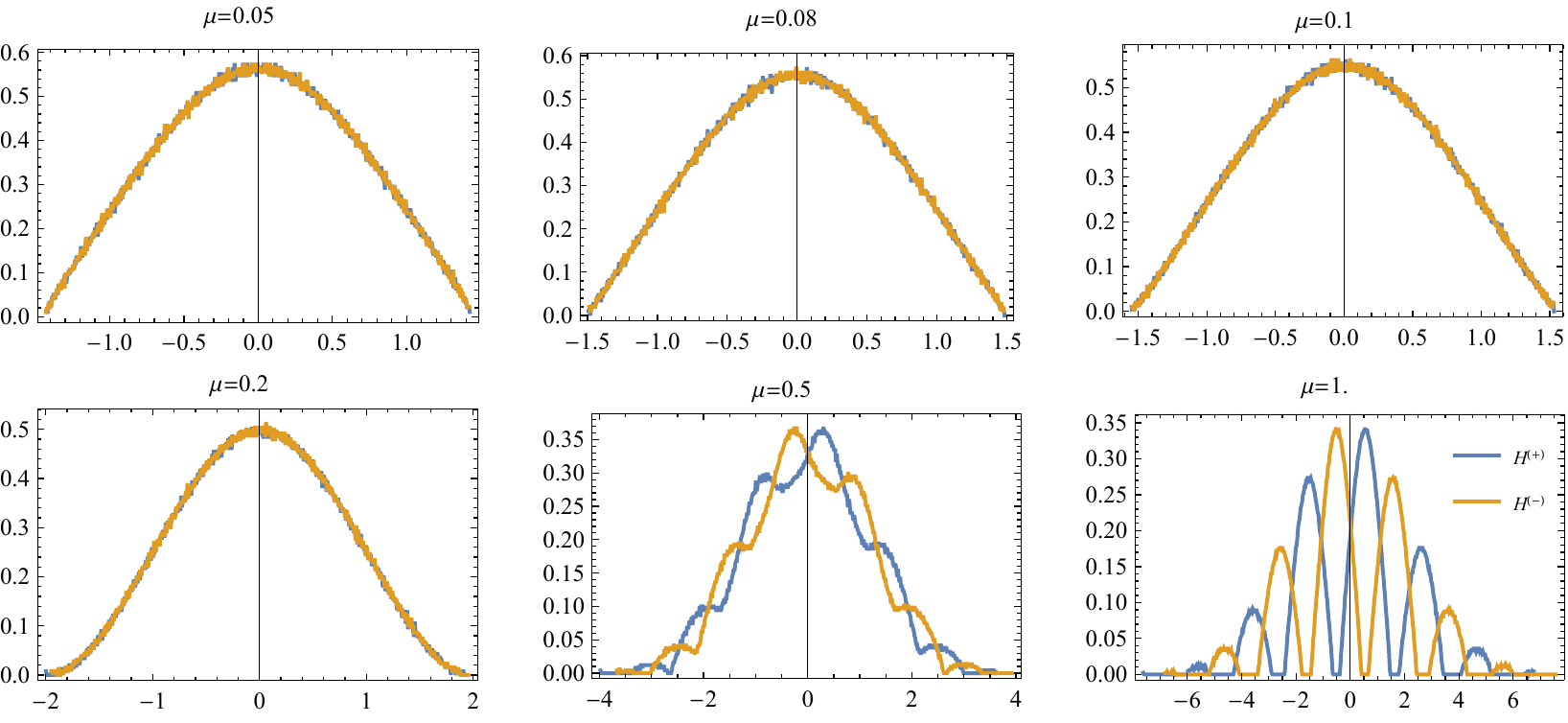}
\caption{
Eigenvalue density of the Hamiltonian $H_\text{def}$ with $q=4$, $J=1$ \eqref{HdefinGamma} and $N=30$, with a single realization of $J_{ijk\ell}$.
We observe that the shape of the eigenvalue density around the ground state exhibits a transition around $\mu\approx 0.1$ from a hard edge to a smooth decay, which is consistent with the behavior of $E_\text{gap}$; for $\mu\lesssim 0.1$ $E_\text{gap}\sim \mu^2$, which is significantly smaller than $E_\text{gap}\sim \mu$.
}
\label{KMspectrum_N30_Mu001to2}
\end{figure}

In Fig.~\ref{KMspectrum_N30_Mu001to2} we display the eigenvalue density of $H_\text{def}$ for $N=30$ and various values of $\mu$.
When $\mu$ is large, $H_\text{def}$ is dominated by $H_M$ where the energy levels are discrete $E_p=\mu(-\frac{N}{4}+p)$ $(p=0,1,\cdots,\frac{N}{2})$ with degeneracies $d_p=\binom{\frac{N}{2}}{p}$.
Though these degeneracies are resolved by $H_\text{SYK}$, the levels at different $E_p$ are not mixed for a sufficiently large $\mu$, hence we obtain a blob structure.

\subsection{Overlap ${}_\beta\langle B_{(\downarrow,\downarrow,\cdots,\downarrow)}|0^{(+)}\rangle$}
In section \ref{section_variational} we have realized that the spin ground state $|B_{(\downarrow,\downarrow,\cdots,\downarrow)}\rangle$ is a good variational ansatz to realize the true ground state energy of $H_\text{def}$ after the Euclidean evolution $e^{-\frac{\beta}{2} H_\text{SYK}}$, with $\beta$ being the variational parameter.
In this section we would like to examine the agreement of these two states more directly, through the overlap of the states
\begin{align}
|{}_\beta\langle B_{(\downarrow,\downarrow,\cdots,\downarrow)}|0^{(+)}\rangle|,
\label{overlap}
\end{align}
where $|0^{(+)}\rangle$ is the ground state of $H^{(+)}$ and $|B_{(\downarrow,\downarrow,\cdots,\downarrow)}\rangle_\beta$ is defined as
\begin{align}
|B_{(\downarrow,\downarrow,\cdots,\downarrow)}\rangle_\beta=\frac{1}{Z(\beta)}e^{-\frac{\beta H_\text{SYK}}{2}}|B_{(\downarrow,\downarrow,\cdots,\downarrow)}\rangle,\quad
Z(\beta)=\langle B_{(\downarrow,\downarrow,\cdots,\downarrow)}|e^{-\beta H_\text{SYK}}|B_{(\downarrow,\downarrow,\cdots,\downarrow)}\rangle,
\end{align}
with $|B_{\bm{s}}\rangle$ defined in \eqref{Bs} and normalized as $\langle B_{\bm{s}}|B_{\bm{s}}\rangle=1$.
Here $\beta$ is chosen for each realization of $J_{ijk\ell}$ such that the overlap \eqref{overlap} is maximized.

\begin{figure}
\includegraphics[width=16cm]{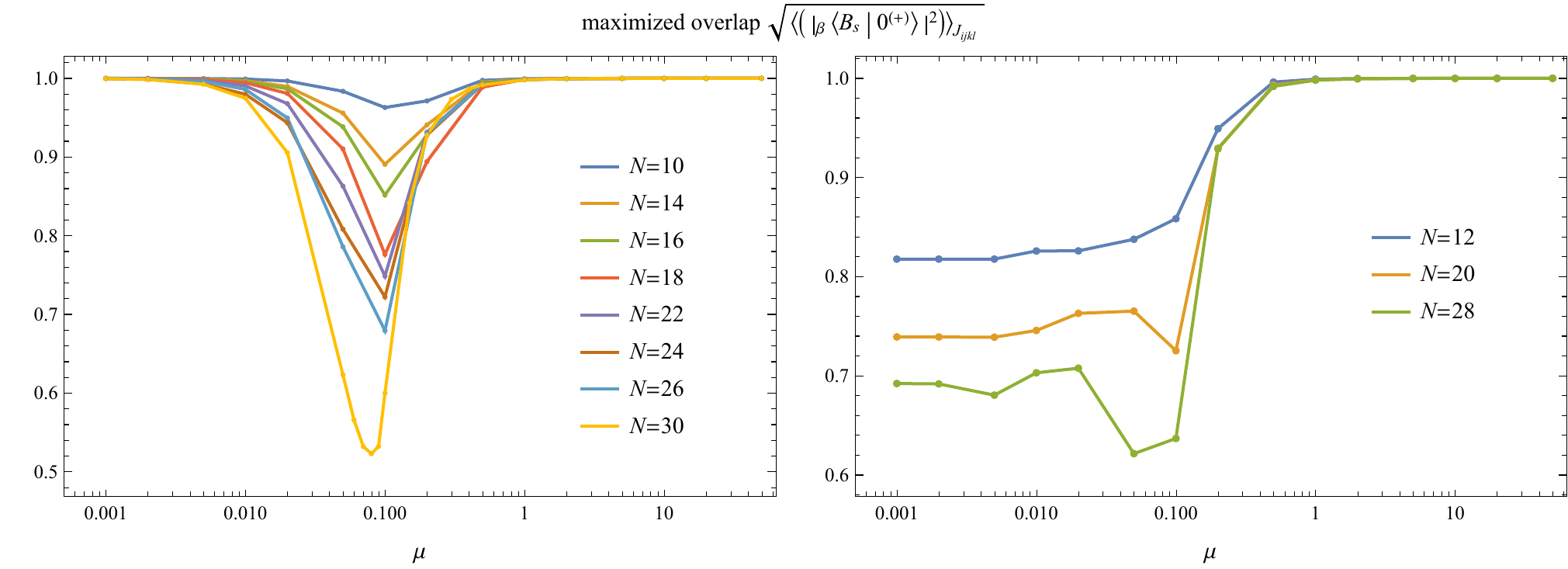}
\caption{The maximized overlap averaged over the realizations of random coupling $J_{ijk\ell}$ as $\sqrt{\bigl\langle|{}_\beta\langle B_{(\downarrow,\downarrow,\cdots,\downarrow)}|0^{(+)}\rangle|^2\bigr\rangle_{J_{ijk\ell}}}$.
Here the horizontal axis is $\mu$.
}
\label{overlapresults}
\end{figure}

\begin{figure}
\begin{center}
\includegraphics[width=8cm]{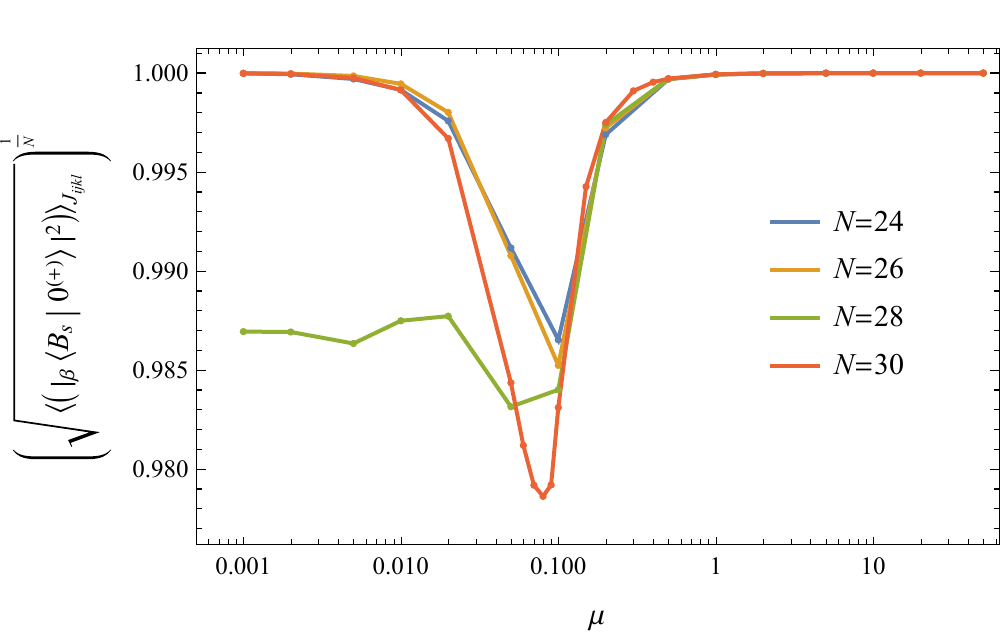}
\end{center}
\caption{The maximized overlap viewed in $\Bigl(\sqrt{\langle|{}_\beta\langle B_{(\downarrow,\downarrow,\cdots,\downarrow)}|0^{(+)}\rangle|^2\rangle_{J_{ijk\ell}}}\Bigr)^{\frac{1}{N}}$, with the horizontal axis $\mu$.}
\label{overlapresultsto1overN}
\end{figure}

Note that $|B_{(\downarrow,\downarrow,\cdots,\downarrow)}\rangle_\beta$ has a definite chirality $\Gamma_c|B_{(\downarrow,\downarrow,\cdots,\downarrow)}\rangle_\beta=+|B_{(\downarrow,\downarrow,\cdots,\downarrow)}\rangle_\beta$ for any values $J_{ijk\ell}$ and $N$.
This follows from the fact $\Gamma_i^{(-)}|B_{(\downarrow,\downarrow,\cdots,\downarrow)}\rangle=0$, where $\Gamma_i^{(\pm)}=\frac{\Gamma_{2i}\pm i\Gamma_{2i-1}}{2}$ the rising/lowering operator for $S_i$, together with the following alternative expression of $\Gamma_c$ \eqref{Gammac}
\begin{align}
\Gamma_c=
(1-2\Gamma_1^{(+)}\Gamma_1^{(-)})
(1-2\Gamma_2^{(+)}\Gamma_2^{(-)})
\cdots
(1-2\Gamma_{\frac{N}{2}}^{(+)}\Gamma_{\frac{N}{2}}^{(-)}),
\end{align}
and the fact that $H_\text{SYK}$ commutes with $\Gamma_c$.
On the other hand, the chirality of the true ground state $|0\rangle$ of $H_\text{def}$ depends on the value of the random coupling $J_{ijk\ell}$, and when $\Gamma_c|0\rangle=-|0\rangle$ the overlap with $|B_{(\downarrow,\downarrow,\cdots,\downarrow)}\rangle_\beta$ is identically zero regardless of the value of $\beta$.
For this reason, in \eqref{overlap} we have used $|0^{(+)}\rangle$ instead of $|0\rangle$ to make the comparison meaningful for all realizations.\footnote{
If one is interested in the overlap between $|B_{(\downarrow,\downarrow,\cdots,\downarrow)}\rangle$ and the true ground state $|0\rangle$, one has just to multiply the ``probability of $|0\rangle$ to have $\Gamma_c=+1$'' to the results displayed in Fig.~\ref{overlapresults}.
Though we do not have an analytic expression, we observe for any $N$ that this probability is almost $1$ for $\mu\ge 0.5$ and not smaller than $0.5$ also for the smaller values of $\mu$.
Especially the difference between $|0\rangle$ and $|0^{(+)}\rangle$ does not matter when we consider $|{}_\beta\langle B_{(\downarrow,\downarrow,\cdots,\downarrow)}|0^{(+)}\rangle|^{\frac{1}{N}}$ (Fig.~\ref{overlapresultsto1overN}) in the large $N$ limit.
}

The results are displayed in Fig.~\ref{overlapresults}.
For large $\mu$, the Hamiltonian is dominated by $H_M$ whose ground state is $|B_{(\downarrow,\downarrow,\cdots,\downarrow)}\rangle$, hence the overlap trivially approaches to $1$.
For small $\mu$, the Hamiltonian is dominated by $H_\text{SYK}$.
Since the Euclidean evolution with $\beta\rightarrow\infty$ is equivalent to the projection onto the ground state of $H_\text{SYK}^{(+)}$, the overlap should again approaches to $1$.
Note, however, that for $N\equiv 4$ mod 8 the spectrum of $H^{(+)}$ is two-hold degenerate.
The degeneracy is resolved by a small perturbation by $H_M$, and at the leading order in $\mu$ the ground state $|0^{(+)}\rangle$ of $H_\text{def}$ is a certain linear combination of the two ground state of $H_\text{SYK}$ which is not necessarily the same linear combination obtained by the projection of $|B_{(\downarrow,\downarrow,\cdots,\downarrow)}\rangle$.
Hence we expect that the overlap is substantially smaller than 1.\footnote{
Though we do not have a clear argument for this effect, we observe that the value of the overlap approaches some finite value as $N$ increases from $N=12$ to $N=28$.
}
The results in Fig.~\ref{overlapresults} are consistent with these expectations.
On the other hand, for intermediate values of $\mu$ we have found that the overlap is not close to 1 any more even for $N\not\equiv 4$ mod 8, and the lowest value around $\mu=0.01$ significantly decreases as $N$ increases.

Note, however, that as the dimension of the Hilbert space increases, the agreement of two vectors $|\phi\rangle$,$|\chi\rangle$ in the sense of $|\langle\phi|\chi\rangle|\approx 1$ becomes less likely to occur.
For example the expectation value of the overlap of two randomly chosen unit vectors in $d$ dimensional space can be evaluated as follows
\begin{align}
\sqrt{\Bigl\langle|\langle e_1|e_2\rangle|^2\Bigr\rangle_{|e_1\rangle,|e_2\rangle\text{: random}}}
&=\sqrt{\int_{\text{U}(d)} dU_1dU_2\langle e|U_1^\dagger U_2|e'\rangle\langle e'|U_2^\dagger U_1|e\rangle}\nonumber \\
&=\sqrt{\frac{1}{d^2}\int_{\text{U}(d)}dU_1dU_2\Tr U_1^\dagger U_2U_2^\dagger U_1}\nonumber \\
&=\sqrt{\frac{1}{d}}
\label{randomoverlap}
\end{align}
where in the second line we have realized the randomness of $|e_1\rangle$,$|e_2\rangle$ as $|e_1\rangle=U_1|e\rangle$, $|e_2\rangle=U_2|e'\rangle$ with random unitary transformations $U_1,U_2$ and an arbitrary pair of fixed unit vectors $|e\rangle$,$|e'\rangle$.
In the third line, taking into account that the result is independent of the choice of $|e\rangle$,$|e'\rangle$, we have further replaced $|e\rangle\langle e|$ and $|e'\rangle\langle e'|$ with $\frac{1}{d}\sum_e|e\rangle\langle e|=\frac{1}{d}$ and $\frac{1}{d}\sum_{e'}|e'\rangle\langle e'|=\frac{1}{d}$.
In the current case, the dimension of the Hilbert space is $d=2^{\frac{N}{2}-1}$, hence $\sqrt{\langle|\langle e_1|e_2\rangle|^2\rangle}\approx e^{-\frac{\log 2}{4}N}$.
The large $N$ calculation of the overlap through the saddle point approximation, which we explain and actually perform for the large $q$ limit in section \ref{section_overlaplargeq}, also suggest that the overlap should behave like $|{}_\beta\langle B_{(\downarrow,\downarrow,\cdots,\downarrow)}|0\rangle|\sim e^{-N\cdot {\cal O}(1)}$.
Hence it would be more reasonable to see $|\langle {}_\beta\langle B_{(\downarrow,\downarrow,\cdots,\downarrow)}|0^{(+)}\rangle|^{\frac{1}{N}}$ instead of $|\langle {}_\beta\langle B_{(\downarrow,\downarrow,\cdots,\downarrow)}|0^{(+)}\rangle|$.
See Fig.~\ref{overlapresultsto1overN}.
The values are always substantially large compared with the case of random overlap $2^{-\frac{1}{4}}=0.841$ \eqref{randomoverlap}, hence we conclude that $|B_{(\downarrow,\downarrow,\cdots,\downarrow)}\rangle_\beta$ is indeed a good approximation to $|0^{(+)}\rangle$ for any values of $\mu$ once $\beta(\mu)$ is chosen appropriately.

\begin{figure}
\begin{center}
\includegraphics[width=8cm]{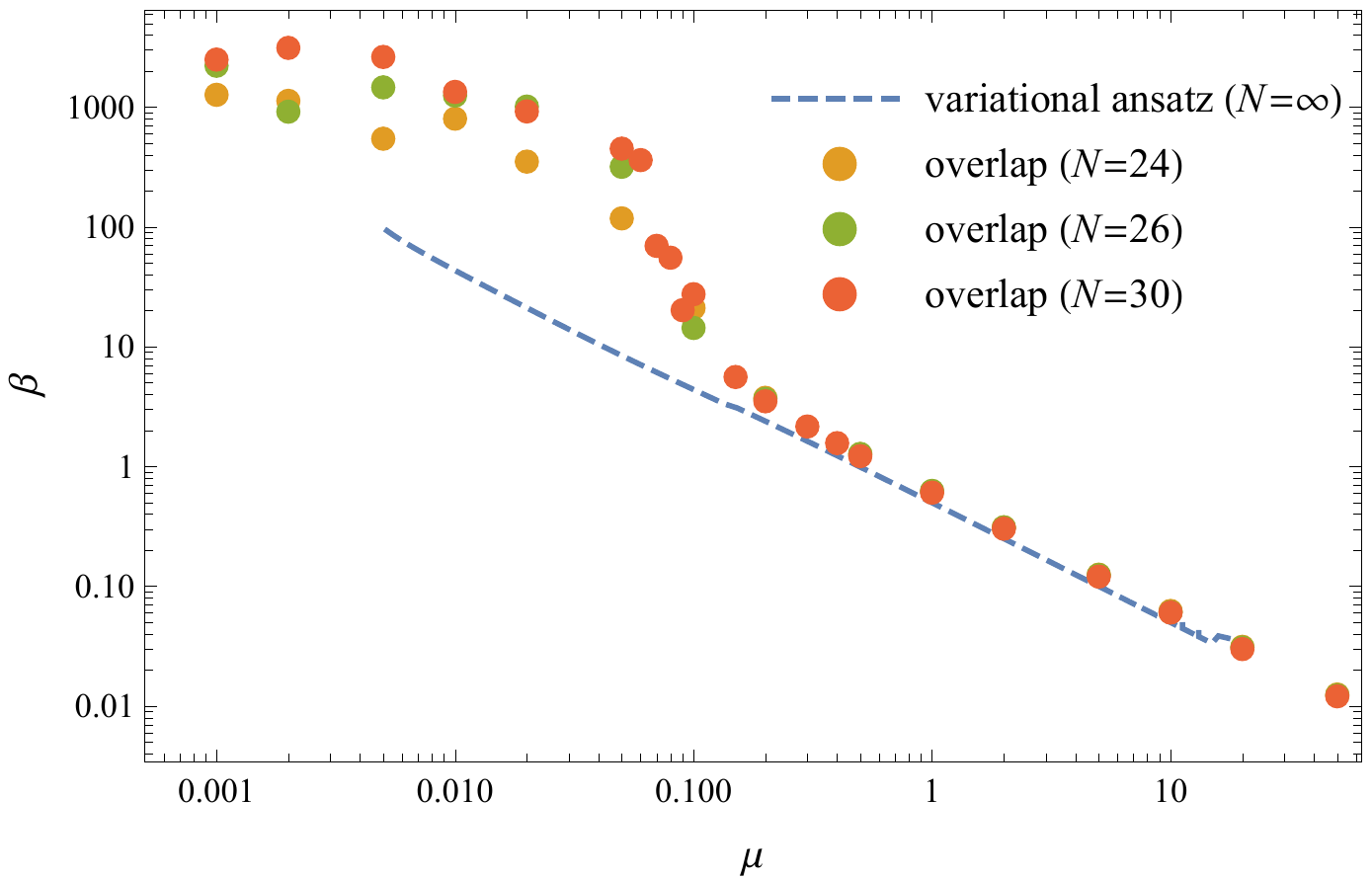}
\end{center}
\caption{The inverse temperature $\beta$ maximizing the overlap \eqref{overlap}, averaged over the ensemble $\langle \beta\rangle_{J_{ijk\ell}}$ compared with the inverse temperature which minimizes the large $N$ variational energy (Blue; see Fig.~\ref{fig:valTvsMuq6} in section \ref{sec_largeNfiniteqanalysis}).
}
\label{betamaximizingoverlap}
\end{figure}

Lastly, the $\beta$ maximizing the overlap at each $\mu$ are obtained as Fig.~\ref{betamaximizingoverlap}.
We found a good agreement for large $\mu$ ($\mu > 0.2$).
On the other hand the two results are significantly different (by factor $\sim 100$) for the smaller $\mu$.
However, it is not necessary to have an agreement in the first place since we have determined $\beta(\mu)$ through the two different quantities.
Indeed, though the variational ansatz reproduced the ground stat energy of the deformed Hamiltonian well, there was a discrepancy in another observable $|\langle S_k\rangle|$ (see Fig.~\ref{fig:ValVsExactq4}; for a possible explanation for the discrepancy, see appendix \ref{sec:q4case}).
This implies that $|B_{\downarrow,\downarrow,\cdots,\downarrow}\rangle_\beta$ with $\beta(\mu)$ determined by minimizing the energy was actually not so a good approximation to the ground state itself.

\subsection{Chaotic property}
\label{sec_chaoticproperty}
In \cite{Garcia-Garcia:2019poj} the authors conjectured that the Hawking-Page like transition of the model \cite{Maldacena:2018lmt} is accompanied with the chaotic/integrable transition.
Here we would like to test this proposal also for the current setup.
In section \ref{section_thermodynamic} we have found that our model does not exhibits a phase transition in $\mu$ or in the temperature $T$.
Hence, if the proposal is correct, our model should not exhibit a chaotic/integrable transition.

As a diagnostics of the quantum chaoticity, in this paper we adopt the level statistics which is relatively easy to study for finite $N$.
It was conjectured that \cite{Bohigas:1983er} if we quantize a classically chaotic system the fluctuation property of the resulting energy spectrum exhibits the same correlation among different levels as in the random matrix theory.
Here the ensemble of the random matrix is determined by the time reversal symmetry of the Hamiltonian of the quantized system.
Though a rigorous proof at fully quantum level is still lacking, this conjecture have been verified in various systems \cite{Guhr:1997ve,1984LNP...209....1B} and also proved at semi-classical level \cite{2004PhRvL..93a4103M,2005PhRvE..72d6207M}.
Hence one may use the presence of the RMT-like level correlation conversely as a reasonable definition of the quantum chaos.

Among various ways to characterize the level correlations, here we adopt the following quantity called the adjacent gap ratio \cite{2007PhRvB..75o5111O,2013PhRvL.110h4101A,2016PhRvB..94n4201B,2015PhRvB..91h1103L}:
\begin{align}
\bar{r}=
\overline{\frac{
\text{min}(E_{i+1}-E_i,E_i-E_{i-1})
}{
\text{max}(E_{i+1}-E_i,E_i-E_{i-1})
}
},
\end{align}
where $\{E_i\}$ is the energy spectrum ($E_i\le E_{i+1}$) and $\overline{(\cdots)}$ in the right-hand side stands for the average over the spectrum.
This quantity is evaluated for the random matrix theories with various type of the ensemble \cite{2013PhRvL.110h4101A} as well as for the Poisson distribution which corresponds to the non-chaotic systems.
By comparing the result obtained from the actual energy spectrum with these known values, one can diagnose whether the systems is chaotic or not.

\begin{figure}
\begin{center}
\includegraphics[width=12cm]{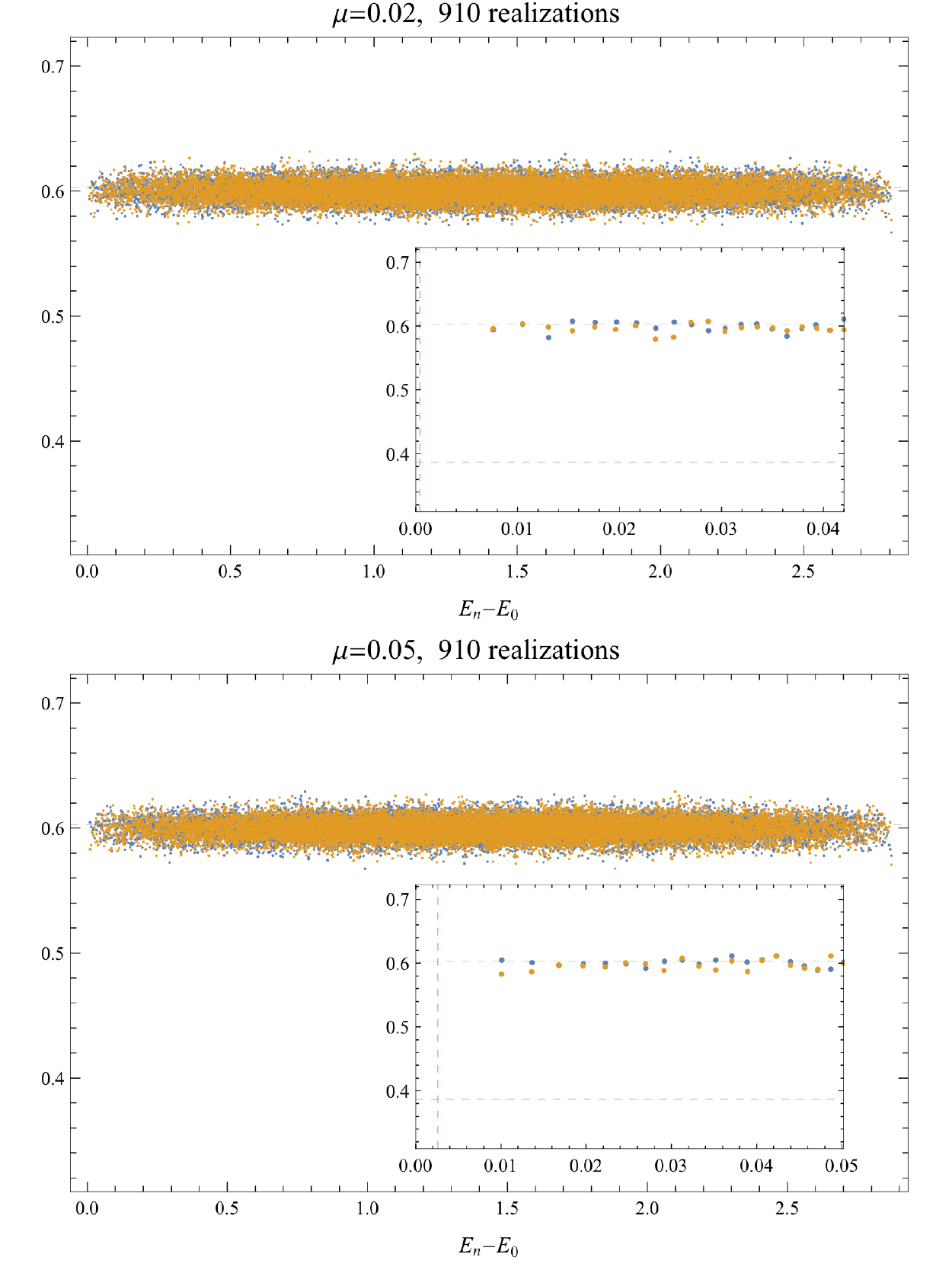}
\end{center}
\caption{Adjacent gap ratio $\langle r_n^{(\pm)}\rangle_{J_{ijk\ell}}$ of $H^{(\pm)}_\text{def}$ for $N=30$.
Here the horizontal axis is $\langle E_n^{(\pm)}\rangle -E_0\rangle_{J_{ijk\ell}}$ with $E_0=\text{min}(E_0^{(+)},E_0^{(-)})$ the energy of the true ground state.
Inset: enlarged view for first 20 levels per each chirality sector, with dashed red line the peak temperature of the specific heat in the large $N$ limit (see Fig.~\ref{SDeq_cheat}) around which we would expect the chaotic/integrable transition if it exists.}
\label{agrMu002and005}
\end{figure}

\begin{figure}
\begin{center}
\includegraphics[width=12cm]{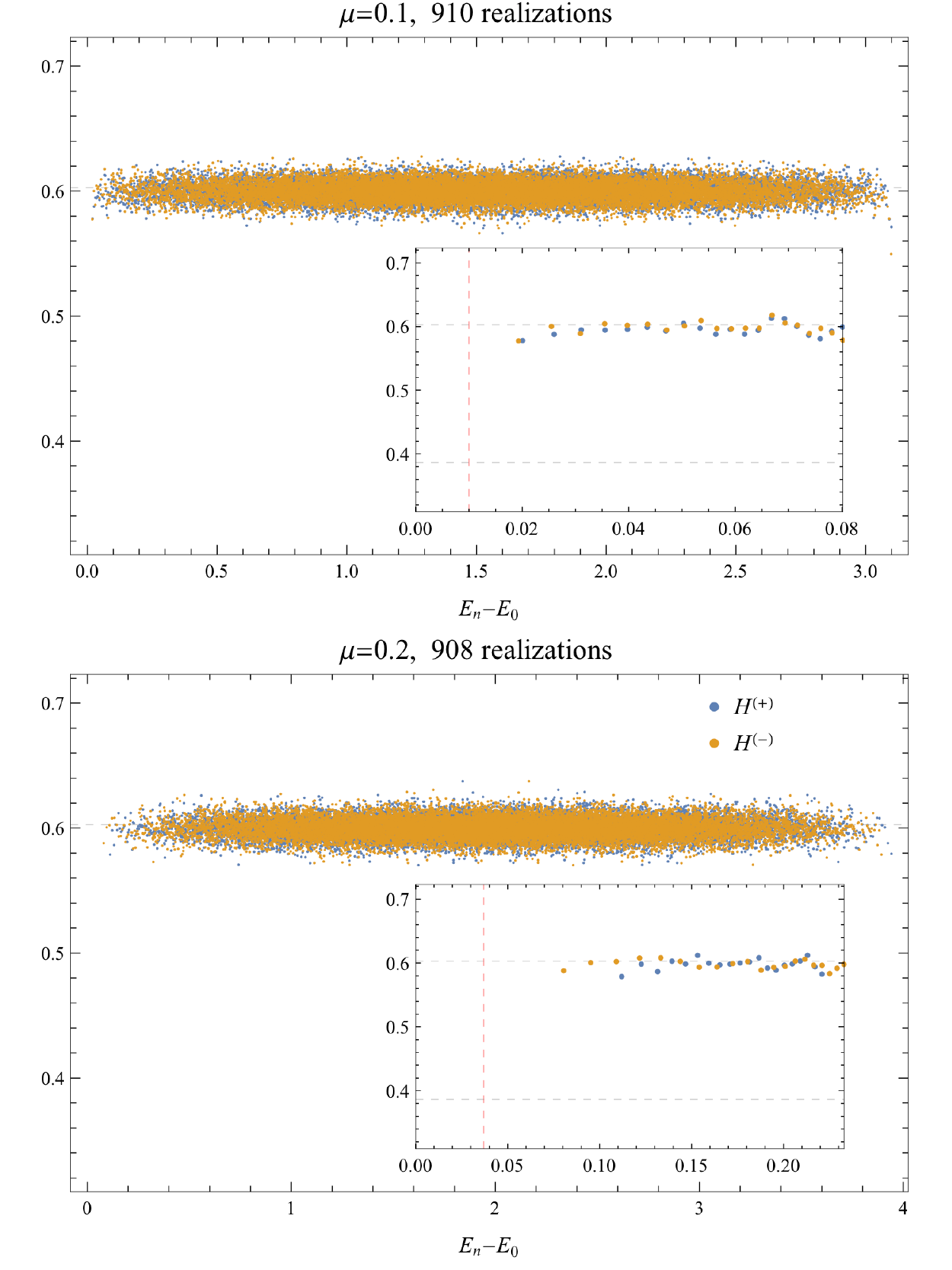}
\end{center}
\caption{Adjacent gap ratio $\langle r_n^{(\pm)}\rangle_{J_{ijk\ell}}$ of $H^{(\pm)}_\text{def}$ for $N=30$.
}
\label{agrMu01and02}
\end{figure}

As the Hamiltonian of our model is trivially separated \eqref{blockdiagonal} due to the conservation of chirality, the adjacent gap ratio should also be defined separately for the spectrum of each of $H^{(\pm)}_\text{def}$ instead of the full spectrum of $H_\text{def}$ \cite{1984LNP...209....1B}

\begin{align}
\langle r_i^{(\pm)}\rangle_{J_{ijk\ell}}=\biggl\langle
\frac{
\text{min}(E_{i+1}^{(\pm)}-E_i^{(\pm)},E_i^{(\pm)}-E_{i-1}^{(\pm)})
}{
\text{max}(E_{i+1}^{(\pm)}-E_i^{(\pm)},E_i^{(\pm)}-E_{i-1}^{(\pm)})
}
\biggr\rangle_{J_{ijk\ell}},
\end{align}
where the spectrum $\{E_i^{(\pm)}\}_{i=1}^{2^{\frac{N}{2}-1}}$ of $H^{(\pm)}_\text{def}$ is sorted such that $E_i^{(\pm)}\le E_{i+1}^{(\pm)}$.
The average is taken over $J_{ijk\ell}$ for each fixed $i$.
Here we do not take the average over the spectrum; in this way we can diagnose the chaoticity of our model at each energy scale separately.
The results are displayed in figures Fig.~\ref{agrMu002and005} and Fig.~\ref{agrMu01and02}.

The time reversal symmetry of one dimensional fermion systems were studied in \cite{PhysRevB.83.075103}.
For $N=30$, $H^{(\pm)}_\text{def}$ has the same time reversal property for both $\mu=0$ and $\mu>0$ which corresponds to the Gaussian unitary ensemble (GUE) \cite{PhysRevB.83.075103,You:2016ldz}, hence we can safely compare our results with the adjacent gap ratio of GUE $r_\text{GUE}=\frac{2\sqrt{3}}{\pi}-\frac{1}{2}$ and that for the Poisson distribution $r_\text{Poisson}=2\log 2-1$.
In contrast to the result obtained in \cite{Garcia-Garcia:2019poj}, here we find that the adjacent gap ratio is close to $r_\text{GOE}$ over whole the spectrum, which implies that the system is chaotic for any values of $\mu$ and the energy scale (temperature); there are no chaotic/integrable transition.
This is consistent with the proposal in \cite{Garcia-Garcia:2019poj}.

\section{Large $N$, large $q$ analysis}
\label{sec_largeq}
In the large $q$ limit, we can study the mass deformed SYK model analytically beyond the low energy approximation.
In this section we study this limit to confirm the validity of the low energy approximation and the observation by the numerical analysis of finite $q$ model in the region where we do not use the low energy approximation.
In the large $q$ limit, the $G,\Sigma$ action reduces to the Liouville action:
\footnote{
%
Note that in the case of the undeformed SYK model, the Liouville action does not capture the infinite number of modes in the OPE expansion with dimension  $h_m = 2m + 1 + \mathcal{O}(1/q)$ \cite{Maldacena:2016hyu} which contribute in the higher point functions even at the leading order in the large $q$ limit \cite{Gross:2017hcz, Gross:2017aos}.
However, this fact does not affect the two point function itself, and hence does not affect the leading part of the free energy in the large $N$ limit where the multi point functions factorize to the two point functions.
We believe the situation is the same also for the mass deformed SYK model.}
\ba
\f{S_E}{N} &=& \f{1}{16q^2} \int d\tau_1 \int d\tau_2 \Big( \partial_{\tau_1} g(\tau_1,\tau_2) \partial_{\tau_2}g(\tau_1,\tau_2) - \partial _{\tau_1 }g_{\text{off}}(\tau_1,\tau_2)\partial_ {\tau_2 }g_{\text{off}}(\tau_1,\tau_2) \Big) \notag \\
&& -\f{\mathcal{J}^2}{4q^2} \int d\tau_1 \int d \tau_2  e^{g(\tau_1,\tau_2)} - \f{\hat{\mu}}{4 q^2}\int d\tau g_{\text{off}}(\tau,\tau),
\label{largeqLiouville}
\ea
with the large $q$ expansion 
\ba
G(\tau) &=& \f{1}{2} \text{sgn}(\tau)\Big( 1 + \f{1}{q}g(\tau) + \cdots \Big) ,\notag \\
G_{\text{off}}(\tau)&=& \f{i}{2} \Big( 1 + \f{1}{q} g_{\text{off}}(\tau) + \cdots \Big) ,
\ea
and we also scale $\mu$ so that $\hat{\mu} = \mu q$ is kept finite in the large $q$ limit.
The derivation is shown in the appendix \ref{appendix_ZinGandSigma}.
At small temperature and the late time of order $\tau \sim q$, this approximation is not valid because of the exponential decay of the correlation functions.
In this case, we also consider the solution in $\tau \gg q $ regime and impose the matching condition between $\tau \ll q$ and $\tau \gg q$ solutions.

\subsection{large $q$ limit at zero temperature}
\begin{figure}[ht]
\begin{minipage}{0.49\hsize}
\begin{center}
\includegraphics[width=7cm]{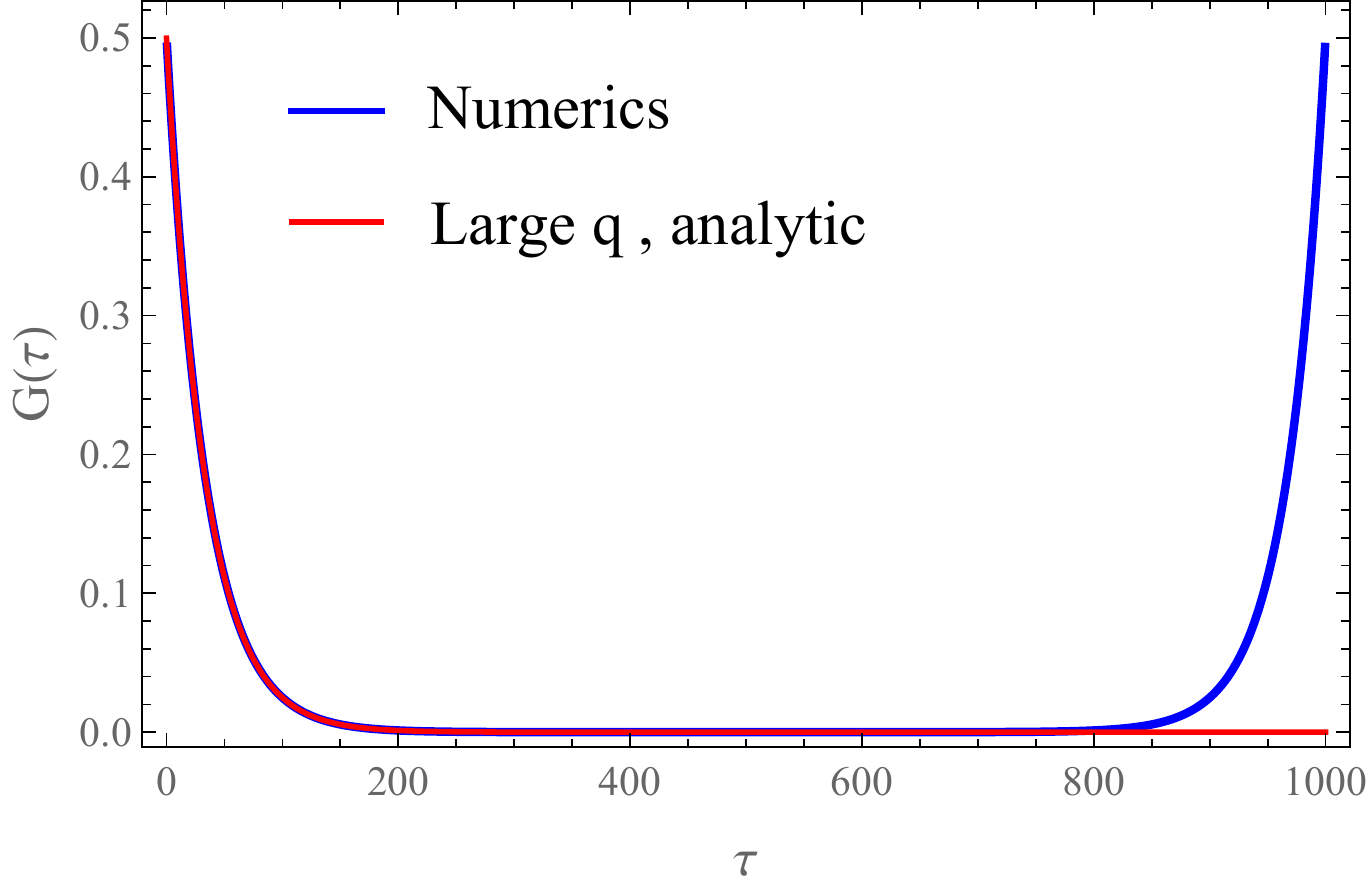} 
\end{center}
\end{minipage}
\begin{minipage}{0.49\hsize}
\begin{center}
\includegraphics[width=7cm]{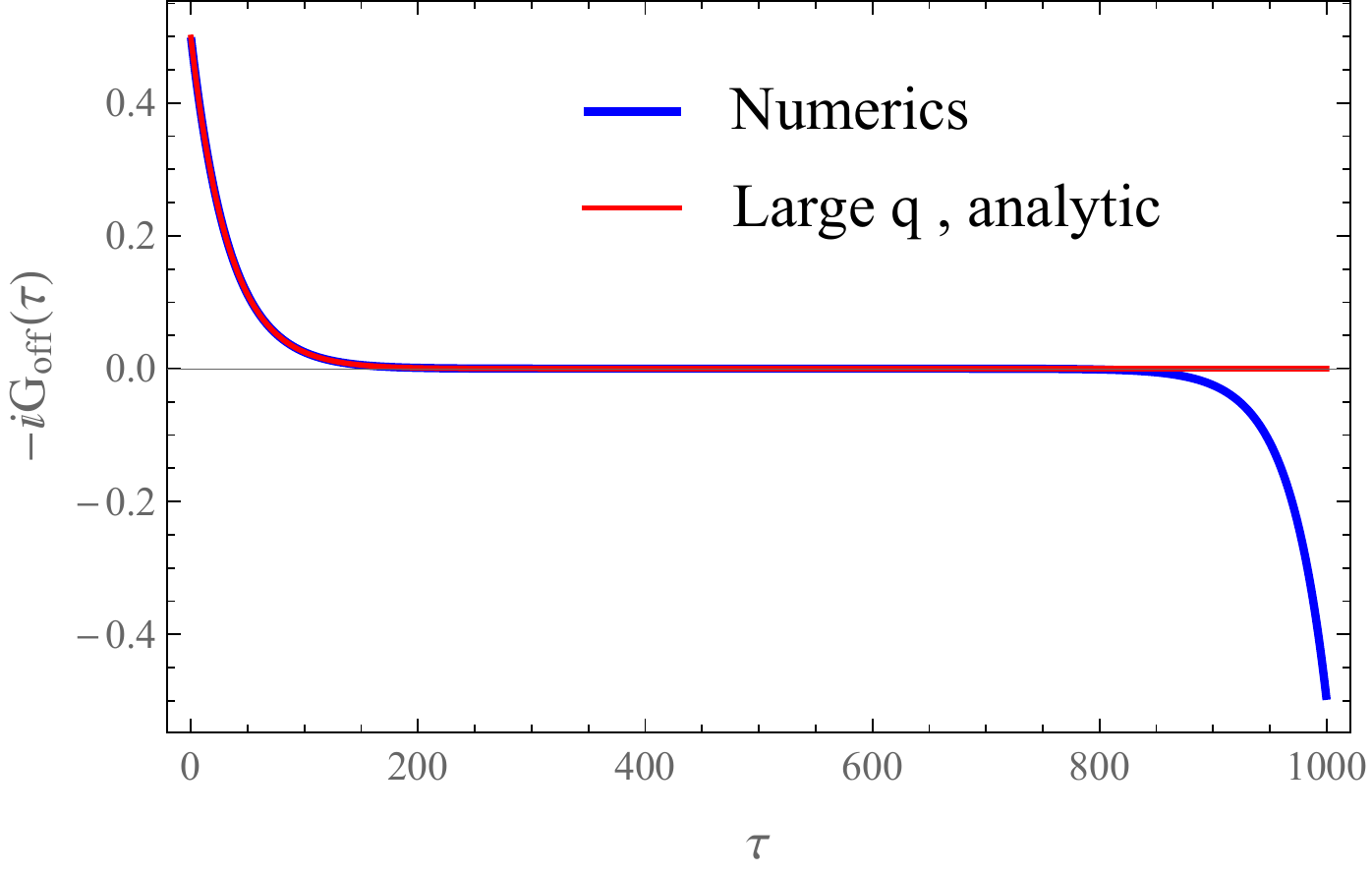}
\end{center}
\end{minipage}
\caption{ The plot of Green functions. For the numerical solution of the Schwinger-Dyson equation, we put $q=96$, $\mathcal{J}=1$ $\mu = 0.03$ and $\beta = 1000$. }  
\label{fig:q96Gplot}
\end{figure}

At large $q$ limit we can write the correlators as 
\ba
G(\tau) &=& \f{1}{2} \text{sgn}(\tau)\Big( 1 + \f{1}{q}g(\tau) + \cdots \Big), \notag \\
G_{\text{off}}(\tau)&=& \f{i}{2} \Big( 1 + \f{1}{q} g_{\text{off}}(\tau) + \cdots \Big).
\ea
In the mass deformed theory, it is convenient to consider to scale the mass term $\mu = \hat{\mu}/q$ and keep $\hat{\mu}$ when $q\to \infty$.
The Schwinger-Dyson equation reduces to the following two equation:
\ba
\partial_{\tau}^2 g(\tau) &=& 2 \mathcal{J}^2e^{g(\tau)}, \qquad (\text{for} \ \ \ \tau > 0) \notag \\
\partial_{\tau}^2 g_{\text{off}}(\tau) &=& -2 \hat{\mu} \delta(\tau), \label{eq:liouville1}
\ea
with the boundary conditions 
\ba
&&g(0) = 0, \qquad  \partial _{\tau} g_{\text{off}}(0^+) = - \hat{\mu}, \notag \\
&&g(\tau)- g_{\text{off}}(\tau)\to 0 , \qquad  \text{as} \ \ \  \tau \to \infty. \label{eq:bdycond1}
\ea
The general solutions of the equations (\ref{eq:liouville1}) become
\ba
e^{g(\tau)} &=& \f{\alpha^2 }{\mathcal{J}^2 \sinh^2(\alpha |\tau| + \gamma)}, \notag \\
e^{g_{\text{off}}(\tau)} &=& \f{4 \tilde{\alpha}^2}{\mathcal{J}^2}e^{-2\tilde{\gamma}}  e^{-2\tilde{\alpha}|\tau|},
\ea
with constants of the integration $\alpha, \tilde{\alpha}, \gamma, \tilde{\gamma}$.
Each boundary condition (\ref{eq:bdycond1}) fixes the constants of integration in a following way
\ba
&&g(0) = 0 \qquad \Rightarrow \qquad  \f{\alpha}{\mathcal{J} \sinh\gamma } = 1 ,\notag \\
&&\partial_{\tau}g_{\text{off}}(0^+) = -\hat{\mu} \qquad  \Rightarrow \qquad   \ 2\tilde{\alpha} =  \hat{\mu}, \notag \\
&&g(\tau)- g_{\text{off}}(\tau)\to 0 , \qquad  \text{as} \ \  \tau \to \infty \qquad  \Rightarrow \qquad  \tilde{\gamma} = \gamma, \qquad \alpha = \tilde{\alpha}.
\ea
This means
\be
4\mathcal{J} \sinh \gamma = 2 \hat{\mu}\qquad  \to \qquad e^{-\gamma} = -\f{\hat{\mu}}{2\mathcal{J}} + \f{\s{\hat{\mu}^2 + 4 \mathcal{J}^2}}{2\mathcal{J}} . \label{eq:gammarelation1}
\ee
This solution for $\tau_1 - \tau_2 > 0$ can be written as 
\be
e^{g(\tau_1,\tau_2)} = \f{h_1'(\tau_1)h_2'(\tau_2)}{\mathcal{J}^2 (h_1(\tau_1) - h_2(\tau_2))^2 }, \qquad e^{g_{\text{off}}(\tau_1,\tau_2)} = f_1(\tau_1)f_2(\tau_2),
\ee
where
\ba
h_1(\tau) &=& \tanh \Big(\alpha \tau + \f{\gamma}{2} \Big) , \qquad h_2(\tau) = \tanh \Big(\alpha\tau - \f{\gamma}{2} \Big) , \notag \\
f_1(\tau) &=& \f{2\alpha}{\mathcal{J}} e^{-\gamma}e^{- 2 \alpha \tau},  \qquad  f_2(\tau) = \f{2\alpha}{\mathcal{J}} e^{-\gamma}e^{ 2 \alpha \tau}.
\ea
We can compare the analytic results here and the numerical solution for the Schwinger-Dyson equation for sufficiently large $q$, and they show good agreement, see Fig.\ref{fig:q96Gplot} .

Using (\ref{eq:gsenergy}), we can compute the ground state energy of the deformed SYK model in the large $q$ limit as follows.
\ba
\f{E}{N}
&=& -\f{1}{2q^2} \f{\hat{\mu}}{\tanh \gamma} - \f{\hat{\mu}}{4q} \Big( 1 - \f{2}{q} + \f{2}{q} \log ( 2 e^{-\gamma} \sinh \gamma) \Big) + \mathcal{O} (q^{-3}) \notag \\
&=& -\f{\mathcal{J}}{q^2} e^{-\gamma} - \f{\hat{\mu}}{4q} \Big( 1 + \f{2}{q} \log ( 2 e^{-\gamma} \sinh \gamma) \Big) + \mathcal{O} (q^{-3}). \label{eq:gdenergylargeq1}
\ea
In small $\hat{\mu}$ limit, we can approximate $\gamma \sim \f{\hat{\mu}}{2 \mathcal{J}} \ll 1$.
In this limit, the ground state energy becomes 
\be
\f{E}{N} = -\f{\mathcal{J}}{q^2} - \f{\hat{\mu}}{4q} + \f{\hat{\mu}}{2 q^2} \Big(1 - \log \f{\hat{\mu}}{\mathcal{J}} \Big).
\ee
The first term is the ground state energy of the SYK model at large $q$ limit.
As a consistency check, this result agrees with the ground state energy obtained in the conformal approximation \eqref{eq:GDEnConf} expanded at $q\gg 1$ with ${\cal J},{\widehat \mu}$ kept fixed.

Given the ground state correlation function, we can compute the several physical observables again.
The spin operator expectation value becomes 
\be
\f{1}{2}\braket{G_{\bm{s}}(\mu)|S_k|G_{\bm{s}}(\mu)} = \f{1}{2}s_k \Big(1 + \f{1}{q} g_{\text{off}}(0) \Big) =\f{1}{2}s_k \Big(1 + \f{2}{q} \log (2 e^{-\gamma}\sinh \gamma) \Big).
\ee
The SYK energy evaluated on the ground state of the deformed Hamiltonian is 
\be
\braket{G_{\bm{s}}(\mu)|H_{SYK}|G_{\bm{s}}(\mu)} = -\f{1}{2q^2} \f{\hat{\mu}}{\tanh \gamma} + \f{\hat{\mu}}{2q^2}.
\ee
In $\hat{\mu} \to 0$ limit, using $\gamma \sim \f{\hat{\mu}}{2\mathcal{J}}$ the first term becomes the SYK ground state energy.
Therefore $\ket{G_{\bm{s}}(\mu)}$ serves an excited state of the SYK model with energy higher than the ground state by $\f{\hat{\mu}}{2q^2}$.
In $\hat{\mu} \to \infty$ limit, $\gamma$ becomes $\infty$ and the $\ket{G_{\bm{s}}(\mu)}$ have the $0$ energy in the SYK Hamiltonian, which is expected to the state $\ket{B_{\bm{s}}}$ \cite{Kourkoulou:2017zaj}.

\subsection{variational approximation in the large $q$ limit}
\label{section_variational}
We can also study the variational approximation of the ground state of the deformed Hamiltonian by the SYK black hole microstate analytically even beyond the low energy approximation.
In large $q$ limit, the trial energy (\ref{eq:trialenergy1}) becomes 
\ba
\f{\braket{H_{SYK} + H_M}_{B_{\bm{s}}}}{N} 
&=& -\f{1}{q^2} \f{\check{\alpha}}{\tan \check{\gamma}} - \f{\hat{\mu}}{4q} \Big( 1 + \f{4}{q} \log \f{\check{\alpha}}{\mathcal{J}}\Big).
\ea
Using $\check{\alpha} = \mathcal{J}\sin \check{\gamma}$, this can be rewritten as 
\be
 \f{\braket{H_{def}}}{N} = -\f{\mathcal{J}}{q^2}\cos \check{\gamma} - \f{\hat{\mu}}{4q} \Big( 1 + \f{4}{q} \log \sin \check{\gamma}\Big). \label{eq:variationalenergylargeq1} 
\ee
Because we are considering the variational method, we minimize the trial energy as a function of $\check{\gamma}$ with fixing $\hat{\mu}$:
\be
\f{\partial}{\partial \check{\gamma}}  \f{\braket{H_{def}}}{N} = -\f{\mathcal{J}}{q^2 } \sin \check{\gamma} + \f{\hat{\mu}}{q^2}\f{1}{\tan \check{\gamma}} = 0.
\ee
This becomes
\be
\f{\hat{\mu}}{\mathcal{J}} = \f{\sin ^2 \check{\gamma}}{\cos \check \gamma}.
\ee
The R.H.S is monotonic on $\check{\gamma} \in [0,\f{\pi}{2}]$ that runs from $0$ to $\infty$ and this have the unique solution.
This is solved as 
\be
\cos \check{\gamma} = -\f{\hat{\mu}}{2\mathcal{J}} + \f{\s{\hat{\mu}^2 + 4 \mathcal{J}^2}}{2\mathcal{J}}.
\ee
Together with the relation with $\gamma$ and $\hat{\mu}$ (\ref{eq:gammarelation1}), we can also write the matching condition as 
\be
e^{-\gamma} = \cos \check{\gamma} \label{eq:matchingcond1}.
\ee
The inverse temperature is given by
\be
\mathcal{J}\beta (\hat{\mu}) = \f{\pi - 2 \check{\gamma}}{\sin \check{\gamma}}.
\ee
For small $\hat{\gamma}$, $\f{\hat{\mu}}{\mathcal{J}} \approx \hat{\gamma}^2$ and the temperature $\beta(\hat{\mu})$ is approximately
\be
\mathcal{J}\beta (\hat{\mu})  \approx \pi \s{\f{\mathcal{J}}{\hat{\mu}}} .
\ee
On the other hand, for large $\hat{\mu}$, we can approximate $\f{\hat{\mu}}{\mathcal{J}} \approx \f{1}{\f{\pi}{2}-\hat{\gamma}}$ and the temperature is approximated as
\be
\mathcal{J}\beta (\hat{\mu}) \approx \f{2\mathcal{J}}{\hat{\mu}}.
\ee
Note that in the large $\beta$ limit, using $c_{\Delta} \approx \f{1}{2}, \alpha_S \approx \f{1}{4q^2}$ and $\Delta = \f{1}{q}$  the low energy approximation the low energy approximation for the relation
$
\f{1}{\beta(\mu)\mathcal{J}} = \f{1}{\pi}\Big( \f{ \mu (c_{\Delta})^2 \Delta}{\mathcal{J}\alpha_S}\Big)^{\f{1}{2(1-2\Delta)}}
$
reduces to 
\be
 \f{1}{\beta(\mu)\mathcal{J}}  \approx \f{1}{\pi} \s{\f{\hat{\mu}}{\mathcal{J}}},
\ee
which completely agrees with the small $\hat{\mu}$ limit of the large $q$ answer.

Using the matching condition (\ref{eq:matchingcond1}), we find that  the exact ground state energy (\ref{eq:gdenergylargeq1}) and the variational energy (\ref{eq:variationalenergylargeq1}) actually exactly agree up to the order of $q^{-2}$.
This means that in large $q$ limit the black hole microstate $\ket{B_{\bm{s}}(\beta)}$ is the same state with the ground state of the deformed Hamiltonian!
Later we will confirm this fact by computing the overlap between $\ket{B_{\bm{s}}(\beta)}$ and $\ket{G_{\bm{s}}(\mu)}$ at large $q$ limit using the Liouville action.

\subsection{large $q$ limit at finite temperature}
In this section we consider the large $q$ limit at finite temperature.
One motivation is to confirm the absence of the Hawking-Page type phase transition in the mass deformation in this paper at large $q$ limit.
In large $q$ limit, $\Sigma$ varies over a relatively short time, which is of order one.
Moreover, (\ref{eq:selfe2}) shows that $\Sigma_{\text{off}}$ is proportional to the delta function.
On the other hand, $G$ and $G_{\text{off}}$ vary with the time scale of order $q$.
Using these separation of the time scales, we can approximate the convolution (\ref{eq:SD2}) as follows.
$\Sigma(\tau)$ is an odd function of $\tau$, and we can approximate $\Sigma(\tau)\sim \delta'(\tau)$.
Therefore, we can approximate the integral 
\ba
\int d\tau'' \Sigma(\tau,\tau'') G(\tau'',\tau')  &\propto&   \partial_{\tau} G(\tau,\tau'),\notag \\
\int d\tau'' \Sigma(\tau,\tau'') G_{\text{off}}(\tau'',\tau') &\propto&  \partial_{\tau} G_{\text{off}}(\tau,\tau').
\ea
However since $\Sigma(\tau,\tau'')$ contains the factor $1/q$ and the equations already contain $\partial_{\tau} G$ and $\partial _{\tau}G_{\text{off}}$, we can ignore the term that contains $\Sigma$.
Because we are considering the large $\tau$ regime, we can also ignore the term $\delta(\tau-\tau')$ in the right hand side of the Schwinger-Dyson equation (\ref{eq:selfe2}).
Therefore, we obtain the equation 
\ba
&&\partial_{\tau} G(\tau,\tau') - i \mu G_{\text{off}}(\tau,\tau')  = 0 ,\notag \\
&&\partial_{\tau} G_{\text{off}}(\tau,\tau') + i \mu G(\tau,\tau') = 0.
\ea
The finite temperature solution is 
\be
G(\tau) = A \cosh [\mu (\beta/2 - \tau)], \qquad G_{\text{off}}(\tau) = i A \sinh[\mu (\beta/2 - \tau)].
\ee
When we expand them in $\tau$, we obtain
\ba
G(\tau) &=& A \cosh \f{\beta\mu}{2} - \mu \tau A \sinh \f{\beta \mu}{2} + \cdots , \notag \\
-i G_{\text{off}}(\tau)&=&   A \sinh \f{\beta\mu}{2} -    \mu \tau A \cosh \f{\beta\mu}{2} + \cdots. \label{eq:qlogqlateearly}
\ea

In the following, we study the thermodynamical properties of the Hamiltonian $H_{def}$ in the large $q$ limit.
We study the inverse temperature regime of order $q\log q, q , \s{q}$ and $1$.
The derivations are skipped here and  shown in appendix \ref{sec:largeqdetail}.
 
\subsubsection{Inverse temperature of order $\beta = q \log q$}
\label{sec_largeqbetaqlogq}
In this regime, it is convenient to use the parameter $\sigma = qe^{-\beta \mu}$, which is of order one quantity in this temperature regime.
In this temperature regime, we can still use the large $q$ expansion $G(\tau) = \f{1}{2}(1 + \f{1}{q} g(\tau) \cdots )$ and $G_{\text{off}}(\tau) = \f{i}{2}(1 + \f{1}{q} g_{\text{off}}(\tau) \cdots )$ at early time.
The solution for $ \tau \ll q$ becomes 
\be
e^{g(\tau)} = \f{\alpha ^2}{\mathcal{J}^2 \sinh^2(\alpha |\tau| + \gamma)} ,\qquad e^{g_{\text{off}}(\tau)} = \f{\tilde{\alpha} ^2}{\mathcal{J}^2 \sinh^2(\tilde{\alpha} |\tau| + \tilde{\gamma})},
\ee
with 
\be
\tilde{\alpha} = \alpha, \qquad \tilde{\gamma} = \gamma+\sigma,\qquad \hat{\mu} = 2\tilde{\alpha}, \qquad  \alpha = \mathcal{J} \sinh \gamma,
\ee
and for $\tau \gg q$
\be
G(\tau) =  \f{1}{2} \cosh \Big[ \mu \Big(\f{\beta}{2} - \tau \Big) \Big], \qquad G_{\text{off}}(\tau) =  \f{i}{2} \sinh \Big[ \mu \Big(\f{\beta}{2} - \tau \Big) \Big].
\ee
The thermal energy, thermal free energy and the thermal entropy are
\begin{align}
\f{E}{N} &= -\f{1}{2q^2} \f{\hat{\mu}}{\tanh \gamma} - \f{\hat{\mu}}{4q}  - \f{\hat{\mu}}{4q} \Big( 1 - \f{2}{q} + \f{2}{q} \log ( \sinh \gamma e^{-\tilde{\gamma}} )\Big), \notag \\ 
-\f{\beta F}{N} &= \f{\beta \hat{\mu}}{2q^2} \Big( \f{q}{2} - 1 +\f{1}{\tanh \gamma} + \log (2\sinh \gamma e^{-\tilde{\gamma}}) + \sigma \Big) + \f{\sigma}{2q}, \notag \\
\f{S}{N} &= \f{1}{2} \f{\sigma}{q} \Big( 1 + \f{q}{\sigma} \Big) = \f{1}{2}e^{-\beta \mu}(1 + \beta \mu). 
\end{align}
We can also rewrite the free energy as 
\be
-\f{\beta F}{N} = \f{\beta\mu}{4}+ \f{e^{-\beta \mu}}{2}  + \f{\beta\mu}{4q} \Big[ \log (2\sinh \gamma) + \f{1}{\tanh \gamma}  -\gamma -1 \Big].\label{eq:Fnqloq}
\ee
where $\hat{\mu} = 2\mathcal{J}\sinh\gamma$ is a function of only $\mu$.
In this expression, it is clear that the free energy is a monotonic, smooth function of $\beta$ in this temperature regime.
This means that there are no phase transition in the large $q$ limit.
We observe the absence of the phase transition numerically in large $N$ finite $q$ case in section \ref{section_thermodynamic}, and the large $q$ analysis here is consistent with this observation.
This is in contrast with the two coupled SYK model \cite{Maldacena:2018lmt} where that model have a phase transition in the same temperature regime.
The main difference from that model is that here the temperature is a monotonic function of $\sigma$.
Because of this, we always have one solution for a given temperature and we do not have phase transition.

\subsubsection{Inverse temperature of order $\beta = q $}
In this order, the off diagonal correlator $|G_{\text{off}}(\tau)|$ is smaller than $1/2$ everywhere and we cannot use the same large $q$ expansion  for the off diagonal correlator as we did in the last subsection.
We can still assume the large $q$ expansion $G(\tau) = \f{1}{2}(1 + \f{1}{q} g(\tau) + \cdots)$ for the diagonal correlation function.
The correlation function for $\tau \ll q$ becomes 
\be
e^{g(\tau)} = \f{\alpha^2}{ \mathcal{J}^2 \sinh^2 (\alpha|\tau| + \gamma)},
\ee
with
\be
\alpha = \f{\hat{\mu}}{2} \tanh \f{\beta \mu}{2}, \qquad \alpha = \mathcal{J} \sinh \gamma,
\ee
and for $\tau \gg q$, 
\be
G(\tau) = \f{1}{2} \f{\cosh [ \mu (\f{\beta}{2} - \tau)]}{\cosh \f{\beta \mu}{2}}, \qquad G_{\text{off}}(\tau) = \f{i}{2} \f{\sinh [ \mu (\f{\beta}{2} - \tau)]}{\cosh \f{\beta \mu}{2}}.
\ee
The free energy becomes 
\be
-\f{\beta F}{N} = \f{1}{2}\log \Bigl(2\cosh \f{\beta \mu}{2}\Bigr) + \f{\beta \mu}{2q} \tanh \f{\beta \mu}{2} \Big[\log (2 \sinh \gamma) + \f{1}{\tanh \gamma} - \gamma -1 \Big]. \label{eq:Fnq}
\ee
The first term, which is of order one, is the same with the free energy of the free fermionic oscillator.
In the small $\beta$ limit, this becomes $\f{1}{2} \log 2$ which is the leading of the thermal entropy of the SYK model at large $q$ limit.
Therefore, in this regime the entropy is increasing from low entropy regime of order $\beta = q \log q$.
In the large $\beta$ limit, this can be expanded as
\be
-\f{\beta F}{N}  \sim \f{\beta \mu}{4} + \f{1}{2} e^{-\beta \mu} + \f{\beta \mu}{2q}\Big[\log (2 \sinh \gamma) + \f{1}{\tanh \gamma} - \gamma -1 \Big],
\ee
which reproduces the free energy (\ref{eq:Fnqloq}) in the order of $\beta = q \log q$.
In the high temperature limit, we can expand $\gamma$ and $F$ as 
\be
\gamma \sim \f{q(\beta\mu)^2}{4  \beta \mathcal{J}}, \qquad  -\f{\beta F}{N} \sim \f{1}{2} \log 2 + \f{(\beta\mu)^2 }{16} + \f{\beta \mathcal{J}}{q^2} + \f{(\beta \mu)^2}{4 q} \log \f{q(\beta \mu)^2}{4 \beta \mathcal{J}}+\cdots.
\ee

\begin{figure}[ht]
\begin{center}
\includegraphics[width=8cm]{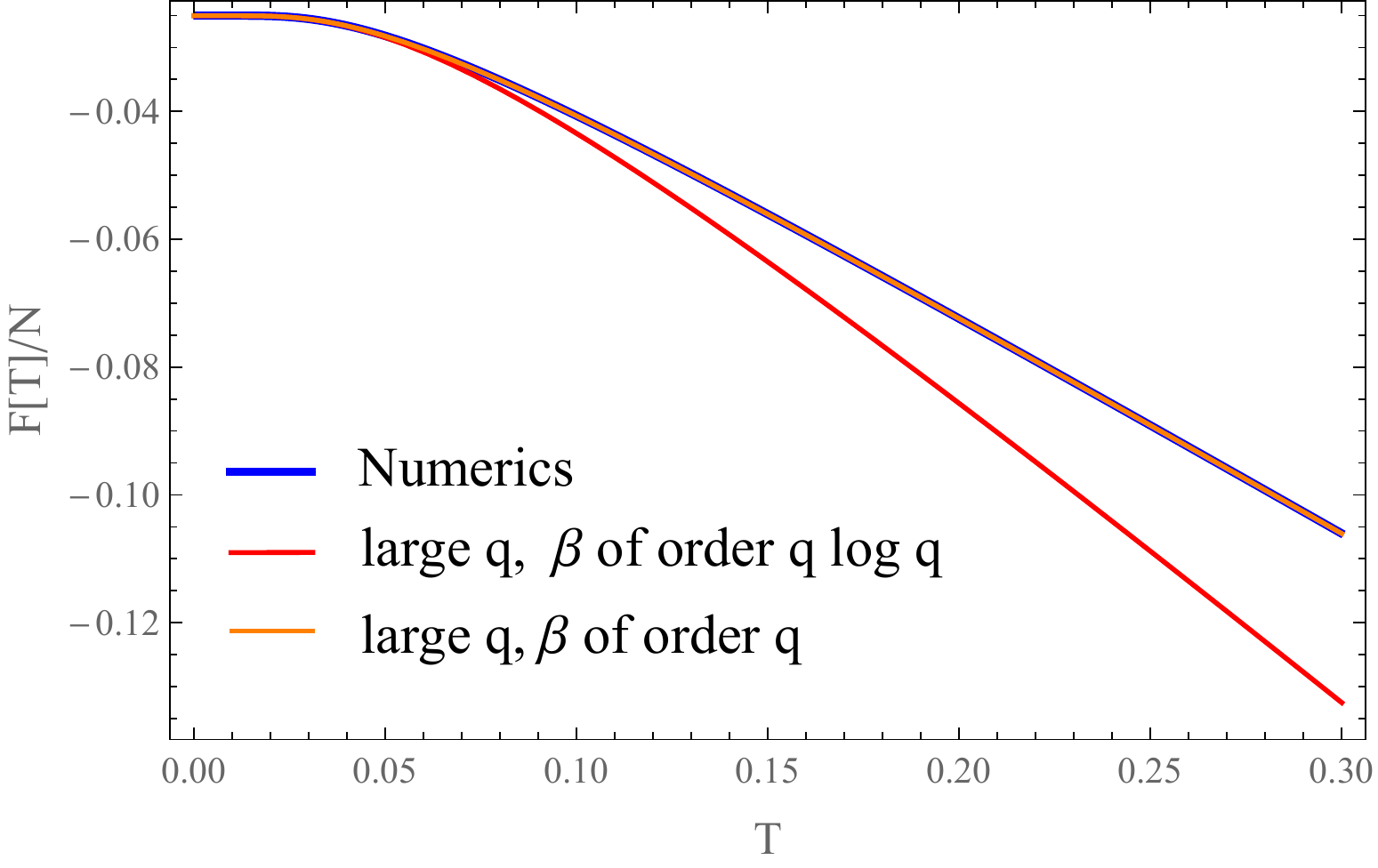}
\caption{Plot of free energy.
The numerical calculation is done for $q =96$, $\mathcal{J} = 1$ and $\mu = 0.1$.
The red and orange lines are analytical results (\ref{eq:Fnqloq}) and (\ref{eq:Fnq}) for the large $q$ limit.
}  
\label{fig:Fnq96plot}
\end{center}
\end{figure}
As a check, we compare the large $q$ results (\ref{eq:Fnqloq}) and (\ref{eq:Fnq}) for free energy with the free energy calculated from the numerical solution for the Schwinger-Dyson equation in Fig.\ref{fig:Fnq96plot}, which shows good agreement.

At this order, we obtain the same results with \cite{Maldacena:2018lmt}.
Actually, we found that both models have the same Schwinger-Dyson equation at this order.
If the temperature is higher than ${\cal O}(q^{-1})$ we still have the same equation of motion and we only reproduce the former results in \cite{Maldacena:2018lmt}, but to make this paper to be self contained, we still continue the finite temperature analysis.

\subsubsection{Inverse temperature of order $\beta = \s{q} $}
In this regime, we can approximate the off diagonal correlation function $G_{\text{off}}$ as 
\be
G_{\text{off}}(\tau) = \f{i}{2} \mu \Big(\f{\beta}{2} - \tau \Big),
\ee
which is of order $\f{1}{\s{q}}$.
The diagonal correlation function $G$ is approximated as $G(\tau) = \f{1}{2} (1 + g(\tau) + \cdots)$ everywhere in $\tau \in [0,\beta]$ and the equation of motion for $g$ becomes 
\be
\partial _{\tau}^2 g (\tau) - 2 \mathcal{J}^2 e^{g(\tau)} - \f{\hat{\mu}^2}{q} = 0.
\ee
The same equation has also appeared  in a different mass deformation of the SYK model \cite{Garcia-Garcia:2017bkg}.
The last term is of order $1/q$, which seems to be ignorable.
But at the time of order $\s{q}$, the other terms are also of the same order.
This can be seen clearly after rescaling as 
\be
x = \f{\tau - \f{\beta}{2}}{\beta} , \qquad e^{\hat{g}} = (\beta \mathcal{J})^2 e^{g}.
\ee
Then, the equation of motion becomes 
\be
\partial_x^2 \hat{g} - 2 e^{\hat{g}} - 2k = 0, \qquad k = \f{q (\mu\beta)^2}{2}.
\ee
The detailed analysis are in appendix \ref{sec:largeqdetail}.

The partition function becomes 
\be
-\f{\beta F}{N} = \f{1}{2}\log 2 + \f{(\beta\mu)^2}{16} + \f{\beta \mathcal{J}}{q^2} - \f{(\beta \mu)^2}{4q} \log (\beta \mathcal{J}) + \f{h(q\beta^2\mu^2)}{q^2},
\ee
where $h(k)$ is a function that we have not determined.

\begin{figure}[ht]
\begin{center}
\includegraphics[width=7cm]{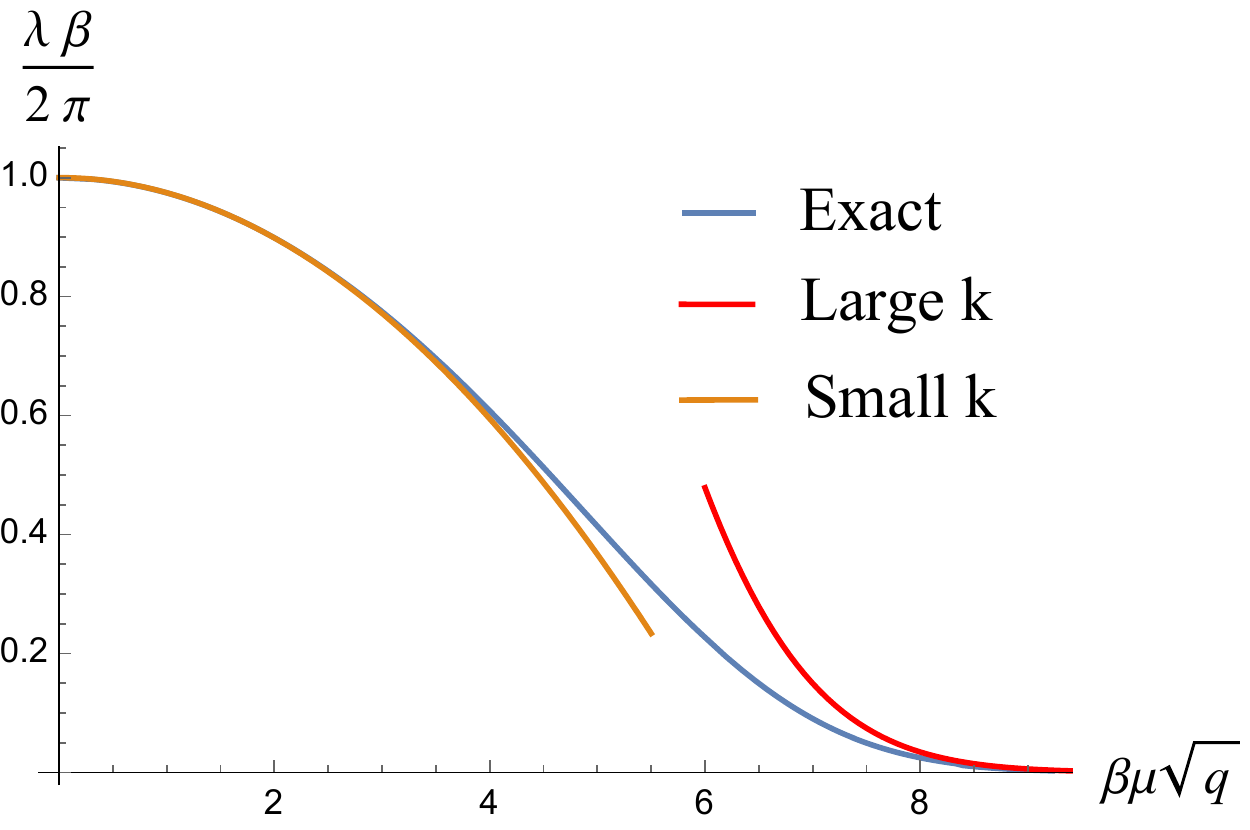}
\caption{Plot of Lyapunov exponent as a function of $\s{q}\beta \mu = \s{2k}$ with exact, small $k$ and large $k$ expansion.}  
\label{fig:lyapunovq1}
\end{center}
\end{figure}

In this regime, the chaos exponent increases from $0$ to the maximal value $\f{2\pi}{\beta}$.
When $k$ is large, the chaos exponent $\lambda$ becomes 
\be
\f{\lambda \beta}{2\pi} \approx \f{1}{\s{\pi}} k^{\f{3}{2}} e^{-\f{1}{4}k},
\ee
and for small $k$ the chaos exponent becomes 
\be
\f{\lambda \beta}{2\pi} \approx 1 - \f{k}{2\pi^2}.
\ee
For finite $k$, we can numerically study the chaos exponent.
The plot is shown in Fig.\ref{fig:lyapunovq1} and the details are shown in the appendix \ref{sec:largeqdetail}.

\subsubsection{Temperature of order $\beta = 1$}
In this limit we can ignore the mass term and we obtain the same physics with the large $q$ SYK model \cite{Maldacena:2016hyu}.
The free energy becomes 
\be
-\f{\beta F}{N} = -\f{\beta F_{SYK}}{N} + \f{(\beta \mu)^2}{16},
\ee
where $F_{SYK}$ is the free energy of the SYK model.
The chaos exponents are maximal when $ 1 \ll \beta \ll \s{q}$ and then decrease to $2 \mathcal{J}$ in the high temperature regime $\beta \ll 1$.

\subsection{Computing the Overlap at large $q$ limit}
\label{section_overlaplargeq}
\begin{figure}[ht]
\begin{minipage}{0.49\hsize}
\begin{center}
\includegraphics[width=4.5cm]{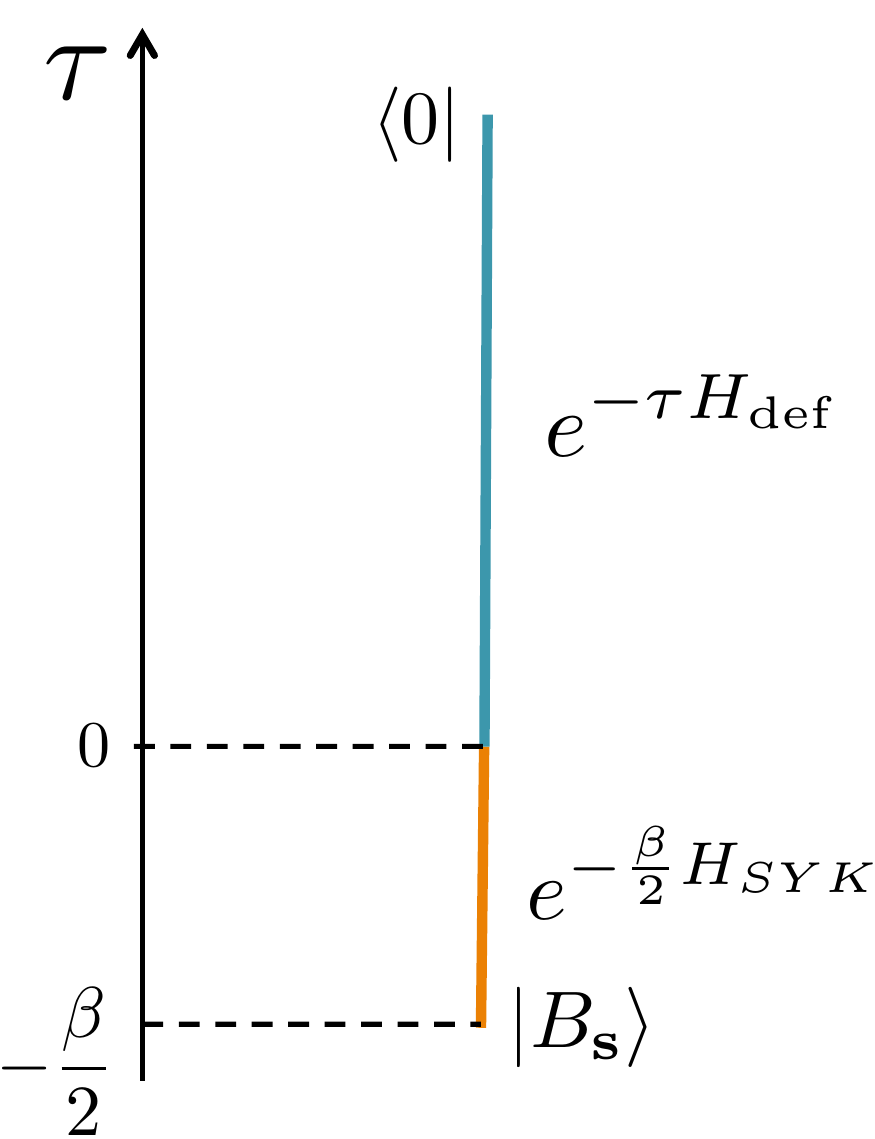} 
\end{center}
\end{minipage}
\begin{minipage}{0.49\hsize}
\begin{center}
\includegraphics[width=6.2cm]{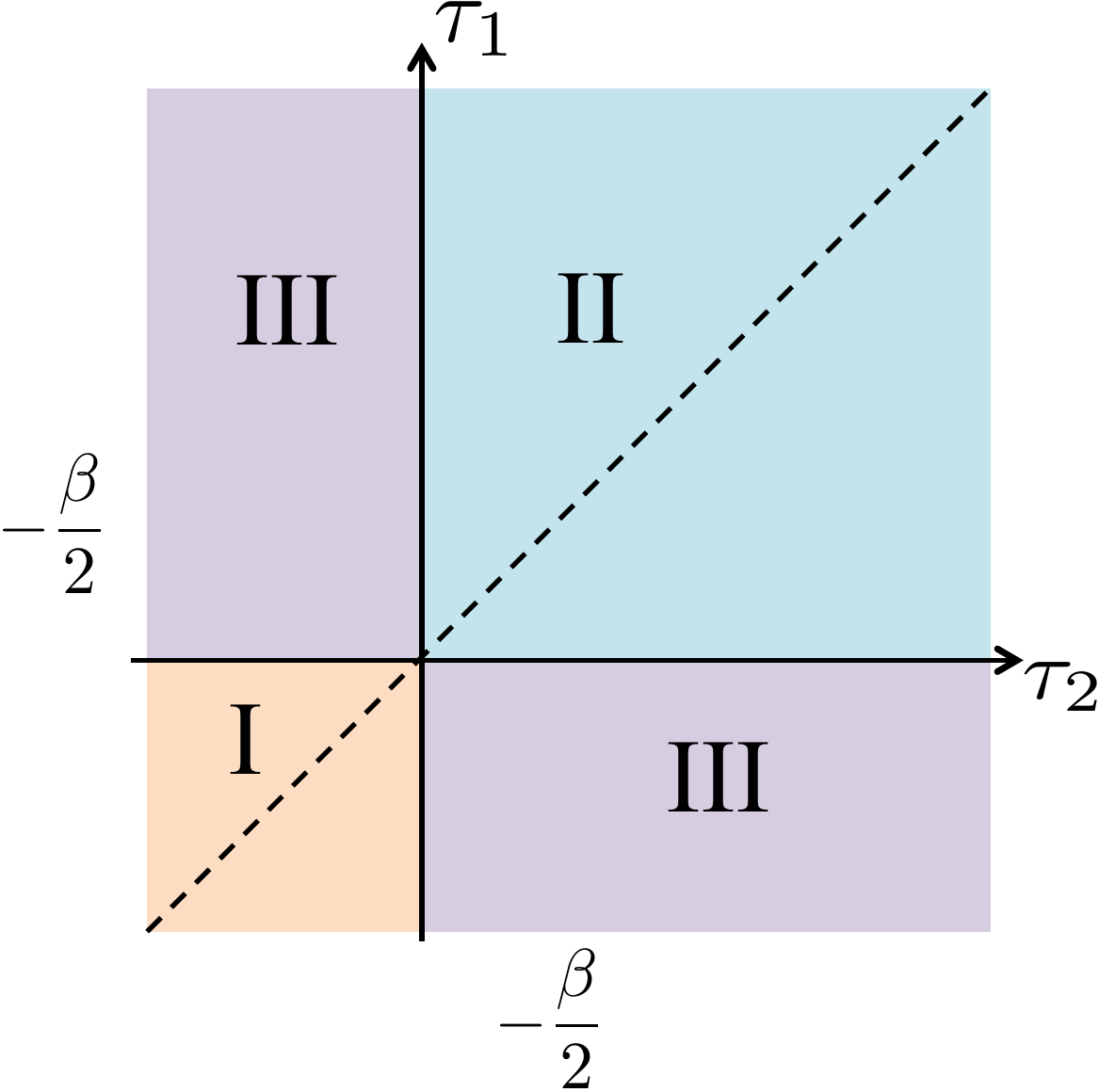}
\end{center}
\end{minipage}
\caption{{\bf Left:} The illustration of the Euclidean path integral for the computation of the overlap.
{\bf Right:} The Euclidean path integral for two time variable.
We divide the region to three parts depending on which Hamiltonian is used in the Euclidean time evolution.}  
\label{fig:overlappathint}
\end{figure}

In the large $q$ limit, we can compute the overlap using the Liouville on shell action.
The strategy is to construct an analog of  ``Janus"  solution \cite{Bak:2003jk} in the large $q$ limit, where a similar holographic computation of the overlap is done in \cite{MIyaji:2015mia}.
The overlap is represented as 
\be
\braket{B_{\bm{s}}(\beta)|G_{\bm{s}}(\mu)} = \lim_{\tau \to \infty} \f{ \braket{B_{\bm{s}} | e^{- \f{\beta}{2}H_{SYK}}e^{-\tau H_{def}}|0}}{\s{\braket{B_{\bm{s}}| e^{- \beta H_{SYK}}| B_{\bm{s}}} } \s{\braket{0 |e^{-2\tau H_{def}}|0 }}     },
\ee
with an initial condition $\ket{0}$, which only changes the normalization constant that should cancel between the numerator and the denominator.
We can treat the Euclidean path integral for the overlap as a Euclidean time dependent coupling where $\tau$ runs in the range $\tau \in [-\f{\beta}{2}, \infty]$ and the time dependent coupling $\mu \theta(\tau)$ as depicted in Fig.\ref{fig:overlappathint}.
After the disorder average, we can again obtain an effective action for $G(\tau_1,\tau_2), \Sigma(\tau_1,\tau_2)$  variables with $\tau_1,\tau_2 \in [-\f{\beta}{2}, \infty]$ and the time dependent mass term $-i \mu \int  _{-\f{\beta}{2}}^{\infty} d\tau \theta(\tau) G_{\text{off}}(\tau,\tau)$.
Because this mass term explicitly depend on the Euclidean time $\tau$, we do not have time translation symmetry and the solution depend on two times $\tau_1 ,\tau_2$.
At $\tau = -\f{\beta}{2}$, the state $\ket{B_{\bm{s}}}$ impose the boundary condition $\psi_{2k-1}\ket{B_{\bm{s}}} = is_k\psi_{2k}\ket{B_{\bm{s}}}$, which leads to the boundary condition
\ba
G\Big(\tau_1 ,-\f{\beta}{2}\Big) = i G_{\text{off}}\Big(\tau_1 ,-\f{\beta}{2}\Big) .
\ea
We also require that the solution approaches to the ground state solution of the deformed Hamiltonian  at  $\tau_1,\tau_2 \to \infty$.

\begin{figure}[ht]
\begin{minipage}{0.40\hsize}
\begin{center}
\includegraphics[width=5.8cm]{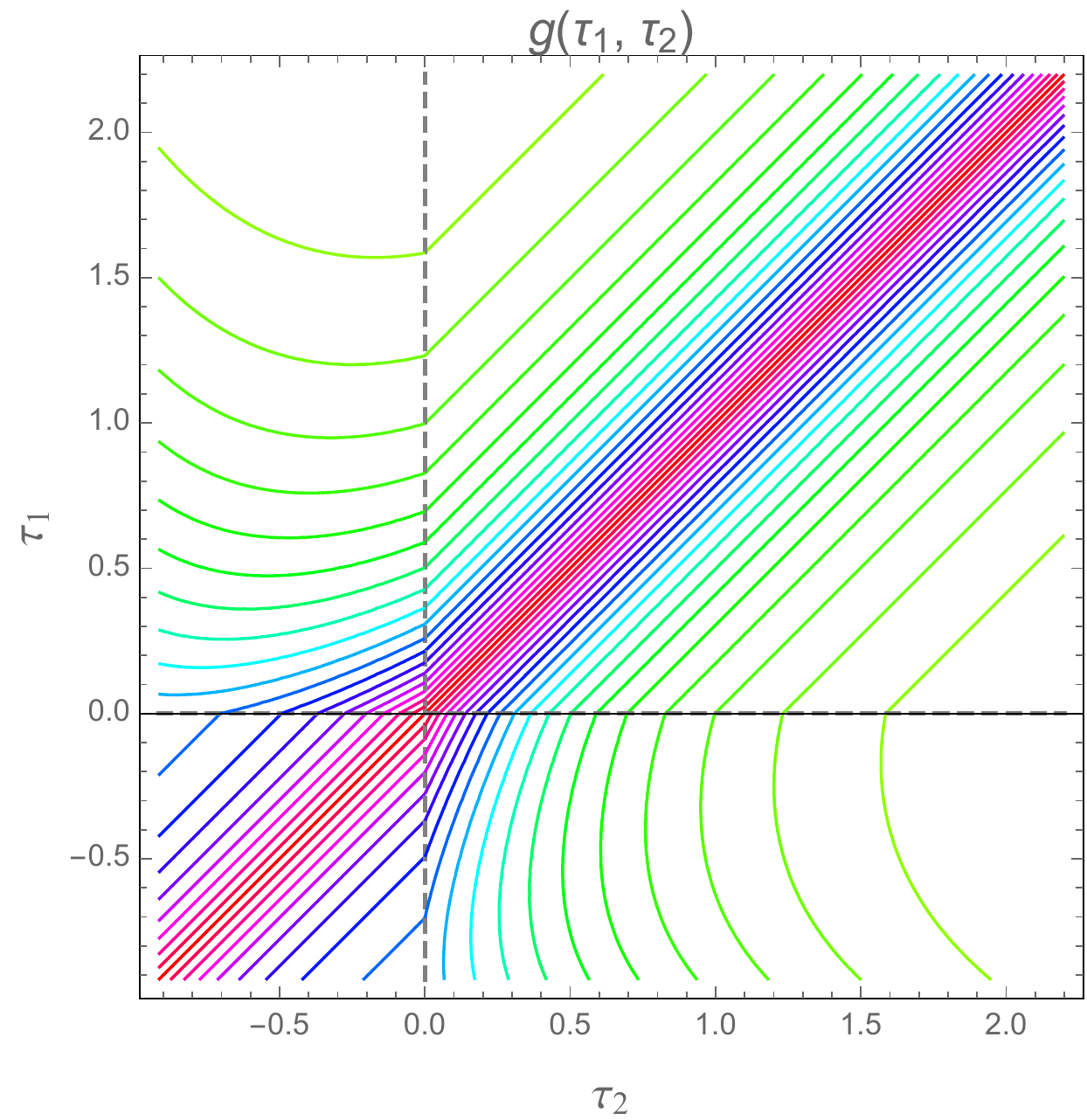}
\end{center}
\end{minipage}
\begin{minipage}{0.40\hsize}
\begin{center}
\includegraphics[width=5.8cm]{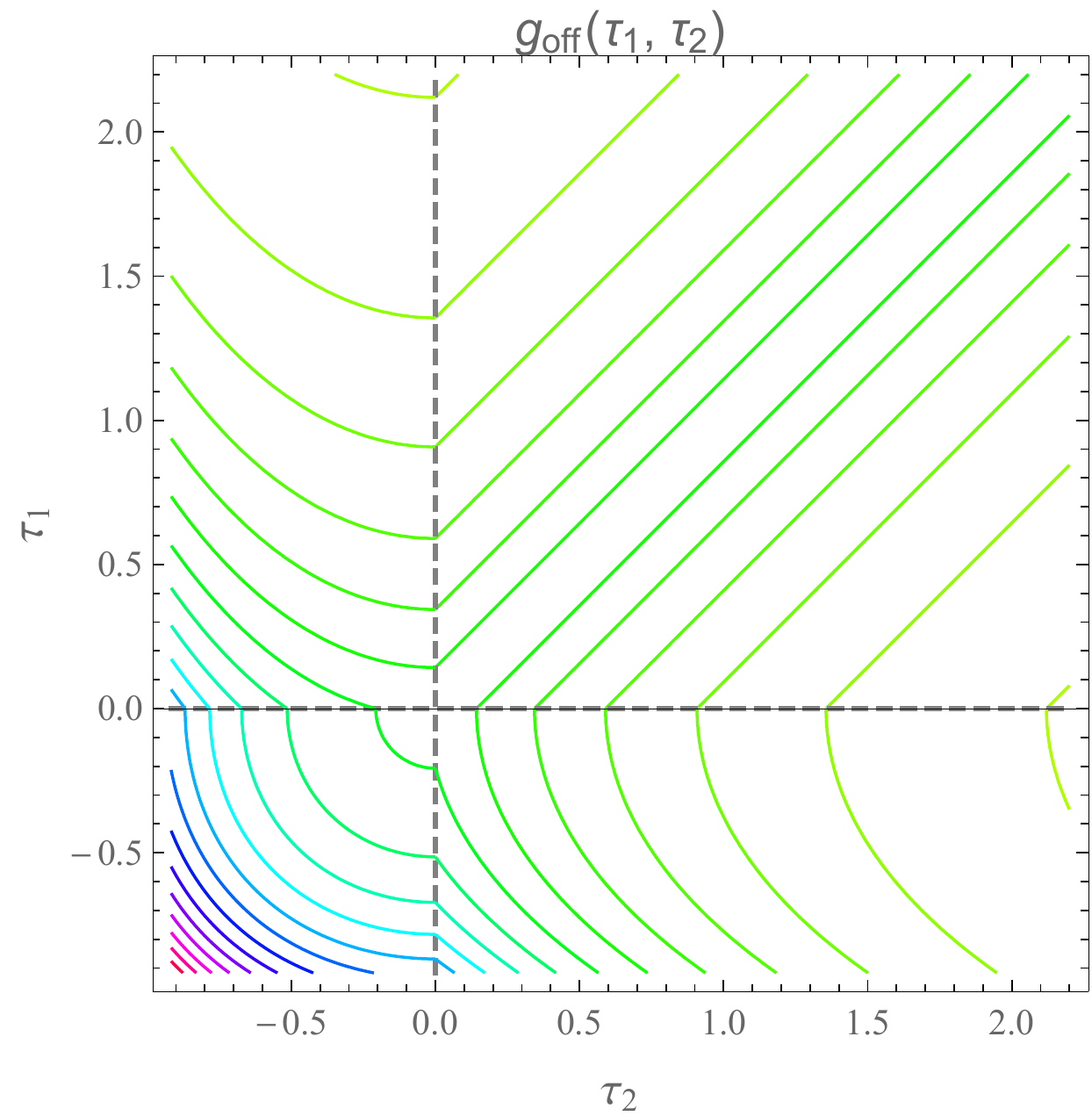}
\end{center}
\end{minipage}
\begin{minipage}{0.15\hsize}
\begin{center}
\includegraphics[width=2.6cm]{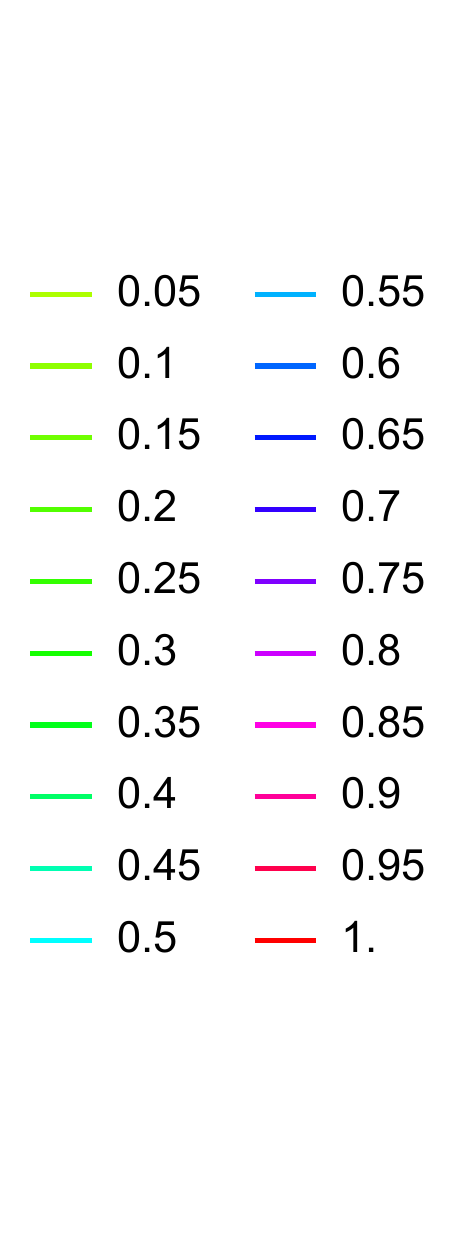}
\end{center}
\end{minipage}
\caption{The contour plot of the two time solutions for the overlap.
In both side we plot $e^{g(\tau_1,\tau_2)}$ and $e^{g_{\text{off}}(\tau_1,\tau_2)}$ as a function of $\tau_1,\tau_2$.
The parameters are taken to be $\check{\gamma} = 0.9$ and $\mathcal{J}= 1$, and the others are determined from $\check{\alpha} = \mathcal{J} \sin \check{\gamma}$, $\alpha = \mathcal{J} \sinh \gamma$ and the matching condition $e^{-\gamma} = \cos \check{\gamma}$. }
\label{fig:twotimesolution}
\end{figure}

In the large $q$ limit, the effective action reduces to the Liouville action.
The difference with the ground state or thermal case is that we do not have time translation symmetry and the Green's functions depend on the two time variables as
\ba
G(\tau_1,\tau_2) &=& \f{1}{2} \text{sgn}(\tau_1-\tau_2) \Big( 1 + \f{1}{q}g(\tau_1,\tau_2) \Big),  \notag \\
G_{\text{off}}(\tau_1,\tau_2) &=& \f{i}{2} \Big( 1 +\f{1}{q} g_{\text{off}}(\tau_1,\tau_2) \Big).
\ea
The field $g$ satisfies the Liouville equation and $g_{\text{off}}$ satisfies the free field equation
\be
\f{\partial^2 g(\tau_1,\tau_2)}{\partial \tau_1\partial \tau_2} = - 2\mathcal{J}^2 e^{g(\tau_1,\tau_2)}, \qquad \f{\partial^2 g_{\text{off}}(\tau_1,\tau_2)}{\partial \tau_1\partial \tau_2} = 0, \qquad \tau_1,\tau_2 \in [-\beta/2,\infty].
\ee
These equation of motion should be satisfied except for the line $\tau_1 = \tau_2$ where we impose 
\be
g(\tau,\tau) = 0 ,\qquad  (\partial _{\tau_1}-\partial _{\tau_2})g_{\text{off}}(\tau_1,\tau_2)|_{\tau_2 \to \tau_1} = -2\hat{\mu}\theta(\tau_1).
\ee
The two time solutions are locally given by
\be
e^{g(\tau_1,\tau_2)} = \f{h_1'(\tau_1)h_2'(\tau_2)}{\mathcal{J}^2 (h_1(\tau_1)-h_2(\tau_2))^2 }, \qquad e^{g_{\text{off}}(\tau_1,\tau_2)} = f_1(\tau_1)f_2(\tau_2).
\ee
The matching condition from the variational method is 
\be
e^{-\gamma} = \cos \check{\gamma}.
\ee
This also gives the relation 
\be
\tanh\f{\gamma}{2} = \tan^2 \f{\check{\gamma}}{2}, \qquad 2e^{-\gamma} \sinh \gamma =  \sin^2 \check{\gamma}.
\ee
Actually we can find the two time solution for maximal overlap. The solution is 
\begin{align}
&\text{I}:
\begin{cases}
\displaystyle
e^{g(\tau_1,\tau_2)} =\f{\check{\alpha}^2 }{\mathcal{J}^2\sin^2 (\check{\alpha}|\tau_1-\tau_2|+\check{\gamma})} \\
\displaystyle
e^{g_{\text{off}}(\tau_1,\tau_2)} = \f{\check{\alpha}^2}{\mathcal{J}^2\cos^2(\check{\alpha}\tau_1)}\f{\check{\alpha}^2}{\mathcal{J}^2\cos^2(\check{\alpha}\tau_2)} \qquad \text{for}\qquad    -\f{\beta}{2}\le\tau_1, \tau_2 \le 0 ,
\end{cases}
\\
\notag
&\text{I\hspace{-.1em}I}:
\begin{cases}
\displaystyle
e^{g(\tau_1,\tau_2)} =\f{\alpha^2 }{\mathcal{J}^2\sinh^2 (\alpha|\tau_1-\tau_2|+\gamma)} \\
\displaystyle
e^{g_{\text{off}}(\tau_1,\tau_2)} = \f{4\alpha^2}{\mathcal{J}^2} e^{-2\gamma} e^{-2\alpha|\tau_1 -\tau_2|} \qquad \text{for} \qquad 0 \le \tau_1, \tau_2 \le \infty,
\end{cases}
\\
&\text{I\hspace{-.1em}I\hspace{-.1em}I}:
\begin{cases}
\displaystyle
e^{g(\tau_1,\tau_2)} = \f{\alpha \check{\alpha}}{\mathcal{J}^2 } \f{\tan  \f{\check{\gamma}}{2}}{\cosh^2 (\alpha \tau_1 +\f{\gamma}{2})  \cos ^2 (\check{\alpha} \tau_2  - \f{\check{\gamma}}{2})} \f{1}{\Big(\tanh(\alpha \tau_1 + \f{\gamma}{2}) - \tan \f{\check{\gamma}}{2} \tan(\check{\alpha}\tau_2 - \f{\check{\gamma}}{2}) \Big)^2} \notag   \\
\displaystyle
e^{g_{\text{off}}(\tau_1,\tau_2)} = \f{2\alpha }{\mathcal{J}} e^{-\gamma} e^{-2\alpha\tau_1} \f{\check{\alpha}^2}{\mathcal{J}^2 \cos^2 (\check{\alpha}\tau_2) } \qquad  \text{for}\qquad   0 \le \tau_1 \le \infty,  -\f{\beta}{2}\le \tau_2 \le 0 .
\end{cases}
\end{align}
In Fig.~\ref{fig:twotimesolution} we have plotted these functions.
More compactly, we can write the solution as 
\be
e^{g(\tau_1,\tau_2)} = \f{h_1'(\tau_1)h_2'(\tau_2)}{\mathcal{J}^2 (h_1(\tau_1)-h_2(\tau_2))^2}, \qquad e^{g_\text{off}(\tau_1,\tau_2)} = f_1(\tau_1)f_2(\tau_2),
\ee
with the region dependent functions
\be
h_1 (\tau) = 
\begin{cases}
\tan\f{\check{\gamma}}{2}\tan(\check{\alpha} \tau + \f{1}{2}\check{\gamma}) \qquad   \tau \in [-\f{\beta}{2},0] \\
\tanh (\alpha \tau + \f{1}{2} \gamma) \qquad \tau>0
\end{cases}
,
h_2 (\tau) = 
\begin{cases}
\tan\f{\check{\gamma}}{2}\tan(\check{\alpha} \tau - \f{1}{2}\check{\gamma}) \qquad \tau \in [-\f{\beta}{2},0] \\
\tanh (\alpha \tau -\f{1}{2} \gamma) \qquad \tau>0
\end{cases},
\ee
\be
f_1 (\tau) = 
\begin{cases}
\displaystyle
\f{\check{\alpha}^2}{\mathcal{J}^2 \cos ^2 (\check{\alpha} \tau )} \qquad   \tau \in [-\tfrac{\beta}{2},0] \\
\f{2\alpha}{\mathcal{J}}e^{-\gamma}e^{-2\alpha\tau} \qquad \tau>0
\end{cases}
,
f_2(\tau) = 
\begin{cases}
\displaystyle
\f{\check{\alpha}^2}{\mathcal{J}^2 \cos ^2 (\check{\alpha} \tau )} \qquad   \tau \in [-\tfrac{\beta}{2},0] \\
\f{2\alpha}{\mathcal{J}}e^{-\gamma}e^{2\alpha\tau} \qquad \tau>0
\end{cases}.
\ee

Now we can compute the overlap in the order of $\f{1}{q^2}$ using the Liouville on shell action.
We denote the on shell action of the Liouville fields for overlap solution as $S$, and use $S_{B_{\bm{s}}}$ for the on shell action for the state $\ket{B_{\bm{s}}(\beta)}$ and $S_{G_{\bm{s}}}$ for the ground state $\ket{G_{\bm{s}}(\mu)}$. 
To compute the overlap, it is convenient to rewrite the Liouville action \eqref{largeqLiouville} using dimensionless coupling $\beta \mathcal{J}$
 and $\beta \mu$.
\ba
\f{S}{N} &=&  \f{1}{8q^2}\int_{-\pi} ^{\infty} d\theta_1 \int _{\theta_1} ^{\infty} d\theta_2  \Big[ (\partial_{\theta_1} g (\theta_1,\theta_2)\partial _{\theta_2} g (\theta_1,\theta_2)- \partial_{\theta_1} g_{\text{off}}(\theta_1,\theta_2)\partial _{\theta_2} g_{\text{off}}(\theta_1,\theta_2) ) \notag \\
&&- \f{(\beta \mathcal{J})^2}{\pi^2} e^{g(\theta_1,\theta_2)} \Big] - \f{\beta \hat{\mu}}{4 \pi q^2} \int _0 ^{\infty} d\theta   g_{\text{off}}(\theta,\theta),
\ea
with $\theta_i = \f{2\pi\tau_i}{\beta}$ for $i=1,2$.
We take the derivative of the action $S$ over $\mathcal{J}$ with $\hat{\mu}$ fixed and the matching condition $\beta =\beta(\mathcal{J},\hat{\mu})$.
Then, we obtain
\ba
&&\f{1}{N} \f{\partial S}{\partial \mathcal{J}}\Bigg|_{\hat{\mu}} \notag \\
 &=& -\f{1}{8\pi^2 q^2 } \f{\partial(\beta\mathcal{J})^2}{\partial \mathcal{J}} \int _{-\pi}^{\infty} d\theta_1 \int _{\theta_1} ^{\infty} d\theta_2 e^{g(\theta_1,\theta_2)} - \f{1}{4\pi q^2} \f{\partial (\beta \hat{\mu})}{\partial \mathcal{J}} \int _0 ^{\infty} d\theta g_{\text{off}}(\theta,\theta) \notag \\
 &=& \f{1}{2\beta\mathcal{J} q^2}\f{\partial(\beta\mathcal{J})}{\partial \mathcal{J}} \int _{-\f{\beta}{2}}^{\infty}d\tau_1 \int _{\tau_1}^{\infty} d\tau_2 \partial_{\tau_1}\partial_{\tau_2}(g(\tau_1,\tau_2) - g_{\text{off}}(\tau_1,\tau_2)) - \f{1}{2\beta q^2} \f{\partial (\beta \hat{\mu})}{\partial \mathcal{J}}  \int _0 ^{\infty} d\tau g_{\text{off}}(\tau,\tau). \notag \\
\ea
Here we again used the fact that we can ignore the contribution from the variation of the field $g,g_{\text{off}}$ because of the equation of motion.
In the third line, we use $\theta_i = \f{2\pi\tau_i}{\beta}$ and the equation of motion for $g,g_{\text{off}}$.
Now, using the property of the two time solution 
\be
\lim_{\tau_2 \to \infty}\partial_{\tau_1} (g(\tau_1,\tau_2)-g_{\text{off}}(\tau_1,\tau_2)  ) = 0,
\ee 
we can integrate over $\tau_2$ in the first term and we obtain 
\ba
\f{1}{N} \f{\partial S}{\partial \mathcal{J}}\Bigg|_{\hat{\mu}} 
&=&-\f{1}{2\beta\mathcal{J} q^2}\f{\partial(\beta\mathcal{J})}{\partial \mathcal{J}} \int _{-\f{\beta}{2}}^{\infty}d\tau_1 \lim_{\tau_2 \to \tau_1+0 } \partial_{\tau_1}(g(\tau_1,\tau_2) - g_{\text{off}}(\tau_1,\tau_2)) \notag \\
&& - \f{1}{2\beta q^2} \f{\partial (\beta \hat{\mu})}{\partial \mathcal{J}}  \int _0 ^{\infty} d\tau g_{\text{off}}(\tau,\tau).
\ea
This means that the derivative of the on shell action only depends on the correlation function on $\tau_1 = \tau_2$ line, and especially that does not depend on the region I\hspace{-.1em}I\hspace{-.1em}I.
Since this two time solution is equal to that of the  boundary state $\ket{B_{\bm{s}}(\beta)}$ in region I and identical to that of the ground state of the deformed Hamiltonian in region I\hspace{-.1em}I, we obtain
\be
\f{\partial}{\partial \mathcal{J}} \log \Bigg[\f{|\braket{B_{\bm{s}}(\beta)|G_{\bm{s}}(\mu)}|}{\s{\braket{B_{\bm{s}}(\beta)|B_{\bm{s}}(\beta)}\braket{G_{\bm{s}}(\mu)|G_{\bm{s}}(\mu)}}}\Bigg]  = \f{\partial}{\partial \mathcal{J}} \Bigg[-S + \f{1}{2}(S_{B_{\bm{s}}} + S_{G_{\bm{s}}}) \Bigg] = 0.
\ee
Since we can explicitly check that the overlap becomes $1$ at $\mathcal{J} = 0$, by integrating the above equation we obtain $|\braket{B_{\bm{s}}(\beta)|G_{\bm{s}}(\mu)}| = 1$ for general $\mathcal{J}$ and $\hat{\mu}$.
Since the Liouville action capture up to $\f{1}{q^2}$ terms in the $\f{1}{q}$ expansion, this overlap computation shows that the overlap behaves as  $e^{-\f{N}{q^3}}$ in large $q$ expansion.
In fact, we observed from the variational approximation that there is a finite difference between $\ket{G_{\bm{s}}(\mu)}$ and $\ket{B_{\bm{s}}(\beta(\mu))}$ even in small $\mu$ regime.

\section{Gravity interpretation}

In this section, we consider the gravity interpretation of the mass deformed SYK model.
Though we do not know the exact dual gravity of the SYK model, we can consider the similar gravity setup as we did for the microstate $\ket{B_{\bm{s}}(\beta)}$ \cite{Kourkoulou:2017zaj}.
Here we take the same approach with  \cite{Kourkoulou:2017zaj} where we consider the gravity configuration with the same symmetry with our SYK setup.
First we consider the ground state $\ket{G_{\bm{s}}(\mu)}$ and its time evolution under the SYK Hamiltonian, and then consider the gravity interpretation.

\subsection{Time evolution under the SYK Hamiltonian}
In this section, we consider the time evolution of the ground state $\ket{G_{\bm{s}}(\mu)}$ under the SYK Hamiltonian $H_{SYK}$.
We can formulate this time evolution as time dependent mass term $H_{\text{def}}(u) = H_{SYK} + \theta(-u)H_M$ where $u$ is the Lorentzian time.
This type of time evolution is called as quantum quench.
A different type of quantum quench and black hole formation was studied in \cite{Maldacena:2018lmt,Chen:2019qqe,Eberlein:2017wah,Bhattacharya:2018fkq}.
The quantum quench with time dependent mass terms are also studied in quantum field theories \cite{Das:2014jna,Caputa:2017ixa}.

We saw that the ground state $\ket{G_{\bm{s}}(\mu)}$ has bigger energy expectation value than the ground state and is an excited state of the SYK model.
Because of the similarity with the state $\ket{B_{\bm{s}}(\beta(\mu))}$, we also expect the similar thermalization for the state $\ket{G_{\bm{s}}(\mu)}$.
We solve this time evolution in the low energy limit where the SYK dynamics is governed by the Schwarzian action.
For $u<0$ with the Lorentzian time $u$, the reparametrization is given by $f(u) = \tan (\alpha(\mu) u )$, which is the Lorentzian version of the reparametrization to obtain the ground state correlation function.
Then, we couple the reparametrization mode $f(u) = \tan (\alpha(\mu) t(u))$ where $t(u)$ is the reparametrization.
For $u >0$, because of the energy conservation, we impose 
\be
E_0 - \f{N \alpha_S}{\mathcal{J}} \{ f(u),u \} = \braket{G_{\bm{s}}(\mu)|H_{SYK}|G_{\bm{s}}(\mu)},
\ee
where $E_0$ is the ground state energy and $-\f{ N \alpha_S}{\mathcal{J}} \{ f(u),u \}$ gives the energy increase from the ground state \cite{Almheiri:2019psf}.
We have already evaluated the right hand side $\braket{G_{\bm{s}}(\mu)|H_{SYK}|G_{\bm{s}}(\mu)}$ in (\ref{eq:SYKEnConf}) and 
the above equation is solved as 
\be
f(u) = \f{a \tanh (\f{\pi}{\beta} u) + b}{c \tanh( \f{\pi}{\beta} u) + d}, \qquad \f{2\pi^2 \alpha_S \mathcal{J}}{ (\beta \mathcal{J})^2} =\alpha(\mu)\f{\Gamma(2\Delta+1)\Gamma(1-\Delta)^2\Gamma(1-4\Delta)}{\Gamma(\Delta)^2\Gamma(1-2\Delta)^3}, 
\ee 
with $\begin{pmatrix}a & b \\ c & d \end{pmatrix} \in SL(2,\mathbb{R})$.
The second equation determines the inverse temperature $\beta$ in terms of $\mu$ \footnote{This relation between $\beta$ and $\mu$ is different from the relation in (\ref{eq:valbetamulow}) though the scaling of $\beta$ with respect to $\mu$ is the same.
This is because here we match the energy in the SYK Hamiltonian $\braket{H_{SYK}}$.
In the large $q$ limit, the relation here and that in (\ref{eq:valbetamulow}) agree.}. 
We can also rewrite $f(u) = A \tanh (\f{\pi}{\beta} u + B) + C $ with three parameters $A,B$ and $C$.
These parameters are fixed by imposing the continuity for $f(u)$ at $u = 0$ up to the second derivative, which becomes 
$f(0) = 0, f'(0) = \alpha(\mu)$ and $f''(0) = 0$.
This condition fixes the reparametrization to be
\be
f(u) = \f{2\alpha_S}{\varepsilon(\Delta)} \f{\pi}{\beta \mathcal{J}} \tanh  \Big(\f{\pi}{\beta} u \Big), \qquad t(u) = \f{1}{\alpha(\mu)} \arctan \Bigg[\f{2\alpha_S}{\varepsilon(\Delta)} \f{\pi}{\beta \mathcal{J}} \tanh  \Big(\f{\pi}{\beta} u \Big) \Bigg]. \label{eq:reparammasstoSYK1}
\ee
Here we defined $\varepsilon(\Delta) = \f{\Gamma(2\Delta+1)\Gamma(1-\Delta)^2\Gamma(1-4\Delta)}{\Gamma(\Delta)^2\Gamma(1-2\Delta)^3}$.
Using the reparametrization (\ref{eq:reparammasstoSYK1}), we can study the time evolution $G^{>}(u_1,u_2) =\braket{G_{\bm{s}}(\mu)|\psi_i(u_1)\psi_j(u_2)|G_{\bm{s}}(\mu)}$ using the reparametrization where $\psi_i(u) = e^{iH_{SYK}u} \psi_i e^{-iH_{SYK}u}$.
The diagonal correlation function becomes
\ba
G^{>}(u_1,u_2) &=& e^{- i \pi \Delta } \Big( \f{ \alpha (\mu)^2  t'(u_1) t'(u_2)}{\mathcal{J}^2 \sin^2 [\alpha(\mu) (t(u_1) -t(u_2) - i\epsilon)]} \Big)^{\Delta} \notag \\
&=& e^{- i \pi \Delta } \Big ( \f{\pi  }{\beta \mathcal{J} \sinh [ \f{\pi}{\beta } (u_1 - u_2 - i\epsilon)] } \Big)^{2\Delta}.
\ea
This is exactly the thermal correlation function in Lorentzian time.
The time evolution of the spin expectation value can be studied from the off diagonal correlation function as $\braket{S_k(u)}
 = -2 i s_k (t'(u))^{2\Delta} G_{\text{off}}(t(u),t(u))$, which becomes 
\ba
\braket{S_k(u)} &=&  4 s_k \alpha(\mu) \mu^{-1} \f{\Gamma(1-\Delta)^2\Gamma(2\Delta)\Gamma(1-4\Delta)}{\Gamma(\Delta)^2\Gamma(1-2\Delta)^3 } (t'(u))^{2\Delta}  \notag \\
&=& \braket{S_k(0)} \Big(\f{1}{1 + (\f{2 \alpha_S }{\varepsilon(\Delta)} \f{\pi}{\beta \mathcal{J}} ) ^2 \tanh^2 (\f{\pi}{\beta} u)} \Big)^{2\Delta} \Big( \f{1}{\cosh \f{\pi}{\beta} u} \Big)^{4\Delta}. \label{eq:expodecayS2}
\ea
The spin operator expectation value decays exponentially at late time.
Therefore, the system loses the initial simple correlation pattern under the SYK time evolution and thermalizes. 
The term $\f{1}{1 + (\f{2 \alpha_S }{\varepsilon(\Delta)} \f{\pi}{\beta \mathcal{J}} ) ^2 \tanh^2 (\f{\pi}{\beta} u)} $ is close to one because $\f{\pi}{\beta \mathcal{J}}$ is very small when $\mu \ll \mathcal{J}$.
Therefore, the time evolution is very close to that in $\ket{B_{\bm{s}}(\beta)}$, which is given in (\ref{eq:expodecayS1})
\footnote{In the two coupled SYK model, similar spin operator is constructed from left and right fermion as $S_i = -2i \psi_i^L \psi_i^R$. Under the decoupled Hamiltonian evolution, this behaves as $\braket{S_i(u)} = \braket{S_i(0)} (\cosh \f{2\pi}{\beta}u)^{-2\Delta}$.
Though this shows the same exponential decay, the early time behavior is different from (\ref{eq:expodecayS1}) and (\ref{eq:expodecayS2}). }.

\subsection{Gravity interpretation}
As it is done in \cite{Kourkoulou:2017zaj}, we can consider the similar gravity configuration of our analysis.
The ground state $\ket{G_{\bm{s}}(\mu)}$ is invariant under the evolution $e^{-i H_{\text{def}} t}$ because it is the ground state of the deformed Hamiltonian $H_{\text{def}}$.
Because $f(\tau) = \tanh (\alpha \tau)$ is the transformation from Poincare coordinate to the global coordinate \cite{Maldacena:2018lmt}, we expect the time translation symmetry in gravity side where the metric in this coordinate is given by
\be
ds_E^2 = \f{d\tau_g^2 + d\sigma^2}{\cos^2 \sigma},\qquad ds_L^2 = \f{-dt_g^2 + d\sigma^2}{\cos^2 \sigma}, \qquad \sigma \in [-\pi/2 , \pi/2].
\ee
Since the system is gapped, we also expect the confined geometry where the emergent direction is capped off at some scale.
Here we simply use the end of the world (EOW) brane picture on which the geometry terminates \cite{Erlich:2005qh,Brower:2006ea,Witten:1998zw,Sakai:2004cn}. 
Because of the time translation symmetry the position of EOW branes should be static under the time translation along global time.
We imagine that we have $N$ bulk fields and at EOW branes we impose the boundary condition $\psi_{2k-1} = i s_k \psi_{2k}$ for the bulk fields as we did in the case of $\ket{B_{\bm{s}}(\beta)}$ states.

\begin{figure}[ht]
\begin{center}
\includegraphics[width=13cm]{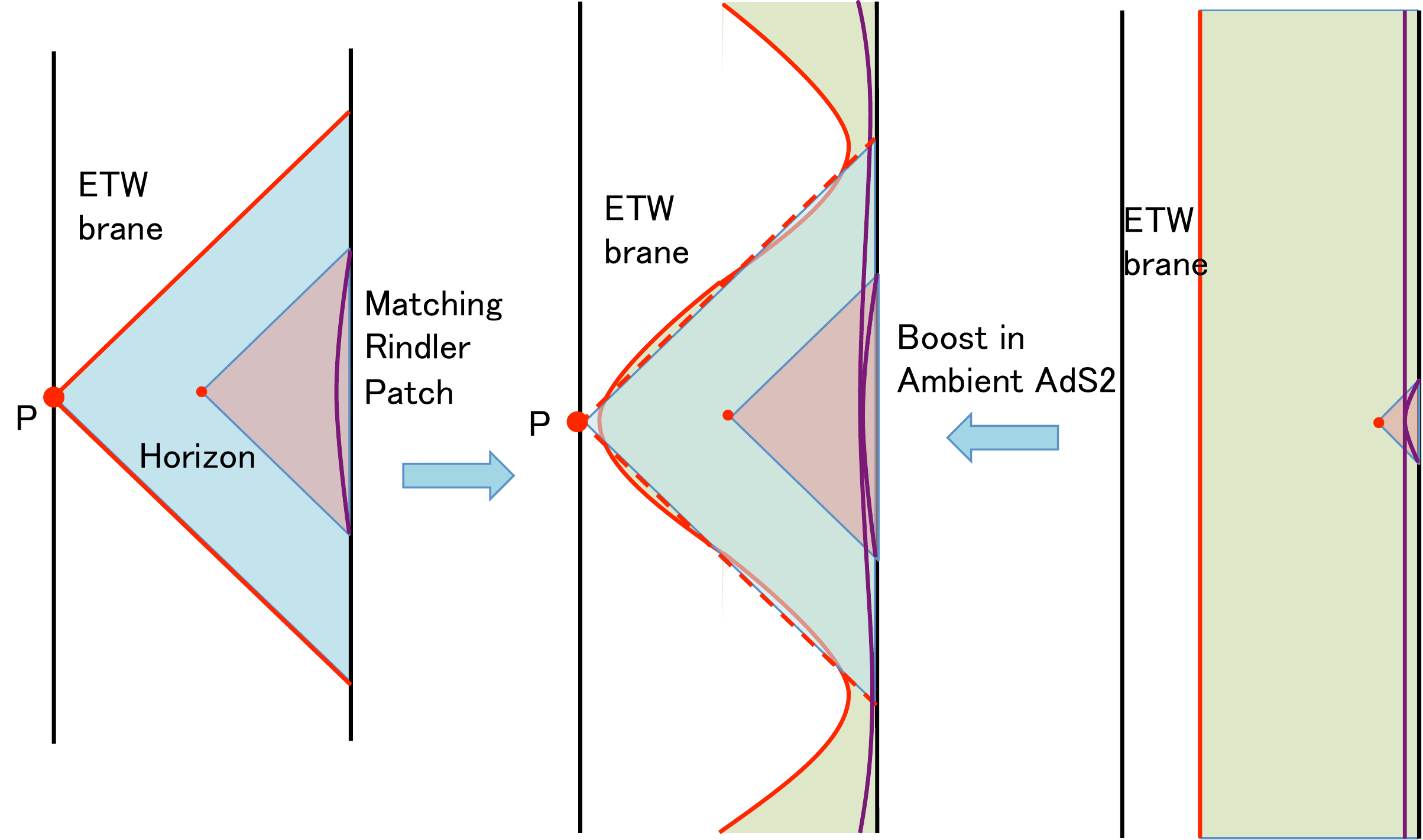}
\caption{A cartoon of the gravity configuration.
The left is the bulk interpretation of the $\ket{B_{\bm{s}}(\beta)}$ and the right is that of the $\ket{G_{\bm{s}}(\mu)}$.
In the middle picture, we compare two geometries matching the Rindler patch of both geometries.
From the Rindler observer, the EOW brane is falling.
The Rindler observer feels the similar falling pattern for the EOW brane.}  
\label{fig:KMgravity1}
\end{center}
\end{figure}

When we evolve the ground state $\ket{G_{\bm{s}}(\mu)}$ by the SYK Hamiltonian, the system thermalizes.
The evolution under the SYK Hamiltonian is given by the reparametrization (\ref{eq:reparammasstoSYK1}).
In gravity picture, this reparametrization gives the transformation from the global coordinate to the Rindler coordinate, which only covers a portion of global AdS$_2$ and has a horizon.
Therefore we obtain the single sided black hole geometry with EOW brane from the ground state of the mass deformed Hamiltonian.

We can also interpret the similarity between $\ket{G_{\bm{s}}(\mu)}$ and $\ket{B_{\bm{s}}(\beta)}$ in gravity.
The symmetry of $\ket{G_{\bm{s}}(\mu)}$ is that in global time whereas the symmetry of  $\ket{B_{\bm{s}}(\beta)}$ is that in Poincare time and EOW branes are static under each symmetry.
We can still match the Rindler patch in both geometries.
Then, the EOW branes are falling from Rindler observer in a similar way, as depicted in Fig.~\ref{fig:KMgravity1}.
In this sense, two geometries are similar.
Especially, we expect that the state $\ket{G_{\bm{s}}(\mu)}$ contains region behind the horizon. 

\begin{figure}[ht]
\begin{center}
\includegraphics[width=13cm]{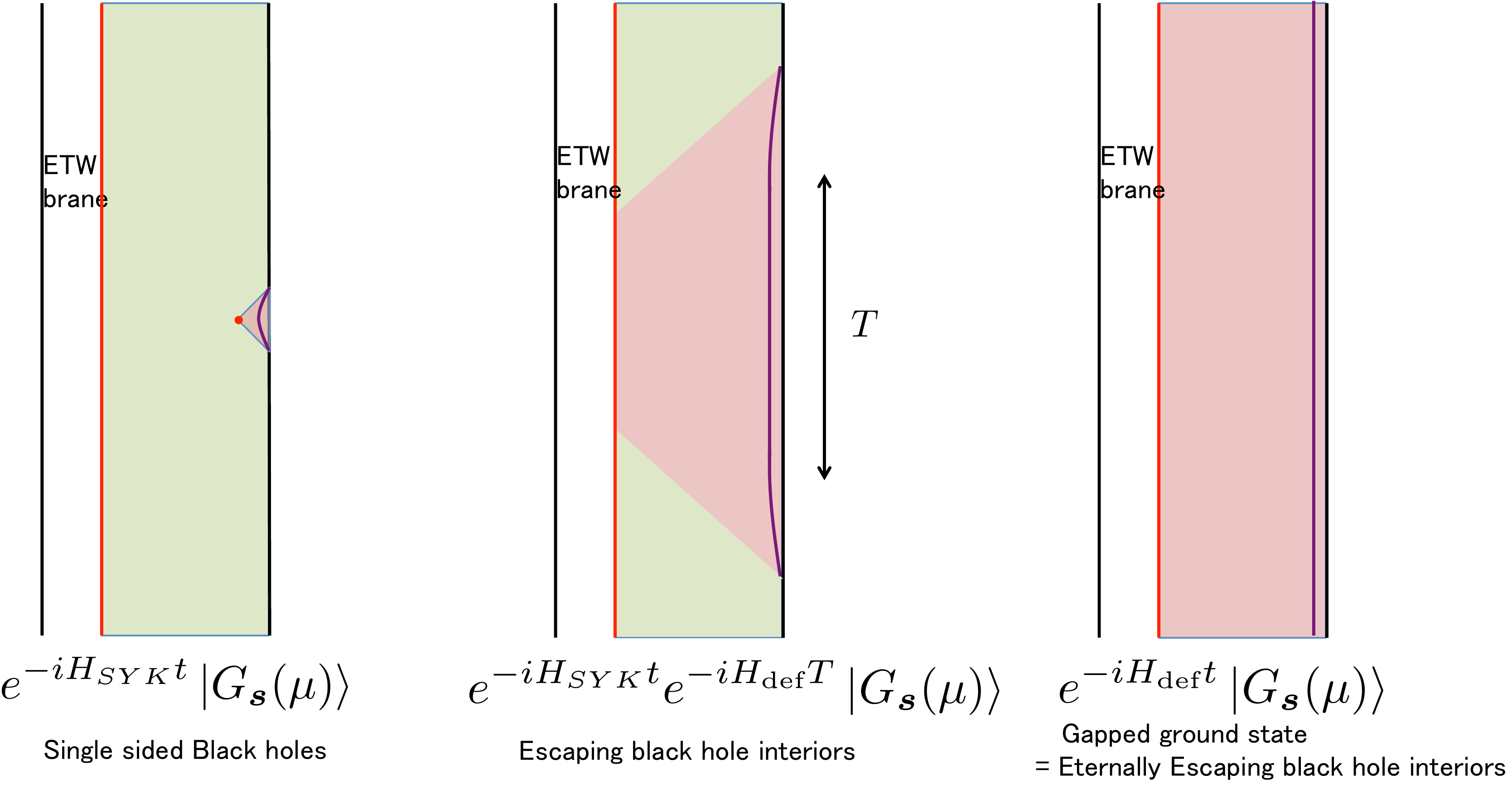}
\caption{A gravity interpretation of the escaping interior protocol on the mass deformed ground state.
{\bf Left:} The SYK evolution, which is interpreted as the evolution without any double trace deformation, makes the black hole with EOW branes. We also evolve in backward by the SYK Hamiltonian.
{\bf Middle:} We apply the escaping interior protocol for finite amount of time $T$ and then evolve by the SYK Hamiltonian.
This is equivalent to shifting the horizon by insert the global AdS$_2$ patch.
{\bf Right:} We apply the escaping interior protocol for eternally.
As a consequence, the horizons are shifted infinitely away from the original horizon.
Finally we recover the global AdS$_2$ with the EOW brane. }  
\label{fig:KMgravity2}
\end{center}
\end{figure}

It is also interesting to consider the protocol to escape the black hole interior \cite{Almheiri:2018ijj,Kourkoulou:2017zaj} of single sided black holes with the black hole microstate $\ket{G_{\bm{s}}(\mu)}$ instead of $\ket{B_{\bm{s}}(\beta)}$.
When we evolve the system by the SYK Hamiltonian, these correspond to single sided black holes.
The escaping protocol \cite{Almheiri:2018ijj} corresponds to evolving the ground state by the deformed Hamiltonian $H_{\text{def}}$.
We can apply the escaping protocol for finite time $T$ and then turn off the mass term.
This corresponds to insert the time evolution by $H_{\text{def}}$ before applying the SYK evolution as $e^{-i H_{SYK} t} e^{-i H_{\text{def}}T}\ket{G_{\bm{s}}(\mu)}$.
Therefore we just delay the black hole formation by inserting global AdS$_2$ region, as depicted in Fig.~\ref{fig:KMgravity2}.
When we apply the escaping protocol eternally, we shift the horizon infinitely and finally we obtain the geometry without horizon.
This corresponds to the evolution $e^{-iH_{\text{def}}t}\ket{G_{\bm{s}}(\mu)}$ and as we observed this corresponds to the global AdS$_2$ patch.
Therefore, in this case after eternally escaping the interiors we obtain the global AdS$_2$ with the EOW brane.
The matching of spins $\bm{s}$ in the state $\ket{G_{\bm{s}}(\mu)}$ and those in the escaping protocol $e^{-i H_{\text{def}}T}$ is important because the mismatch of the spins gives excited states of the $H_{\text{def}}$.
As we saw in the finite temperature analysis of the $H_{\text{def}}$, high energy behavior is similar to that of the SYK model and chaotic.
Therefore,  when we have mismatch for order $N$ spins, we expect that this mismatch leads to the black hole formation and failure of the escaping protocol.
Therefore the state dependent deformation is important\footnote{It is also important to choose the correct pair of fermions to make a spin operator.} to avoid the black hole generation.
In this way, we can clearly understand the escaping protocol starting from the special microstates $\ket{G_{\bm{s}}(\mu)}$.

\section{Discussion}
\label{sec_discussion}

\subsection{Similarities and differences compared with Maldacena-Qi model}

Because the model is similar to that of the eternal traversable model \cite{Maldacena:2018lmt}, it is good to compare with that.
The Hamiltonian of the eternal traversable model is given by
\be
H_{ETW} =  i^{\f{q}{2}}\sum_{i_1 < \cdots <i_q} J^L_{i_1 \cdots i_q} \psi_{i_1}^L \cdots \psi_{i_q}^L + (-i)^{\f{q}{2}}\sum_{i_1 < \cdots <i_q} J^R_{i_1 \cdots i_q} \psi_{i_1}^R \cdots \psi_{i_q}^R + i \mu \sum_{i=1}^N \psi_i ^L \psi_i ^R,
\label{HMQ}
\ee
with $J^L_{i_1i_2\cdots i_q}=J^R_{i_1i_2\cdots i_q}$.
Here we introduce two copies of Majorana fermions $\psi_i^L$ and $\psi_i^R$ which satisfy the canonical commutation relation.

The similar thing is that both systems are gapped systems.
This is natural because in both models we explicitly introduce the mass term in the Lagrangian.
Both systems can be analyzed using conformal symmetry and the ground state has the same time translation symmetry that corresponds to the global time in AdS$_2$.
In the large $q$ limit, the finite temperature behavior beyond the order of $\beta \sim q$ is the same with that of Maldacena-Qi two coupled model because we obtain the same equations.

When we consider the gravity interpretation, it is more surprising.
In the traversable wormhole case the two side are connected in the deep interior.
On the other hand, in our case the geometry is lost at the mass gap scale, which should happen in duals of confining phase \cite{Erlich:2005qh,Brower:2006ea,Sakai:2004cn,Witten:1998zw}.
This suggests that we may be able to understand the spacetime connectivity in a similar way to understand the confined geometry.

There are differences even in qualitative levels.
The first big difference is the absence of the Hawking-Page like transition.
There are many examples of mass deformation of the SYK model, tensor models or matrix models that show the Hawking-Page like transition \cite{Maldacena:2018lmt,Azeyanagi:2017drg,Ferrari:2019ogc} in the large $N$ limit and it is surprising that we have not Hawking-Page like transition even at small mass range.
We expect that this is reminiscent of the higher spin like nature of the SYK model, which suppress the order of transition.

Another difference is the size of the mass gap in the theory at low energy.
In the two coupled SYK model, the physical mass gap is much larger than the parameter $\mu$ in the Lagrangian in small $\mu$ limit.
Therefore the chaos helps to open a gap \cite{Cottrell:2018ash}.
On the other hand, in our case the mass gap is much smaller than the naive gap $\mu$.
In our model the chaos suppress the mass gap, which seems to be more natural.
We expect this is related to the absence of the Hawking-Page like transition.
We will revisit this problem in the future \cite{RosaNosakaNumasawa}.

\subsection{Comparison with the Complex SYK model}
It is also good to compare with the complex SYK model \cite{Sachdev:2015efa,Bulycheva:2017uqj,Gu:2019jub} because this model also takes the similar form of Hamiltonian.
In the complex SYK model, the Hamiltonian is written in terms of the Dirac fermions $c_i$, $i = 1, \cdots ,N$ as 
\be
H_{cSYK} = \sum _{j_1 < \cdots < j_{q/2}, \  k_1 < \cdots < k_{q/2} } J_{j_1\cdots j_{q/2};k_1\cdots k_{q/2} } \mathcal{A} \Big\{ c_{j_1}^{\dagger}\cdots c_{j_{q/2}}^{\dagger} c_{k_1}\cdots  c_{k_{q/2}} \Big\} - \mu \sum_{i=1}^N c_i^{\dagger} c_i.
\ee
Here $\mathcal{A}\{ \cdots \}$ is the antisymmetrization and the couplings $J_{j_1\cdots j_{q/2};k_1\cdots k_{q/2} }$ are independent complex variables with zero mean and the variance $\braket{|J_{j_1\cdots j_{q/2};k_1\cdots k_{q/2} }|^2} = J^2\f{(q/2)!((q/2)-1)!}{N^{q-1}}$.
The last term comes from the chemical potential $\mu$  for the generator of the  global U(1) symmetry  $\hat{Q} = \sum_i c_i^{\dagger} c_i - N/2$.
When we rewrite the Dirac fermion by two Majorana fermions as $c_i = \f{1}{\s{2}}(\psi_{2i-1} - i\psi_{2i})$, the chemical potential term takes the same form with the mass term in the mass deformed SYK (\ref{Hdef}) with $s_k =1 $ for all $k$ \cite{Bhattacharya:2017vaz}.

The main difference is the existence of the U(1) symmetry.
The complex SYK model have a soft mode that is associated to the U(1) symmetry whereas the SYK model do not have such a mode.
The mass deformed SYK model has always a mass gap at zero temperature but the complex SYK model has a gapless excitation\footnote{However, there is also an observation that the complex SYK model also have a gapped phase \cite{Sorokhaibam:2019qho}. }.
The chaos exponents are also studied in the complex model in the large $q$ limit \cite{Bhattacharya:2017vaz} and the $\mu$ dependence of the chaos exponent is different from the mass deformed SYK model in Fig.\ref{fig:lyapunovq1}.

One similarity is the specific charge $\mathcal{Q} = \braket{Q}/N$ in the complex SYK model and the spin operator expectation value.
In the complex SYK model, a natural correlation function is 
\be
G_{cSYK}(\tau_1,\tau_2) = - \f{1}{N} \sum_{i=1}^N\braket{c_i(\tau_1)c_i^{\dagger}(\tau_2)}.
\ee
The specific charge is encoded in the correlation function as $\lim _{\tau \to 0^+} G(\tau,0) = -\f{1}{2} + \mathcal{Q}$.
By decomposing the Dirac fermion $c_i = \f{1}{\s{2}}(\psi_{2i-1} - i\psi_{2i})$, in terms of the Majorana fermion correlation function $G(\tau_1 ,\tau_2) = \braket{\psi_{i}(\tau_1)\psi_i(\tau_2)}$ and $G_{\text{off}}(\tau_1,\tau_2) = \braket{\psi_{2k-1}(\tau_1)\psi_{2k}(\tau_2)}$ the correlation function becomes $G_{cSYK}(\tau_1,\tau_2) = - G(\tau_1,\tau_2) - i G_{\text{off}}(\tau_1,\tau_2)$.
Therefore, we can think of the specific charge $\mathcal{Q}$ as a counterpart of the spin operator expectation value $ \braket{S_k} = -i G_{\text{off}}(0)$ in the mass deformed SYK model.
A quantitative difference is that the specific charge in the complex SYK is not fixed in the IR \cite{Gu:2019jub}, whereas the spin operator expectation value in the mass deformed SYK is determined by the IR conformal field theory data as (\ref{eq:SpinExConf}) in small $\mu$ limit. 

\subsection{Possible microstates from the mass deformation}
We show that we can prepare the $2^{\f{N}{2}}$ states of the form $\ket{G_{\bm{s}}(\mu)}$ from the mass deformation $H_{\text{def}}$.
In this paper we focus on the spin operator $S_k = -2i \psi_{2k-1}\psi_{2k}$ that is constructed from an even index fermion and the odd index fermion. 
The way to construct the spin operator is not restricted to this form.
For example, we can shuffle the index of even fermion as $2k \to 2 \sigma(k)$ where $\sigma \in \mathcal{S}_\f{k}{2}$ is the element of the permutation group $\mathcal{S}_\f{k}{2}$, and then construct the spin operator $S_k' = -2i\psi_{2k-1}\psi_{2\sigma(k)}$.
The mass deformation with $S_k'$ gives a different set of states where the states have a spin operator expectation value in different directions.
We can also construct with a pair of even index fermions.
In this way, we can prepare many set of states as ground states of the mass deformed SYK in this paper.

\subsection{Future problems}
There are several future problems.

In this work  we study the chaos exponent only at large $q$ limit.
It is interesting to do this at finite $q$ numerically.
We study the quantum quench problem in the small $\mu$ limit.
At infinite $\mu$, the ground state reduces to the infinite temperature boundary state $\ket{B_{\bm{s}}}$ and in this regime  real time evolutions are studied in \cite{Kourkoulou:2017zaj,Numasawa:2019gnl} at finite $N$.
It is also interesting future problem to study the real time evolution in finite $\mu$ both in large $N$ and finite $N$.

In this paper we mainly study the SYK model side.
Recently Jackiw Teitelboim (JT) gravity with EOW brane is studied \cite{Penington:2019kki}.
It is a good problem to analyze the Jackiw Teitelboim gravity $+$ matter theory with EOW brane and introduce the double trace deformation.
When the brane is tensionless, JT $+$ matter with EOW brane system just reduces to the orbifold of the traversable wormholes \cite{Maldacena:2018lmt}.
The analysis with the non zero tension EOW brane may lead to the bulk understanding of (the absence of) the Hawking Page like transition.

We did not find any energy/$\mu$-dependence of the adjacent gap ratio for our model \eqref{eq:IntroHdef}; there are no chaotic/integrable transition.
This result is in contrast to the observation in \cite{Garcia-Garcia:2019poj} for the two coupled SYK model \cite{Maldacena:2018lmt}.
Indeed in the two coupled SYK model \eqref{HMQ} the level correlation is qualitatively different in the two extreme regime $\mu\rightarrow 0$ and $\mu\rightarrow \infty$.
In the limit $\mu\rightarrow 0$ the energy spectrum becomes a direct product of the energy spectrum of two SYK models $\{E_m+E_n\}_{m,n\ge 0}$.
When the spectrum enjoys such direct product structure and there are no hierarchy between the level spacings of the two system (which is true in the current case), the two spectrums are completely mixed up.
Hence there are no level repulsion between the adjacent levels even if each system has the RMT-like level correlations.
In the limit of $\mu\rightarrow\infty$ the Hilbert space effectively splits into the eigenspaces of $S$.
Within each eigenspace the direct product structure of the Hamiltonian is lost, and the levels have the RMT-like correlation.
Hence one can expect the transition as $\mu$ increases.
In our model \eqref{eq:IntroHdef}, on the other hand, the picture at $\mu\rightarrow\infty$ is same as the two coupled SYK model while in the limit $\mu\rightarrow 0$ the system reduces to a single SYK model which is again chaotic.

To gain more insight on the mechanism of the Hawking-Page like transition and the chaotic/integrable transition (or their absence) and on how these two phenomena can be correlated, it would be very useful to repeat the same analysis for a generalization of the two coupled SYK model \cite{Maldacena:2018lmt} such that the left coupling $J^L_{ijk\ell}$ and the right coupling $J^R_{ijk\ell}$ are chosen independently to each other.
From the viewpoint of our model, this model is obtained by stating from the Hamiltonian \eqref{eq:IntroHdef} and then omitting all terms in $H_\text{SYK}$ which mix $\psi_{2i-1}$'s and $\psi_{2i}$'s.
This model share the same features of both of the two coupled SYK model and our model.
By rewriting the partition function in the large $N$ limit by using the bi-local fields, one finds that the large $N$ partition function is completely identical to the partition function of our model.
On the other hand, the Hamiltonian of this model has the structure of direct product in the limit $\mu\rightarrow 0$ similar to the two-coupled SYK model, which strongly suppress the RMT-like level correlation in the small $\mu$ regime.
It is worthwhile to test whether this model actually exhibits a chaotic/integrable transition at some finite $\mu$ or not.
One can further consider an interpolation of the two coupled SYK model and this model by tuning the independentness of $J^L_{ijk\ell}$ and $J^R_{ijk\ell}$ continuously, where we observed that the Hawking-Page like transition disappears at some intermediate point before the two couplings become completely independent with each other.
It would be interesting to clarify how the chaotic property as well as the other thermodynamic quantities behaves around this point.
We would like to report these results in \cite{RosaNosakaNumasawa}.

Note that it is subtle whether we should really classify a model which is almost the tensor product of two chaotic system as ``integrable'' although the nearest-neighbor level repulsions are highly suppressed.
To clarify this point, it is worth to study other diagnoses of the quantum chaos such as the spectral rigidity or the spectral form factor (i.e. the long range correlation of the level fluctuations) and the OTOCs.
Especially, while in the analysis of the level statistics one always has to take into account the finite $N$ artifact, the OTOCs allow a direct large $N$ evaluation \cite{Maldacena:2016hyu} which would be more appropriate for the purpose of comparing the chaotic property with the large $N$ Hawking-Page like transition.

\section*{Acknowledgements}
We thank Juan Maldacena, Ahmed Almheiri, Shinsei Ryu, Masahiro  Nozaki, Mao Tian Tan and Jonah Kudler-Flam for useful discussions.
T.~Nosaka thanks Masanori Hanada, Dario Rosa, Masaki Tezuka and Jacobus Verbaarschot for valuable discussions.
T.Numasawa is supported by the Simons Foundation through the ``It from Qubit" Collaboration.
Part of the numerical analyses in this work was carried out at KIAS Center for Advanced Computation Abacus System and also at the Yukawa Institute Computer Facility.
We are
grateful to the conference Quantum Information and String
Theory 2019 in YITP.

\appendix

\section{A derivation of the Large $N$ equations}
\label{appendix_ZinGandSigma}
In this appendix, we give a derivation of the large $N$ effective action  and the Schwinger-Dyson equation of mass deformed SYK model.
The deformed Hamiltonian is 
\be
H_{def} = i^{\f{q}{2}} \sum_{i_1 < \cdots <i_q } J_{i_1\cdots i_q} \psi_{i_1}\cdots \psi_{i_q} + i \mu  \sum_{k=1}^{\f{N}{2}} s_k \psi_{2k-1}\psi_{2k},
\ee
with mean $\braket{J_{i_1\cdots i_q}} = 0$ and variance $\braket{J_{i_1\cdots i_q}^2} = \f{J^2}{N^{q-1}}(q-1)!  = \f{1}{q} \f{\mathcal{J}^2 (q-1)!}{(2N)^{q-1}}$.
By shifting the sign of $\psi_i$ and $J_{i_1 \cdots, i_q}$, we can set $s_k = 1$ for any $k = 1, \cdots N/2$ in the following derivation.
The partition function becomes
\ba
Z&=
&\int \prod_{i_1<\cdots <i_q} dJ_{i_1\cdots i_q} \prod_{i,\tau}\mathcal{D}\psi_{i}(\tau)  \exp \Big[{- \f{N^{q-1}}{2J^2 (q-1)!} \sum_{i_1<\cdots<i_q} J_{i_1\cdots i_q}^2} - \int d \tau \f{1}{2} \sum_{i=1}^N \psi_i (\tau) \partial_{\tau}\psi_i(\tau) \notag \\
&-& i^{\f{q}{2}} \sum_{i_1 < \cdots <i_q } J_{i_1\cdots i_q}  \int d \tau \psi_{i_1}(\tau)\cdots \psi_{i_q}(\tau ) - i\mu \int d \tau  \sum_{k=1}^{\f{N}{2}} \psi_{2k-1}(\tau)\psi_{2k}(\tau) \Big]. 
\ea
The integral over $J_{i_1\cdots i_q} $ is 
\ba
&&\int \prod_{i_1<\cdots <i_q} dJ_{i_1\cdots i_q} \exp\Big[ {- \f{N^{q-1}}{2J^2 (q-1)!} \sum_{i_1<\cdots<i_q} J_{i_1\cdots i_q}^2} - i^{\f{q}{2}} \sum_{i_1 < \cdots <i_q } J_{i_1\cdots i_q}  \int d \tau \psi_{i_1}(\tau)\cdots \psi_{i_q}(\tau ) \notag \\
&=& \exp \Big[ \f{J^2 (q-1)!}{2 N^{q-1}}  (-1)^\f{q}{2} \sum_{i_1<\cdots<i_q}\int d\tau \psi_{i_1}(\tau)\cdots \psi_{i_q}(\tau )  \int d\tau' \psi_{i_1}(\tau')\cdots \psi_{i_q}(\tau ') \Big] \notag \\
&=& \exp \Big[ \f{J^2 (q-1)!}{2 N^{q-1}}  (-1)^\f{q}{2} \f{1}{q!}\sum_{1\le i_1,\cdots ,i_q \le N} \int d\tau \psi_{i_1}(\tau)\cdots \psi_{i_q}(\tau )  \int d\tau' \psi_{i_1}(\tau')\cdots \psi_{i_q}(\tau ') \Big]  \notag \\
&=& \exp \Big[ \f{J^2 }{2q N^{q-1}}  (-1)^\f{q}{2} (-1)^{\sum_{l=1}^q(q-l)} \sum_{1\le i_1,\cdots ,i_q \le N} \int d\tau \int d \tau' \big(\psi_{i_1}(\tau)\psi_{i_1}(\tau' ) \big) \cdots \big(\psi_{i_q}(\tau)\psi_{i_q}(\tau' ) \big)\Big] \notag \\
&=& \exp \Big[ \f{J^2 }{2q N^{q-1}}   \int d\tau \int d \tau'\Big( \sum_{i=1}^N \psi_i(\tau)\psi_i(\tau ') \Big)^q\Big].
\ea
In the second line, the phase factor $(-1)^q$ appears from  $((i)^\f{q}{2})^2$.
In the third line, we extend the sum from $\sum_{i_1<\cdots<i_q} $to $\sum_{1\le i_1,\cdots ,i_q \le N}$.
Because $\psi_i(\tau)$ is a Grassmann number,  $\psi_i(\tau)^2 = 0$ and the sum $\sum_{1\le i_1,\cdots ,i_q \le N}$ survives when all of  $i_1,\cdots, i_q$ are different.
There are $q!$ same contributions, we divide by $q!$ and then the sum reduces to the sum in the second line. 
In the fourth line, we reorder the fermions and we get the sign $ (-1)^{\sum_{l=1}^q(q-l)}$, which becomes $(-1)^\f{q(q-1)}{2}$.
The phase $(-1)^{\f{q}{2}}(-1)^{\f{q(q-1)}{2}} = (-1)^{\f{q^2}{2}}$ becomes $1$ because $q$ is an even number.
The partition function now becomes 
\ba
Z &=& \int \prod_{i,\tau}\mathcal{D}\psi_{i}(\tau)  \exp \Big[ - \int d \tau \f{1}{2} \sum_{i=1}^N \psi_i (\tau) \partial_{\tau}\psi_i(\tau) \notag \\
 &&+\f{J^2 }{2q N^{q-1}}   \int d\tau \int d \tau'\Big( \sum_{i=1}^N \psi_i(\tau)\psi_i(\tau ') \Big)^q
 - i\mu \int d \tau  \sum_{k=1}^{\f{N}{2}} \psi_{2k-1}(\tau)\psi_{2k}(\tau) \Big] . \label{eq:fermionpath2}
\ea
 
Next, we further rewrite the partition function in terms of the correlation function $G(\tau,\tau')$ and the self energy $\Sigma(\tau,\tau')$.
First we insert the delta functional
\be
\int \prod_{\tau>\tau'} \mathcal{D}G(\tau ,\tau') \prod_{\tau>\tau'} \delta\big(\sum_{i=1}^N \psi_i(\tau)\psi_i(\tau') - N G(\tau,\tau')\big)  = 1,
\ee
to (\ref{eq:fermionpath2}):
\ba
Z &=& \int \prod_{i,\tau}\mathcal{D}\psi_{i}(\tau) \prod_{\tau>\tau'} \mathcal{D}G(\tau ,\tau')\prod_{\tau>\tau'} \delta\big(\sum_{i=1}^N \psi_i(\tau)\psi_i(\tau') - N G(\tau,\tau')\big)  \notag \\
 &&\times  \exp \Big[ - \int d \tau \f{1}{2} \sum_{i=1}^N \psi_i (\tau) \partial_{\tau}\psi_i(\tau) 
 +\f{J^2 }{2q N^{q-1}}   \int d\tau \int d \tau'\Big( \sum_{i=1}^N \psi_i(\tau)\psi_i(\tau ') \Big)^q \notag \\
 &&- i\mu \int d \tau  \sum_{k=1}^{\f{N}{2}} \psi_{2k-1}(\tau)\psi_{2k}(\tau) \Big] \notag \\
  &=& \int \prod_{i,\tau}\mathcal{D}\psi_{i}(\tau) \prod_{\tau>\tau'} \mathcal{D}G(\tau ,\tau')\prod_{\tau>\tau'} \delta\big(\sum_{i=1}^N \psi_i(\tau)\psi_i(\tau') - N G(\tau,\tau')\big)  \notag \\
 &&\times  \exp \Big[ - \int d \tau \f{1}{2} \sum_{i=1}^N \psi_i (\tau) \partial_{\tau}\psi_i(\tau) 
 +\f{J^2 N}{2q}   \int d\tau \int d \tau'G(\tau,\tau')^q \notag \\
 &&- i\mu \int d \tau  \sum_{k=1}^{\f{N}{2}} \psi_{2k-1}(\tau)\psi_{2k}(\tau) \Big] .
\ea
In the 2nd line, we replace the factor $\Big( \sum_{i=1}^N \psi_i(\tau)\psi_i(\tau ') \Big)^q$ by $N^qG(\tau,\tau')^q$ because we have the delta functional that relates them.
Next, we represent the delta functional as the following integral
\footnote{Strictly speaking, we need to take the correct contour to make the integral convergent.}
:
\ba
&&\prod_{\tau>\tau'}\delta\big(\sum_{i=1}^N \psi_i(\tau)\psi_i(\tau') - N G(\tau,\tau')\big) \notag \\
&=&\int  \prod_{\tau>\tau'} \mathcal{D}\Sigma(\tau ,\tau') \exp \Big[\f{1}{2} \int d \tau \int d\tau' \Sigma(\tau,\tau')\Big(\sum_{i=1}^N\psi_i(\tau)\psi_i(\tau ')  - NG_(\tau,\tau') \Big) \Big].
\ea
Using this expression for the delta functional, we obtain
\begin{align}
Z&=\int \prod_{i,\tau}\mathcal{D}\psi_{i}(\tau) \prod_{\tau>\tau'} \mathcal{D}G(\tau ,\tau')  \prod_{\tau>\tau'} \mathcal{D}\Sigma(\tau ,\tau')  \notag \\
&\quad \times \exp \Big[ - \int d \tau \f{1}{2} \sum_{i=1}^N \psi_i (\tau) \partial_{\tau}\psi_i(\tau) +  \f{1}{2} \int d \tau \int d\tau' \Sigma(\tau,\tau') \sum_{i=1}^N\psi_i(\tau)\psi_i(\tau ') \notag \\
 &\quad -\f{N}{2} \int d \tau \int d\tau' \Sigma(\tau,\tau')  G(\tau,\tau') +\f{J^2 N}{2q}   \int d\tau \int d \tau'G(\tau,\tau')^q - i\mu \int d \tau  \sum_{k=1}^{\f{N}{2}} \psi_{2k-1}(\tau)\psi_{2k}(\tau) \Big]. 
\end{align}
Until now we do exactly the same transformation with that of the ordinary SYK model.
From now, we further introduce the additional delta functional 
\be
\int \prod_{\tau,\tau'} \mathcal{D}G_{\text{off}}(\tau ,\tau') \prod_{\tau,\tau'} \delta\big(\sum_{k=1}^{\f{N}{2}}  \psi_{2k-1}(\tau)\psi_{2k}(\tau') - \f{N}{2}G_{\text{off}}(\tau,\tau') \big) = 1.
\ee
With this delta functional, we can replace the fermions in the mass term by $G_{\text{off}}(\tau,\tau')$:
\begin{align}
Z&=\int \prod_{i,\tau}\mathcal{D}\psi_{i}(\tau) \prod_{\tau>\tau'} \mathcal{D}G(\tau ,\tau')  \prod_{\tau>\tau'} \mathcal{D}\Sigma(\tau ,\tau')\prod_{\tau,\tau'} \mathcal{D}G_{\text{off}}(\tau ,\tau')\nonumber \\
&\quad  \prod_{\tau,\tau'} \delta\big(\sum_{k=1}^{\f{N}{2}} \psi_{2k-1}(\tau)\psi_{2k}(\tau') - \f{N}{2}G_{\text{off}}(\tau,\tau') \big)   \notag \\
&\quad  \times \exp \Big[ - \int d \tau \f{1}{2} \sum_{i=1}^N \psi_i (\tau) \partial_{\tau}\psi_i(\tau) +  \f{1}{2} \int d \tau \int d\tau' \Sigma(\tau,\tau') \sum_{i=1}^N\psi_i(\tau)\psi_i(\tau ') \notag \\
 &\quad -\f{N}{2} \int d \tau \int d\tau' \Sigma(\tau,\tau')  G(\tau,\tau') +\f{J^2 N}{2q}   \int d\tau \int d \tau'G(\tau,\tau')^q - i\mu\f{N}{2} \int d \tau G_{\text{off}}(\tau,\tau) \Big].
\end{align}
Next, we represent the delta functional as 
\begin{align}
&\prod_{\tau,\tau'} \delta\big(\sum_{k=1}^{\f{N}{2}} \psi_{2k-1}(\tau)\psi_{2k}(\tau') - \f{N}{2}G_{\text{off}}(\tau,\tau') \big) \notag \\
=& \int \prod_{\tau,\tau'} \mathcal{D}\Sigma_{\text{off}}(\tau,\tau')  \exp\Big[\int d\tau \int d\tau' \Sigma_{\text{off}}(\tau,\tau') \big( \sum_{k=1}^{\f{N}{2}} \psi_{2k-1}(\tau)\psi_{2k}(\tau') - \f{N}{2}G_{\text{off}}(\tau,\tau') \big) \Big].
\end{align}
Using this expression for the delta functional, we get
\begin{align}
Z&= \int \prod_{i,\tau}\mathcal{D}\psi_{i}(\tau) \prod_{\tau>\tau'} \mathcal{D}G(\tau ,\tau')  \prod_{\tau>\tau'} \mathcal{D}\Sigma(\tau ,\tau')\prod_{\tau,\tau'} \mathcal{D}G_{\text{off}}(\tau ,\tau')  \prod_{\tau,\tau'} \mathcal{D}\Sigma_{\text{off}}(\tau,\tau') \exp \Bigg[
 \notag \\
&\quad  - \int d \tau \f{1}{2} \sum_{i=1}^N \psi_i (\tau) \partial_{\tau}\psi_i(\tau) +  \f{1}{2} \int d \tau \int d\tau' \Sigma(\tau,\tau') \sum_{i=1}^N\psi_i(\tau)\psi_i(\tau ')\nonumber \\
&\quad +\int d\tau \int d\tau' \Sigma_{\text{off}}(\tau,\tau')  \sum_{k=1}^{\f{N}{2}}\psi_{2k-1}(\tau)\psi_{2k}(\tau')
  \notag \\
 &\quad -\f{N}{2} \int d \tau \int d\tau' \Sigma(\tau,\tau')  G(\tau,\tau') - \f{N}{2} \int d\tau \int d\tau' \Sigma_{\text{off}}(\tau,\tau') G_{\text{off}}(\tau,\tau') \notag \\
 &\quad  +\f{J^2 N}{2q}   \int d\tau \int d \tau'G(\tau,\tau')^q - i\mu\f{N}{2} \int d \tau G_{\text{off}}(\tau,\tau) \Bigg] .
\end{align}
The fermion path integral gives the following functional determinant:
\ba
&&\int \prod_{i,\tau}\mathcal{D}\psi_{i}(\tau) 
\exp \Bigg[ - \int d \tau \f{1}{2} \sum_{i=1}^N \psi_i (\tau) \partial_{\tau}\psi_i(\tau)  \notag \\
&+&  \f{1}{2} \int d \tau \int d\tau' \Sigma(\tau,\tau') \sum_{i=1}^N\psi_i(\tau)\psi_i(\tau ') +\int d\tau \int d\tau' \Sigma_{\text{off}}(\tau,\tau')  \sum_{k=1}^{\f{N}{2}}\psi_{2k-1}(\tau)\psi_{2k}(\tau') \notag \\
&=& \int \prod_{i,\tau}\mathcal{D}\psi_{i}(\tau) 
\exp \Bigg[ - \f{1}{2}\int d \tau \int d\tau' \sum_{k=1}^{N/2}  \notag \\
&&\begin{pmatrix} \psi_{2k-1}(\tau) & \psi_{2k}(\tau) \end{pmatrix}
\Bigg[\begin{pmatrix} 1 & 0 \\ 0  &1 \end{pmatrix} \partial_{\tau} \delta(\tau - \tau') -
\begin{pmatrix} \Sigma(\tau ,\tau') & \Sigma_{\text{off}}(\tau,\tau') \\ -\Sigma_{\text{off}}(\tau',\tau)  & \Sigma(\tau ,\tau') \end{pmatrix} \Bigg]
\begin{pmatrix} \psi_{2k-1}(\tau') \\ \psi_{2k}(\tau') \end{pmatrix}
\Bigg] \notag \\
&=& \Bigg[ \text{Pf}\Bigg( 
\begin{pmatrix} 1 & 0 \\ 0  &1 \end{pmatrix} \partial_{\tau}-
\begin{pmatrix} \Sigma& \Sigma_{\text{off}} \\ -\Sigma_{\text{off}}^T  & \Sigma \end{pmatrix} 
\Bigg]^{\f{N}{2}} \notag \\
&=&\exp \Bigg[\f{N}{2}
\log \text{Pf} 
\Bigg( 
\begin{pmatrix} 1 & 0 \\ 0  &1 \end{pmatrix} \partial_{\tau}-
\begin{pmatrix} \Sigma& \Sigma_{\text{off}} \\ -\Sigma_{\text{off}}^T  & \Sigma \end{pmatrix} 
\Bigg)
\Bigg].
\ea 
Then, we get the effective action in terms of the $G,\Sigma$ variables:
\ba
Z  
&=& \int\prod_{\tau>\tau'} \mathcal{D}G(\tau ,\tau')  \prod_{\tau>\tau'} \mathcal{D}\Sigma(\tau ,\tau')\prod_{\tau,\tau'} \mathcal{D}G_{\text{off}}(\tau ,\tau')  \prod_{\tau,\tau'} \mathcal{D}\Sigma_{\text{off}}(\tau,\tau')  \notag \\
&&\exp \f{N}{2} \Bigg[
\log \text{Pf} 
\Bigg( 
\begin{pmatrix} 1 & 0 \\ 0  &1 \end{pmatrix} \partial_{\tau}-
\begin{pmatrix} \Sigma& \Sigma_{\text{off}} \\ -\Sigma_{\text{off}}^T  & \Sigma \end{pmatrix} 
\Bigg) \notag \\
&&- \int d \tau \int d\tau' \Sigma(\tau,\tau')  G(\tau,\tau') -  \int d\tau \int d\tau' \Sigma_{\text{off}}(\tau,\tau') G_{\text{off}}(\tau,\tau') \notag \\
&&+\f{J^2 }{q}   \int d\tau \int d \tau'G(\tau,\tau')^q - i\mu \int d \tau G_{\text{off}}(\tau,\tau) \Bigg] .
\ea

\subsection{large $q$ expansion and Liouville action}
In this section we derive the Liouville action at large $q$ limit.
The original Euclidean action is 
\begin{align}
&-\f{S_E}{N} \notag \\
&=
\f{1}{2} \log \text{Pf} \Big(
\begin{pmatrix}
1 & 0 \\
0 & 1
\end{pmatrix}
\partial_{\tau}  - 
\begin{pmatrix}
\Sigma & \Sigma_{\text{off}} \\
-\Sigma_{\text{off}}^T & \Sigma
\end{pmatrix}
\Big) \notag \\
&\quad -\f{1}{2} \int d\tau \int d\tau'
\Big[ \f{1}{2}\Tr \Big[
\begin{pmatrix}
\Sigma(\tau,\tau') & \Sigma_{\text{off}}(\tau,\tau') \\
-\Sigma_{\text{off}}(\tau',\tau) & \Sigma(\tau,\tau')
\end{pmatrix}
\begin{pmatrix}
G(\tau,\tau') & -G_{\text{off}}(\tau',\tau) \\
G_{\text{off}}(\tau,\tau') & G(\tau,\tau')
\end{pmatrix}\Big]\nonumber \\
&\quad - \f{J^2}{q} G(\tau,\tau')^q  \Big]
- \f{1}{2} i\mu \int d \tau G_{\text{off}}(\tau,\tau)
\notag \\
&=
\f{1}{2} \log \text{Pf} \Big(
\begin{pmatrix}
1 & 0 \\
0 & 1
\end{pmatrix}
\partial_{\tau}  - 
\begin{pmatrix}
\Sigma & \Sigma_{\text{off}} \\
-\Sigma_{\text{off}}^T & \Sigma
\end{pmatrix}
\Big) \notag \\
&\quad -\f{1}{2} \int d\tau_1 \int d\tau_2
\Big( G(\tau_1,\tau_2)\Sigma(\tau_1,\tau_2) + G_{\text{off}}(\tau_1,\tau_2)\Sigma_{\text{off}}(\tau_1,\tau_2) - \f{\mathcal{J}^2}{2q^2} (2G(\tau_1,\tau_2))^q  \Big)\notag \\
&\quad - \f{i}{2} \f{\hat{\mu}}{q} \int d \tau_1 G_{\text{off}}(\tau_1,\tau_1).
\end{align}
We define
\ba
G(\tau_1,\tau_2) &=& G_0(\tau_1,\tau_2) \Big(1 + \f{1}{q} g(\tau_1,\tau_2) \Big), \notag \\
G_{\text{off}}(\tau_1,\tau_2) &=& G_{0 \text{off}}(\tau_1,\tau_2) \Big(1 + \f{1}{q} g_{\text{off}}(\tau_1,\tau_2) \Big).
\ea
where $G_0(\tau_1,\tau_2) = \f{1}{2}\text{sgn}(\tau_1-\tau_2)$ and $G_{0 \text{off}}(\tau_1,\tau_2) = \f{i}{2} \text{sgn}(\hat{\mu})$ is the two point function of free fermion with a Hamiltonian $H = i \mu \sum_k \psi_{2k-1}\psi_{2k}$ with $\mu \to 0$ limit.
In large q limit, they satisfy
\ba
\begin{pmatrix}
1 & 0 \\
0 & 1
\end{pmatrix}
\partial_{\tau}  
\begin{pmatrix}
G_0(\tau,\tau') & -G_{0 \text{off}}(\tau',\tau) \\
G_{0 \text{off}}(\tau,\tau') & G_0(\tau,\tau') 
\end{pmatrix}
&=&\begin{pmatrix}
1 & 0 \\
0 & 1
\end{pmatrix}
\partial_{\tau}  
\begin{pmatrix}
 \f{1}{2}\text{sgn}(\tau-\tau') & - \f{i}{2} \text{sgn}(\hat{\mu})\\
\f{i}{2} \text{sgn}(\hat{\mu}) & \f{1}{2}\text{sgn}(\tau-\tau')
\end{pmatrix}
\notag \\
&=& \begin{pmatrix}
\delta(\tau-\tau') & 0 \\
0 & \delta(\tau-\tau')
\end{pmatrix}.
\ea
Then, we can write them as 
\be
G_{0 ab} =  \begin{pmatrix}
G_0(\tau,\tau') & -G_{0 \text{off}}(\tau',\tau) \\
G_{0 \text{off}}(\tau,\tau') & G_0(\tau,\tau') 
\end{pmatrix}
, \qquad [G_0]^{-1}_{ab} = \begin{pmatrix}
1 & 0 \\
0 & 1
\end{pmatrix}
\partial_{\tau} .
\ee
We can expand the Pfaffian as 
\ba
\log \text{Pf} (G_0^{-1} -\Sigma) &=& \log \text{Pf}(G_0^{-1}(1 - G_0 * \Sigma)) \notag \\
&=& \log \text{Pf} (G_0^{-1}) - \f{1}{2} \Tr (G_0*\Sigma) - \f{1}{4} \Tr (G_0*\Sigma*G_0*\Sigma ) + \cdots.
\ea
Then, the action becomes
\ba
\f{S_E}{N} &\approx& - \f{1}{2} \log \text{Pf} (G_0^{-1})  + \f{1}{4}\f{|\hat{\mu}|}{q}  \int d\tau +\f{1}{8} \Tr (G_0*\Sigma*G_0*\Sigma )  \notag \\
&& + \f{1}{2q} \int d\tau_1 \int d\tau_2 \Bigg( \Sigma(\tau_1,\tau_2) G_0(\tau_1,\tau_2) g(\tau_1,\tau_2) +\Sigma_{\text{off}}(\tau_1,\tau_2) G_{0\text{off}}(\tau_1,\tau_2) g_{\text{off}}(\tau_1,\tau_2)  \Bigg)  \notag \\
&& -\f{\mathcal{J}^2}{4 q^2} \int d\tau_1 \int d\tau_2 e^{g(\tau_1,\tau_2)} - \f{|\hat{\mu}|}{4q^2} \int d\tau g_{\text{off}}(\tau,\tau) \notag \\
&=&- \f{1}{2} \log \text{Pf} (G_0^{-1})  + \f{1}{4}\f{|\hat{\mu}|}{q}  \int d\tau +\f{1}{8} \Tr (G_0*\Sigma*G_0*\Sigma )  \notag \\
&& + \f{1}{4q} \int d\tau_1 \int d\tau_2 \sum_{ab}\Sigma_{ab}(\tau_1,\tau_2) G_{0ab}(\tau_1,\tau_2) g_{ab}(\tau_1,\tau_2) \notag \\
&& -\f{\mathcal{J}^2}{4 q^2} \int d\tau_1 \int d\tau_2 e^{g(\tau_1,\tau_2)} - \f{|\hat{\mu}|}{4q^2} \int d\tau g_{\text{off}}(\tau,\tau).
\ea
To integrate out $\Sigma$ field, it is helpful to introduce
\be
\Phi_{ab} (\tau_1,\tau_2) = [G_0* \Sigma]_{ab}(\tau_1,\tau_2) =  \int d\tau G_{0ac}(\tau_1,\tau)\Sigma_{cb}(\tau,\tau_2).
\ee
Then, this satisfies
\be
\Sigma_{ab} (\tau_1,\tau_2) = \partial _{\tau_1} \Phi_{ab}(\tau_1,\tau_2).
\ee
Then, the effective action becomes 
\ba
S_E/N &\approx&- \f{1}{2} \log \text{Pf} (G_0^{-1})  + \f{1}{4}\f{|\hat{\mu}|}{q}  \int d\tau +\f{1}{8} \Tr (G_0*\Sigma*G_0*\Sigma )  \notag \\
&& + \f{1}{4q} \int d\tau_1 \int d\tau_2  \sum_{ab}\Sigma_{ab}(\tau_1,\tau_2) G_{0ab}(\tau_1,\tau_2) g_{ab}(\tau_1,\tau_2) \notag \\
&& -\f{\mathcal{J}^2}{4 q^2} \int d\tau_1 \int d\tau_2 e^{g(\tau_1,\tau_2)} - \f{|\hat{\mu}|}{4q^2} \int d\tau g_{\text{off}}(\tau,\tau) \notag \\
&\approx&- \f{1}{2} \log \text{Pf} (G_0^{-1})  + \f{1}{4}\f{|\hat{\mu}|}{q}  \int d\tau +\f{1}{8} \Tr (\Phi*\Phi)  \notag \\
&& - \f{1}{4q} \int d\tau_1 \int d\tau_2  \sum_{ab} \Phi_{ab}(\tau_1,\tau_2) \partial_{\tau_1}(G_{0ab}(\tau_1,\tau_2) g_{ab}(\tau_1,\tau_2)) \notag \\
&& -\f{\mathcal{J}^2}{4 q^2} \int d\tau_1 \int d\tau_2 e^{g(\tau_1,\tau_2)} - \f{|\hat{\mu}|}{4q^2} \int d\tau g_{\text{off}}(\tau,\tau) .
\ea
By integrating out $\Phi$, we obtain the effective action as
\ba
S_E/N &=& \f{1}{8q^2} \int d\tau_1 d\tau_2 \sum_{ab} \partial_{\tau_1}(G_{0ab}(\tau_1,\tau_2) g_{ab}(\tau_1,\tau_2)) \partial_{\tau_2}(G_{0ab}(\tau_1,\tau_2) g_{ab}(\tau_1,\tau_2))\notag \\
 && -\f{\mathcal{J}^2}{4 q^2} \int d\tau_1 \int d\tau_2 e^{g(\tau_1,\tau_2)} - \f{|\hat{\mu}|}{4q^2} \int d\tau g_{\text{off}}(\tau,\tau). 
\ea
There is a nontrivial Jacobian when we change the integration variable from $\Sigma$ to $\Phi$, but this is $g_{ab}$ independent.
We also omit the other terms that are $g_{ab}$ independent.
Because $G_{0ab}$ are constants except for $\tau_1 = \tau_2$ and $g(\tau,\tau) =0$, the effective action now becomes
\ba
\f{S_E}{N} &=& \f{1}{16q^2} \int d\tau_1 \int d\tau_2 \Big( \partial_{\tau_1} g(\tau_1,\tau_2) \partial_{\tau_2}g(\tau_1,\tau_2) - \partial _{\tau_1 }g_{\text{off}}(\tau_1,\tau_2)\partial_ {\tau_2 }g_{\text{off}}(\tau_1,\tau_2) \Big) \notag \\
&& -\f{1}{4q^2} \int d\tau_1 \int d \tau_2 \mathcal{J}^2 e^{g(\tau_1,\tau_2)} - \f{|\hat{\mu}|}{4 q^2}\int d\tau g_{\text{off}}(\tau,\tau).
\ea

\section{Numerical Solution to the Schwinger-Dyson equations}
\label{sec_solveSDnumerical}
After the Wick rotation and the compactification of $\tau$ direction $\tau\sim \tau+\beta$ with $\psi_i(\tau+\beta)=-\psi_i(\tau)$, we can rewrite the partition function of the mass deformed SYK model as
\begin{align}
Z=\biggl\langle
\int {\cal D}\psi e^{-\int d\tau (\frac{1}{2}\psi\partial_\tau\psi+H_{def})}
\biggr\rangle_{J_{ijk\ell}}
=
\int {\cal D}G{\cal D}\Sigma {\cal D}G_{\text{off}}{\cal D}\Sigma_{\text{off}}
e^{-NS_{\text{eff}}(G,\Sigma,G_\text{off},\Sigma_\text{off})},
\end{align}
where
\begin{align}
S_\text{eff}=S_{\text{eff}}^{(1)}+S_{\text{eff}}^{(2)}+S_{\text{eff}}^{(3)}+S_{\text{eff}}^{(4)},
\end{align}
with
\begin{align}
S_\text{eff}^{(1)}&=-\frac{1}{2}\log\Pf
\begin{pmatrix}
-\frac{1}{2}\delta(\tau-\tau')\partial_{\tau'}+\frac{1}{2}\Sigma(\tau,\tau')&\frac{1}{2}\Sigma_\text{off}(\tau,\tau')\\
-\frac{1}{2}\Sigma_\text{off}(\tau',\tau)&-\frac{1}{2}\delta(\tau-\tau')\partial_{\tau'}+\frac{1}{2}\Sigma(\tau,\tau')
\end{pmatrix},\nonumber \\
S_\text{eff}^{(2)}&=\frac{1}{2}\int d\tau d\tau'(\Sigma(\tau,\tau')G(\tau,\tau')+\Sigma_\text{off}(\tau,\tau')G_\text{off}(\tau,\tau')),\nonumber \\
S_\text{eff}^{(3)}&=-\frac{J^2}{2q}\int d\tau d\tau' G(\tau,\tau')^q,\nonumber \\
S_\text{eff}^{(4)}&=\frac{i\mu}{2}\int d\tau G_\text{off}(\tau,\tau),
\label{Seff1to4}
\end{align}
as explained in the appendix \ref{appendix_ZinGandSigma}.
In the limit of $N\rightarrow\infty$, we can evaluate the integrations over the bi-local fields by the saddle point approximation
\begin{align}
Z\approx e^{-NS_\text{eff}(G,\Sigma,G_\text{off},\Sigma_\text{off})},
\label{saddleapprox}
\end{align}
with $G,\Sigma,G_\text{off},\Sigma_\text{off}$ satisfying the saddle point equations (namely, the equation of motion for $S_\text{eff}$)
\begin{align}
\frac{\delta S_\text{eff}}{\delta G(\tau,\tau')}=
\frac{\delta S_\text{eff}}{\delta \Sigma(\tau,\tau')}=
\frac{\delta S_\text{eff}}{\delta G_\text{off}(\tau,\tau')}=
\frac{\delta S_\text{eff}}{\delta \Sigma_\text{off}(\tau,\tau')}=0.
\end{align}

In this section we explain how to solve the saddle point equations and evaluate $S_\text{eff}$ over the solution numerically.
First we assume that the solution depends on $\tau,\tau'$ only through the difference $\tau-\tau'$ and satisfies the anti-periodicity $G(\tau-\tau'+\beta)=-G(\tau-\tau')$ reflecting the anti-periodicity of $\psi_i$, so that we can expand $G,\Sigma,G_\text{off},\Sigma_\text{off}$ in discrete fourier series $G(\tau)=\frac{1}{\beta}\sum_{n=-\infty}^{\infty}e^{-i\omega_n\tau} {\widetilde G}(\omega_n)$ with $\omega_n=\frac{2\pi}{\beta}(n+\frac{1}{2})$.
Now, for the numerical computation, let us further discretize $\tau$ coordinate as
\begin{align}
\tau=\frac{\beta m}{2\Lambda},\quad (m=0,1,\cdots,2\Lambda-1),
\end{align}
which is equivalent to introducing an UV cutoff to $\omega_n$ as $-\Lambda\le n\le \Lambda-1$, so that each of $G,\Sigma,G_\text{off},\Sigma_\text{off}$ is a finite $(2\Lambda)$ dimensional vector both in $\omega$-space and $\tau$-space and related as
\begin{align}
G_m=G\Bigl(\tau=\frac{\beta}{2\Lambda}m\Bigr)=\frac{1}{\beta}\sum_{n=-\Lambda}^{\Lambda-1}e^{-\frac{\pi im}{\Lambda}(n+\frac{1}{2})}{\widetilde G}_n,\quad
{\widetilde G}_n={\widetilde G}(\omega_n)=\frac{\beta}{2\Lambda}\sum_{m=0}^{2\Lambda-1}e^{\frac{\pi im}{\Lambda}(n+\frac{1}{2})}G_m,
\label{fourier}
\end{align}
and the same for $\Sigma,G_\text{off},\Sigma_\text{off}$.
With \eqref{fourier}, each term in the effective action \eqref{Seff1to4} reduces to a discrete summation\footnote{
Here we have renormalized the functional determinant $S_\text{eff}^{(1)}$ such that the partition function correctly reproduces the partition function of $N$ free fermions in the limit of $J,\mu\rightarrow 0$, as explained also in \cite{Maldacena:2016hyu}.
}
\begin{align}
S_\text{eff}^{(1)}&=
-\frac{1}{4}\sum_{n=-\Lambda}^{\Lambda-1}
\log\Bigl[
\Bigl(
1+\frac{{\widetilde \Sigma}_n}{i\omega_n}
\Bigr)^2
-\frac{{\widetilde\Sigma}_{\text{off},n}{\widetilde\Sigma}_{\text{off},-n-1}}{\omega_n^2}
\Bigr]-\frac{\log 2}{2},\nonumber \\
S_\text{eff}^{(2)}&=\frac{\beta^2}{4\Lambda}\sum_{m=0}^{2\Lambda-1}(\Sigma_mG_m+\Sigma_{\text{off},m}G_{\text{off},m})
=\frac{1}{2}\sum_{n=-\Lambda}^{\Lambda-1}({\widetilde\Sigma}_n{\widetilde G}_{-n-1}+{\widetilde\Sigma}_{\text{off},n}{\widetilde G}_{\text{off},-n-1}),\nonumber \\
S_\text{eff}^{(3)}&=-\frac{\beta^2 J^2}{4\Lambda q}\sum_{m=0}^{2\Lambda-1}(G_m)^q,\nonumber \\
S_\text{eff}^{(4)}&=\frac{i\mu}{2}\sum_{n=-\Lambda}^{\Lambda-1}{\widetilde G}_{\text{off},n}.
\label{F1to4disc}
\end{align}
and the saddle point equations are given by the ordinal derivatives of these terms by either the $\tau$-components or the $\omega$-components of $G,\Sigma,G_\text{off},\Sigma_\text{off}$.
It is convenient to perform $G$-derivative by $\tau$-components and the derivative in $\Sigma,G_\text{off},\Sigma_\text{off}$ by $\omega$-components and we obtain
\begin{align}
\frac{\partial S_\text{eff}}{\partial G_m}=0&\rightarrow \Sigma_m=J^2(G_m)^{q-1},\nonumber \\
\frac{\partial S_\text{eff}}{\partial {\widetilde \Sigma}_n}=0&\rightarrow {\widetilde G}_n+\frac{i\omega_n-{\widetilde \Sigma}_{-n-1}}{(i\omega_n-{\widetilde\Sigma}_{-n-1})^2+{\widetilde\Sigma}_{\text{off},n}{\widetilde\Sigma}_{\text{off},-n-1}}=0,\nonumber \\
\frac{\partial S_\text{eff}}{\partial {\widetilde G}_{\text{off},n}}=0&\rightarrow {\widetilde \Sigma}_{\text{off},n}=-i\mu,\nonumber \\
\frac{\partial S_\text{eff}}{\partial {\widetilde \Sigma}_{\text{off},n}}=0&\rightarrow {\widetilde G}_{\text{off},n}=\frac{{\widetilde\Sigma}_{\text{off},n}}{2}\Bigl[
\frac{1}{(i\omega_n-{\widetilde\Sigma}_{-n-1})^2+{\widetilde\Sigma}_{\text{off},n}{\widetilde\Sigma}_{\text{off},-n-1}}
+\frac{1}{(i\omega_n+{\widetilde\Sigma}_n)^2+{\widetilde\Sigma}_{\text{off},n}{\widetilde\Sigma}_{\text{off},-n-1}}
\Bigr].
\label{SDeq}
\end{align}

From the fourth equation in \eqref{SDeq} we find that ${\widetilde G}_{\text{off},n}$ satisfies the following symmetry property:
\begin{align}
{\widetilde G}_{\text{off},-n-1}={\widetilde G}_{\text{off},n}.
\end{align}
For ${\widetilde G}_n$ and ${\widetilde \Sigma}_n$, we find that we can consistently impose the following symmetry property
\begin{align}
{\widetilde G}_{-n-1}=-{\widetilde G}_{n},\quad
{\widetilde \Sigma}_{-n-1}=-{\widetilde \Sigma}_n,
\label{reflectionproperty}
\end{align}
though they may not be satisfied for general solutions.
In the same way we can also impose the following reality relations:
\begin{align}
{\widetilde G}_n^*=-{\widetilde G}_n,\quad
{\widetilde \Sigma}_n^*=-{\widetilde \Sigma}_n,\quad
{\widetilde G}_{\text{off},n}^*=-{\widetilde G}_{\text{off},n} .
\label{reality}
\end{align}
If we impose these symmetry properties, the Schwinger-Dyson equations \eqref{SDeq} finally simplifies into the following pair of equations
\begin{align}
{\widetilde G}_n+\frac{i\omega_n+{\widetilde\Sigma}_n}{(i\omega_n+{\widetilde\Sigma}_n)^2-\mu^2}=0,\quad
\Sigma_m=J^2(G_m)^{q-1},
\label{eqsforGandSigma_fornumerics}
\end{align}
together with
\begin{align}
{\widetilde G}_{\text{off},n}=\frac{-i\mu}{(i\omega_n+{\widetilde\Sigma}_n)^2-\mu^2},\quad
{\widetilde\Sigma}_{\text{off},n}=-i\mu.
\end{align}
The equation for $G$ and $\Sigma$ \eqref{eqsforGandSigma_fornumerics} can be solved numerically by using the same iteration technique as exploited in the undeformed SYK model \cite[appendix G]{Maldacena:2016hyu}.
Once we obtain a set of solution $(G,\Sigma)$ the large $N$ partition function, or the large $N$ free energy $F=-\frac{1}{\beta}\log Z$, can be evaluated from \eqref{saddleapprox} with \eqref{Seff1to4} as
\begin{align}
F\approx \frac{NS_\text{eff}}{\beta}
=-\frac{1}{4\beta}\sum_{n=-\Lambda}^{\Lambda}\log\Bigl[\Bigl(1+\frac{{\widetilde\Sigma}_n}{i\omega_n}\Bigr)^2+\frac{\mu^2}{\omega_n^2}\Bigr]
-\frac{1}{2\beta}\sum_{n=-\Lambda}^{\Lambda}{\widetilde\Sigma}_n{\widetilde G}_n
-\frac{\beta J^2}{4\Lambda q}\sum_{m=0}^{2\Lambda-1}(G_m)^q
-\frac{\log 2}{2\beta},
\label{free_numerical_final}
\end{align}
where we have also used the symmetry property of $G$ \eqref{reflectionproperty}.

Note that in the above formulation $G_\text{off}$ is merely an auxiliary field which does not contribute to the partition function, but just play a role to fix ${\widetilde\Sigma}_\text{off}$ as ${\widetilde\Sigma}_{\text{off},n}=-i\mu$.
Nevertheless $G_\text{off}$ itself is a physically meaningful observable $G_\text{off}(\tau)=\langle\psi_{2i-1}(\tau)\psi_{2i}(0)\rangle$ and useful for the consistency check of the different approaches of the computations.

\section{Detail of the large $q$ finite temperature analysis \label{sec:largeqdetail}}
In this appendix we give a detail of the large $q$ analysis.
As it was done in \cite{Maldacena:2018lmt}, we divide the range of the inverse temperature into four region that consists from the inverse temperature of order $q \log q$, $q$ , $\s{q}$ and $1$.
\subsection{Inverse temperature of order $\beta = q \log q$}
In this order, we fix 
\be
\sigma = q e^{-\beta \mu}.
\ee
The general solution at early time is given by
\ba
e^{g(\tau)} &=& \f{\alpha^2}{\mathcal{J}^2 \sinh ^2(\alpha |\tau| + \gamma)}, \notag \\
e^{g_{\text{off}}(\tau)} &=& \f{4\tilde{\alpha}^2}{\mathcal{J}^2} e^{-2\tilde{\gamma}} e^{ - 2 \tilde{\alpha}|\tau| },\label{eq:qlogqgeneqrly}
\ea
with the boundary conditions
\be
g(0) = 0, \qquad \partial_{\tau}g_{\text{off}} (0_+) = - \hat{\mu}, \label{eq:bdat0finT1}
\ee
and at $\tau \to \infty$ we impose that (\ref{eq:qlogqgeneqrly}) matches with the early time expansion (\ref{eq:qlogqlateearly})
\ba
\f{1}{2} + \f{1}{q} \log \f{2 \alpha}{\mathcal{J}} - \f{\gamma}{q} - \f{\alpha}{q} \tau  + \cdots = A \cosh \f{\beta\mu}{2} - \mu \tau A \sinh \f{\beta \mu}{2} + \cdots, \notag \\
\f{1}{2} + \f{1}{q} \log \f{2\tilde{\alpha}}{\mathcal{J} } - \f{\tilde{\gamma}}{q}-\f{\tilde{\alpha}}{q} \tau + \cdots = A \sinh \f{\beta\mu}{2} -    \mu \tau A \cosh \f{\beta\mu}{2}+ \cdots. \label{eq:bdatinffinT1}
\ea
The condition at $\tau=0$ (\ref{eq:bdat0finT1}) gives 
\be
\alpha = \mathcal{J} \sinh \gamma , \qquad 2 \tilde{\alpha} = \hat{\mu}.
\ee
The condition at $\tau \to \infty$ gives 
\ba
\f{1}{2} + \f{1}{q} \log \f{2\alpha}{\mathcal{J}} - \f{\gamma}{q} &=& A \cosh \f{\beta \mu}{2 }, \notag \\
\f{1}{2} + \f{1}{q} \log \f{2\tilde{\alpha}}{\mathcal{J}} - \f{\tilde{\gamma}}{q}&=& A \sinh \f{\beta \mu}{2 },
\ea
and 
\ba
\f{\alpha}{q} &=& \mu A \sinh \f{\beta \mu }{2}, \notag \\
\f{\tilde{\alpha}}{q} &=& \mu A \cosh \f{\beta \mu }{2},
\ea
which lead to
\be
\alpha = \tilde{\alpha}, \qquad \tilde{\gamma} = \gamma + \sigma.
\ee
The parameter $A$ is also determined as 
\be
A = e ^{-\f{\beta \mu}{2}}.
\ee
We ignored subleading terms in large $q$ expansion\footnote{For example, we find that $\f{\alpha}{\tilde{\alpha}} = \tanh \f{\beta \mu}{2} = \f{q^2 - \sigma^2}{q^2 + \sigma^2} = 1 - \f{2\sigma^2}{q^2} + \cdots$.
This can be approximated by $1$ in the large $q$ limit. }.

The energy is given by
\ba
\f{E}{N} &=&  \f{1}{q} \partial_{\tau} G(\tau,0)\Big |_{\tau = 0^+}  + i\f{\mu}{2}\Big( 1 -\f{2}{q} \Big)G_{\text{off}}(0,0) \notag \\
&=& \f{1}{q^2} \f{1}{2}\partial_{\tau}g(\tau,0) \Big |_{\tau \to 0} + i\f{\hat{\mu}}{2q} \Big( 1 - \f{2}{q} \Big) \f{i}{2} \Big( 1 + \f{1}{2} g_{\text{off}} (0)\Big) \notag \\
&=& -\f{1}{2q^2} \f{\hat{\mu}}{\tanh \gamma} - \f{\hat{\mu}}{4q}  - \f{\hat{\mu}}{4q} \Big( 1 - \f{2}{q} + \f{2}{q} \log ( \sinh \gamma e^{-\tilde{\gamma}} )\Big)  \notag \\
&=& -\f{1}{2q^2} \f{\hat{\mu}}{\tanh \gamma} - \f{\hat{\mu}}{4q} \Big( 1 - \f{2}{q} + \f{2}{q} \log \f{\hat{\mu}}{\mathcal{J}}e^{-\tilde{\gamma}} \Big) + \mathcal{O} (q^{-3}).
\ea
In terms of $\gamma, \sigma$ variables, it can be written as 
\ba
 \beta\f{\partial l }{\partial \beta} &=& -\beta E \notag \\
 &= & -\f{\beta\hat{\mu}}{2q^2}\Big[-\f{q}{2} + 1 -\f{1}{\tanh \gamma} - \log (2\sinh \gamma e^{-\hat{\gamma}}) \Big]  \notag \\
 &=& - \f{\log \f{q}{\sigma}}{2 q}\Big[-\f{q}{2} + 1 -\f{1}{\tanh \gamma} - \log (2\sinh \gamma e^{-\hat{\gamma}}) \Big].
\ea
Here we defined $l \equiv \f{\log Z}{N} = -\f{\beta F}{N}$.
To derive the free energy, it is convenient to take the derivative of the partition function:
\be
\mathcal{J}\f{\partial l }{\partial \mathcal{J}} = \f{\beta}{q^2} \int _0 ^\beta  d\tau  \mathcal{J}^2 e^{g(\tau)}    =  \f{\beta\hat{\mu}}{2q^2} \Big[\f{1}{\tanh \gamma} - 1 \Big],
\ee
\ba
\mathcal{\mu}\f{\partial l }{\partial \mu} = -\f{i}{2}\beta \mu G_{\text{off}}(0,0) 
= \f{\beta \hat{\mu}}{4q} \Big[1 + \f{2}{q} \log (2\sinh \gamma e^{-\tilde{\gamma}}) \Big] .
\ea
Note that 
\be
\mathcal{J}\f{\partial l }{\partial \mathcal{J}} + \mathcal{\mu}\f{\partial l }{\partial \mu} = \f{\beta \hat{\mu}}{2q^2} \Big[\f{1}{\tanh \gamma} - 1 + \f{q}{2} + \log (2\sinh\gamma e^{-\tilde{\gamma}})  \Big ]  = \beta \f{\partial l}{\partial \beta}.
\ee
This is because the partition function is the function of dimension less quantity: $l = l(\beta \mu,\beta\mathcal{J})$.
The free energy is given by
\ba
l(\gamma, \sigma) &=& \f{\log \f{q}{\sigma}}{2q} \Big( \f{q}{2} - 1 +\f{1}{\tanh \gamma} + \log (2\sinh \gamma e^{-\tilde{\gamma}}) + \sigma \Big) + \f{\sigma}{2q} \notag \\
&=& \f{\log \f{q}{\sigma}}{2q} \Big( \f{q}{2} - 1 +\f{1}{\tanh \gamma} + \log (2\sinh \gamma e^{-(\gamma+ \sigma) }) + \sigma \Big) + \f{\sigma}{2q} .
\ea
As a function of $\beta$ and $\mu$, the free energy becomes
\be
l(\beta,\mu) = \f{\beta\mu}{4}+ \f{e^{-\beta \mu}}{2}  + \f{\beta\mu}{4q} \Big[ \log (2\sinh \gamma) + \f{1}{\tanh \gamma}  -\gamma -1 \Big],
\ee
where $\hat{\mu} = 2\mathcal{J}\sinh\gamma$ is a function of only $\mu$.
The thermal entropy is given by
\be
S /N =\f{l + \beta E}{N} =  l - \beta\f{\partial l }{\partial \beta} = \f{\sigma}{q} \Big(1 + \log \f{q}{\sigma} \Big) = e^{-\beta \mu} (1 + \beta\mu).
\ee

\subsubsection{ Temperature of order $\beta = q$}
In this regime $\f{\sigma}{q} = e^{-\beta \mu} = e^{-\f{\beta}{q} \hat{\mu}}$ is finite.
In other word, $\sigma$ is of order $q$, if we extrapolate the large $q$ expansion
the off diagonal correlation function  becomes
\ba
G_{\text{off} }(\tau) &=& \f{i}{2} \Big(1 + \f{1}{q} g_{\text{off}}(\tau) + \cdots \Big) \notag \\
&=& \f{i}{2} \Big(1 + \f{1}{q}(2\log \f{\hat{\mu}}{\mathcal{J}} - 2 \tilde{\gamma} - \hat{\mu}\tau ) + \cdots\Big) \notag \\
&=& \f{i}{2} \Big(1 - \f{2\sigma }{q} + o(q^{-1})\Big) .
\ea
Therefore, even at $\tau = 0$ the leading of $q$ expansion of $G_{\text{off}}(\tau)$ becomes smaller than $\f{1}{2}$.
Especially, we do not expect the expansion $G_{\text{off}}(\tau) \sim \f{i}{2} (1 + \f{1}{q} g_{\text{off}}(\tau) + \cdots)$.
We still have the long time expansion
\be
G(\tau) = A \cosh [\mu (\beta/2 - \tau)], \qquad G_{\text{off}}(\tau) = i A \sinh[\mu (\beta/2 - \tau)],
\ee
and early time $1/q$ expansion for diagonal correlator
\be
G(\tau) = \f{1}{2}(1 + \f{1}{q}g(\tau) + \cdots ), \qquad e^{g(\tau)} = \f{\alpha^2 }{\mathcal{J}^2 \sinh^2 (\alpha \tau + \gamma)}.
\ee
Matching the late time and early time correlator for $G(\tau)$ gives 
\be
\f{1}{2} + \f{1}{q} \log \f{2 \alpha}{\mathcal{J}} - \f{\gamma}{q} - \f{\alpha}{q} \tau  + \cdots = A \cosh \f{\beta\mu}{2} - \mu \tau A \sinh \f{\beta \mu}{2} + \cdots, 
\ee
which leads to 
\ba
&&\f{1}{2} + \f{1}{q} \log \f{2 \alpha}{\mathcal{J}} - \f{\gamma}{q} = A \cosh \f{\beta\mu}{2},  \notag \\
&&\f{\alpha}{q} =  \mu A\sinh \f{\beta \mu}{2}.
\ea
The first equation gives 
\be
A \cosh \f{\beta \mu}{2} = \f{1}{2} + o(q^{-1}),
\ee
which determines 
\be
A = \f{1}{2 \cosh \f{\beta \mu}{2}} + o(q^{-1}) .
\ee
The second equation with $A = \f{1}{2 \cosh \f{\beta \mu}{2}} + o(q^{-1})$ gives
\be
\alpha = \f{\hat{\mu}}{2} \tanh \f{\beta \mu}{2}.
\ee
The initial condition gives $g(0) = 0$, which gives
\be
\alpha = \mathcal{J} \sinh \gamma.
\ee
This determines $\gamma$ as 
\be
\sinh \gamma = \f{\hat{\mu}}{2 \mathcal{J}}\tanh \f{\beta \mu}{2}.
\ee

Therefore, the correlation functions become
\ba
G(\tau) &=& \f{1}{2} \f{\cosh \mu (\f{\beta}{2}-\tau)}{\cosh \f{\mu\beta}{2}} + o(1/q), \notag \\
G_{\text{off}}(\tau) &=& \f{i}{2} \f{\sinh \mu (\f{\beta}{2}-\tau)}{\cosh \f{\mu\beta}{2}} + o(1/q).
\ea
To evaluate the partition function, it is convenient to use 
\ba
\mathcal{J}\partial_{\mathcal{J}} l &=& \f{\beta \mu}{2q} \tanh \f{\beta \mu}{2} \Big[ \f{1}{\tanh \gamma} - 1 \Big], \notag \\
\mathcal{\mu}\partial_{\mathcal{\mu}} l &=& \f{\beta \mu}{4} \tanh \f{\beta \mu}{2} + o(1/q),
\ea
and use $\mathcal{J}\partial_{\mathcal{J}} l + \mathcal{\mu}\partial_{\mathcal{\mu}} l - \mathcal{\beta}\partial_{\mathcal{\beta}} l=0$.
The integral becomes \footnote{This only reproduce $\mathcal{\mu}\partial_{\mathcal{\mu}} l = \f{\beta \mu}{4} \tanh \f{\beta \mu}{2} $, which is order one in $1/q$ expansion.}
\be
l = \f{1}{2}\log (2\cosh \f{\beta \mu}{2}) + \f{\beta \mu}{2q} \tanh \f{\beta \mu}{2} \Big[\log (2 \sinh \gamma) + \f{1}{\tanh \gamma} - \gamma -1 \Big]. 
\ee
In the high temperature limit, we can expand $\gamma$ and $l$ as 
\be
\gamma \sim \f{q(\beta\mu)^2}{4  \beta \mathcal{J}}, \qquad l \sim \f{1}{2} \log 2 + \f{(\beta\mu)^2 }{16} + \f{\beta \mathcal{J}}{q^2} + \f{(\beta \mu)^2}{4 q} \log \f{q(\beta \mu)^2}{4 \beta \mathcal{J}}+\cdots.
\ee

\subsection{Temperature of order $\beta \sim \s{q}$}
We study the temperature of order $\beta \sim \s{q}$.
In this regime we can approximate $G(\tau) = \f{1}{2}(1+\f{g(\tau)}{q})$ everywhere in $\tau \in [0,\beta]$.
The Schwinger-Dyson equation becomes 
\be
0 = \partial_{\tau} G_{\text{off}}(\tau) - \int d\tau' \Sigma(\tau-\tau')G_{\text{off}}(\tau') + i\mu G(\tau) \sim \partial_{\tau} G_{\text{off}}(\tau)  + i\f{\mu}{2}.
\ee
Here we ignore the term that contains $\Sigma$ because $\Sigma$ is of order $q$ which can be ignored at the leading of the $1/q$ expansion.
We also approximate $G(\tau) \sim \f{1}{2}$, which is the leading of the $1/q$ expansion that we mentioned above.
Then we can solve this equation with the condition $G_{\text{off}}(\f{\beta}{2} + \tau) = -G_{\text{off}}(\f{\beta}{2} -\tau) $ as
\be
G_{\text{off}}(\tau) = \f{i}{2}\mu \Big( \f{\beta}{2}-\tau \Big).
\ee
The equation for $G(\tau)$ becomes 
\be
0 = \partial_{\tau}[\partial_{\tau} G(\tau) - \int d\tau' \Sigma(\tau-\tau')G(\tau') + i\mu G_{\text{off}}(\tau)].
\ee
Using the expansion $G(\tau) = \f{1}{2}(1 + \f{g(\tau)}{q})$ and $G_{\text{off}}(\tau) = \f{i}{2}\mu ( \f{\beta}{2}-\tau)$, we obtain the Liouville like equation
\be
\partial_{\tau}^2 g(\tau) - 2\mathcal{J}^2 e^{g(\tau)} - q\mu^2 = 0. \label{eq:Liouvillesq}
\ee
We can further change the variables as
\be
x = \f{\tau - \f{\beta}{2}}{\beta} , \qquad e^{\hat{g}} = (\beta \mathcal{J})^2 e^{g},
\ee
where $x \in [-\f{1}{2},\f{1}{2}]$.
Then, the Liouville like equation becomes 
\be
\partial_x^2 \hat{g} - 2 e^{\hat{g}} - 2k = 0, \qquad k = \f{q (\mu\beta)^2}{2}.
\ee
$k$ is now finite in our parameter regime $\beta \sim \s{q}, \mu \sim 1/q$.
The boundary condition for $\hat{g}$ is
\be
e^{\hat{g}(\pm \f{1}{2})} = (\beta \mathcal{J})^2,
\ee
which is $\infty$ in the leading of $1/q$ expansion.
Therefore, we should seek the solution which diverges at $x = \pm \f{1}{2}$.
The first integral of the equation (\ref{eq:Liouvillesq}) is 
\be
\Big(\f{d\hat{g}}{dx}\Big)^2 - 4 e^{\hat{g}} - 4 \hat{g}k = \text{const} = - 4 e^{\hat{g}_0} - 4 \hat{g}_0k,
\ee
where we defined $\hat{g}_0 \equiv \hat{g}(0)$ with $\hat{g}'(0) = 0$.
Then, we can take the integral of this first integral as
\be
2x = \int _{\hat{g}_0}^{\hat{g}} \f{dg}{\s{e^g - e^{\hat{g}_0}+ k (g - \hat{g}_0)} }.
\ee
The condition $\hat{g}(\pm \f{1}{2}) = \infty$ determines $\hat{g}_0$ as a function of $k$ through the integral equation 
\be
1 = \int _{\hat{g}_0}^{\infty} \f{dg}{\s{e^g - e^{\hat{g}_0}+ k (g - \hat{g}_0)} }, \label{eq:fistintg0}
\ee
or 
\be
e^{\f{\hat{g}_0}{2}} = \int _0^{\infty} \f{dg}{\s{e^g -1 + \tilde{k}g}}, 
\ee
where we defined $\tilde{k} = k e^{-\hat{g}_0}$.

As a check, we can consider the $k=0$ case.
In this case, we obtain 
\be
e^{\hat{g}_0} = \pi^2,\qquad  e^{\hat{g}(x)} = \f{\pi^2}{\cos^2 \pi x},
\ee
which is consistent with the $\beta \to \infty$ limit of the large $q$ SYK model with keeping $\f{\tau}{\beta}$ finite.
The free energy in this regime is given by
\be
l = \f{1}{2}\log 2 + \f{(\beta\mu)^2}{16} + \f{\beta \mathcal{J}}{q^2} - \f{(\beta \mu)^2}{4q} \log (\beta \mathcal{J}) + \f{h(q\beta^2\mu^2)}{q^2},
\ee
with an undetermined function $h(k)$.
The $\mathcal{J}$ dependence is determined from the derivative of the free energy
\ba
\mathcal{J}\f{\partial l}{\partial \mathcal{J}}  = \f{1}{2q^2} \int_{-\f{1}{2}}^{\f{1}{2}} dx \ e^{\hat{g}(x)} =  \f{1}{2q^2} \int_{-\f{1}{2}}^{\f{1}{2}} \Big(\f{1}{2}\f{\partial^2 \hat{g}}{\partial x^2} - k \Big) 
= \f{1}{2q^2}( \hat{g}'(1/2) - k ).
\ea
Here we use the equation of motion $e^{\hat{g}} = \f{1}{2}\f{\partial^2 \hat{g}}{\partial x^2}  - k$.
Using the first integral, we can express the derivative $g'(x)$ for $x>0$ region as 
\be
\hat{g}'(x) = 2\s{e^{\hat{g}} -e^{\hat{g}_0} + k(\hat{g} - \hat{g}_0 ) }.
\ee
Therefore, in the regime of $\beta \sim \s{q}$, we can approximate $g'(1/2)$ as 
\be
g'(1/2) \sim 2 \mathcal{J}\beta.
\ee
In this way we obtain the $\mathcal{J}$ dependent term in the free energy $l$, but this method leaves $\mu$ dependence unfixed.
The leading $\mu$ dependent term comes from 
\be
\f{\partial l}{\partial \mu} = -\f{i}{2}\beta G_{\text{off}}(0) = \f{\beta^2 \mu}{8}.
\ee

\subsubsection{chaos exponent at order of $\beta \sim \s{q}$}
Here we consider the out of time ordered four point function 
\be
F(t_1,t_2) =\f{1}{N^2}\sum_{i,j=1}^N \Tr [\rho(\beta/4)\psi_i(t_1)\rho(\beta/4)\psi_j(0)\rho(\beta/4)\psi_i(t_2)\rho(\beta/4)\psi_j(0)],
\ee
where $\rho(\beta/4) = (e^{-\beta H_{def}}/Z(\beta))^{\f{1}{4}}$ and  we study exponential growth of this correlator.
We can study the growth rate from the retarded kernel, which satisfies the following equation at large $q$ limit:
\be
\partial_{t_1}\partial_{t_2}K_R(t_1,t_2;t_3,t_4) = 2q \delta(t_{13})  \delta(t_{24}) \Sigma( \f{\beta}{2} +it_{34}).
\ee
Here we neglected the contribution from $\Sigma_{\text{off}}$.
The chaos exponent is given by the eigenstate $K_R * \psi = \psi$ with the form of $\psi(t_1,t_2) = e^{\lambda(t_1+t_2) }\chi(t_1-t_2)$.
Then, $\chi$ satisfies the equation 
\be
-\partial_y^2\chi(y) -2 e^{\hat{g}_l(y)}\chi(y) = -\Big(\f{\lambda \beta}{2}\Big)^2 \chi(y),
\ee
where we defined the Lorentzian continuation $\hat{g}_l(y) \equiv \hat{g}(iy)$.
Therefore by studying the bound state in this Schr\"{o}dinger type equation with the potential $-e^{\hat{g}_l(y)}$, we obtain the chaos exponent.
The function $\hat{g}_l(y)$ has a maximum at $y=0$ and decreases as $|y | \to \infty$.
The second excited state is given by $\chi = \partial_y \hat{g}_l(y)$, with the eigenvalue $0$.
Because of this the above Schr\"{o}dinger equation has only one negative energy state.
This equation can be studied analytically in the large $k$ or small $k$ limit and numerically for general $k$.

\subsubsection{large $k$ limit}
In the large $k$ limit, we can solve the equation (\ref{eq:fistintg0}) as
\be
\hat{g}_0 \approx -\f{k}{4} + 2 \log k .
\ee
From the Schr\"{o}dinger equation 
\be
\f{d^2\hat{g}}{dx^2 } - 2 e^{\hat{g}} - 2k = 0,
\ee
we know the second derivative as 
\be
\f{d^2\hat{g}}{dx^2 } \Big |_{x = 0}= 2 e^{\hat{g}_0} + 2k = 2k^2 e^{-\f{k}{4}} + 2k \sim 2 k.
\ee
Therefore, $\hat{g}_l(y)  = \hat{g}(i y)$ becomes 
\be
\hat{g}_l(y) = \hat{g}_0 - \f{1}{2} \f{d^2\hat{g}}{dx^2 } \Big |_{x = 0} y^2 + \mathcal{O}(y^3) = \log(k^2 e^{-\f{k}{4}}) - k y^2 + \mathcal{O}(y^3).
\ee
The potential for chaos exponent is given by
\be
e^{\hat{g}_l(y)} \approx k^2 e^{-\f{k}{4}} e^{- k y^2}.
\ee
Because the potential is very narrow for $k \to \infty$, we can approximate this as 
\be
e^{\hat{g}_l(y)} \approx k^2 e^{-\f{k}{4}} \s{\f{2\pi}{2k} } \delta (y) = \s{\pi} k^{\f{3}{2}} e^{-\f{1}{4}k}\delta(y),
\ee
where we use 
\be
\f{1}{\s{2\pi\sigma^2}} e^{-\f{x^2 }{2 \sigma^2}}  \approx \delta (x),
\ee
for small $\sigma$. In our case , $\sigma = 1/\s{2k}$.
The chaos exponent is now derived as 
\be
- \f{(\lambda \beta)^2}{4}\chi(y) = [-\partial_{y}^2  - 2\s{\pi} k^{\f{3}{2}} e^{-\f{1}{4}k}\delta(y)] \chi(y).
\ee 
The bound state of delta function potential 
\be
  -\partial _x^2 \psi(x) - V_0 \delta(x) \psi(x) = E_0 \psi(x),
\ee
is given by
\be
E_0 = -\f{V_0^2}{4},
\ee
with the ground state wavefunction
\be
\psi_0(x) = \s{\f{V_0}{2}} e^{-\f{V_0}{2}|x|}.
\ee
Therefore, we obtain 
\be
\lambda \beta = 2\s{\pi} k^{\f{3}{2}} e^{-\f{1}{4}k}.
\ee

\subsubsection{small $k$ limit}
At $k=0$, we obtain $e^{\hat{g}_l(y)} = \f{\pi^2}{\cosh^2 \pi y}$ and this gives $\lambda = \f{2\pi}{\beta}$, which is the maximal chaos exponent.
For small $k$, we can approximate $\hat{g}(x) = \hat{g}_{(0)}(x) + \hat{g}_{(1)} (x)+ \cdots$ where $ e^{\hat{g}_{(0)}(x)}  = \f{\pi^2}{\cos ^2\pi x}$ is the $k=0$ case and $\hat{g}_{(1)}$ is the first order correction in $k$.
After Wick rotation, we obtain the potential $V(y)= -2 e^{\hat{g}(iy)} = -\f{2\pi^2 }{\cosh ^2\pi y} (1 + \hat{g}_{(1)}(iy))$.
Therefore, this gives a shift of the potential $\delta V(y) =-2 e^{\hat{g}(iy)}\hat{g}_{(1)}(iy) $.
The shift of ground state energy $\delta E = \bra{\chi_0}\delta V\ket{\chi_0}$ gives the shift of the chaos exponent.
$\chi_0(y)$ is the ``scramblon"  wave function at $k = 0$ that is given by $\braket{y|\chi_0} = \chi_0(y) = \s{\f{\pi}{2}}\f{1}{\cosh\pi y}$.

The equation of motion for $\hat{g}_{(1)}(x)$ becomes
\be
\partial_x^2 \hat{g}_{(1)}(x)  - \f{2 \pi^2}{\cos^2\pi x} \hat{g}_{(1)}(x) - 2k = 0.
\ee
The solution is given \cite{Garcia-Garcia:2017bkg} by
\be
\f{\pi^2}{2k} \hat{g}_{(1)}(x) = -1 + \log( 2\cos\pi x) + \f{i}{4}(\text{Li}_2(-e^{2\pi ix})-\text{Li}_2(-e^{-2\pi ix}) ) \tan \pi x .
\ee
After Wick rotation, we obtain
\be
\f{\pi^2}{2k}\hat{g}_{(1)}(iy) = -1 + \log (2\cosh\pi y)  +\f{1}{4}(\text{Li}_2(-e^{2\pi y})-\text{Li}_2(-e^{-2\pi y}) )\tanh\pi y.
\ee
Then, the shift of the ``ground state energy" is given by
\be
\bra{\chi_0}\delta V\ket{\chi_0} = \int _{-\infty}^{\infty} dy \f{\pi}{2} \f{2\pi^2}{ \cosh ^4\pi y} \hat{g}_{(1)} (iy) =  k.
\ee
The ``ground state energy" at $k = 0$ is 
\be 
E_{(0)} = - \Big(\f{\lambda_L \beta }{2} \Big)^2 = - \pi^2.
\ee
Therefore, the ground state energy shift is 
\be
E  = E_{(0)} + E_{(1)} + \cdots = -\pi^2 + k + \cdots.
\ee
This gives the leading correction to the chaos exponent as
\be
\f{\lambda_L \beta}{2\pi} = \s{1 - \f{k}{\pi^2}} = 1 - \f{k}{2\pi^2} + \cdots.
\ee

\section{Comments on $q=4$ case}
\label{sec:q4case}

\begin{figure}[ht]
\includegraphics[width=5.65cm]{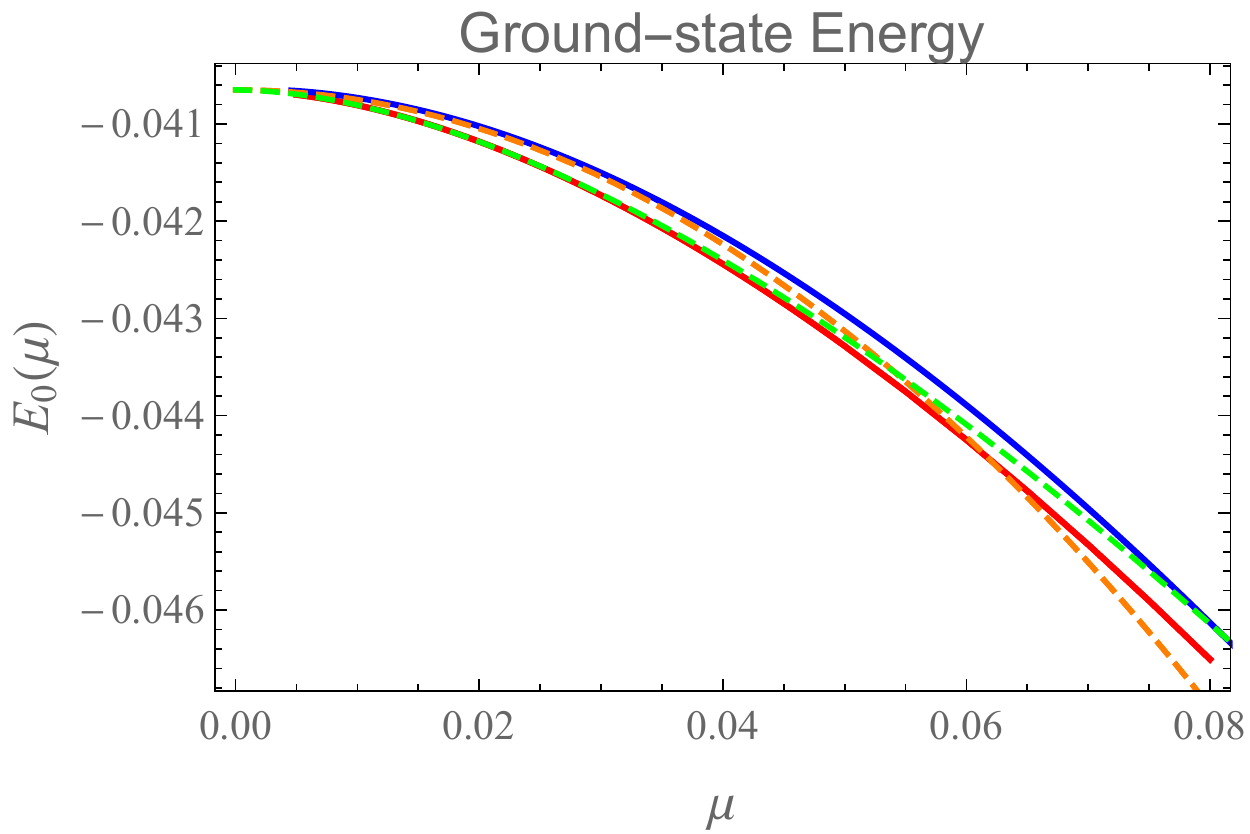} 
\includegraphics[width=5.4cm]{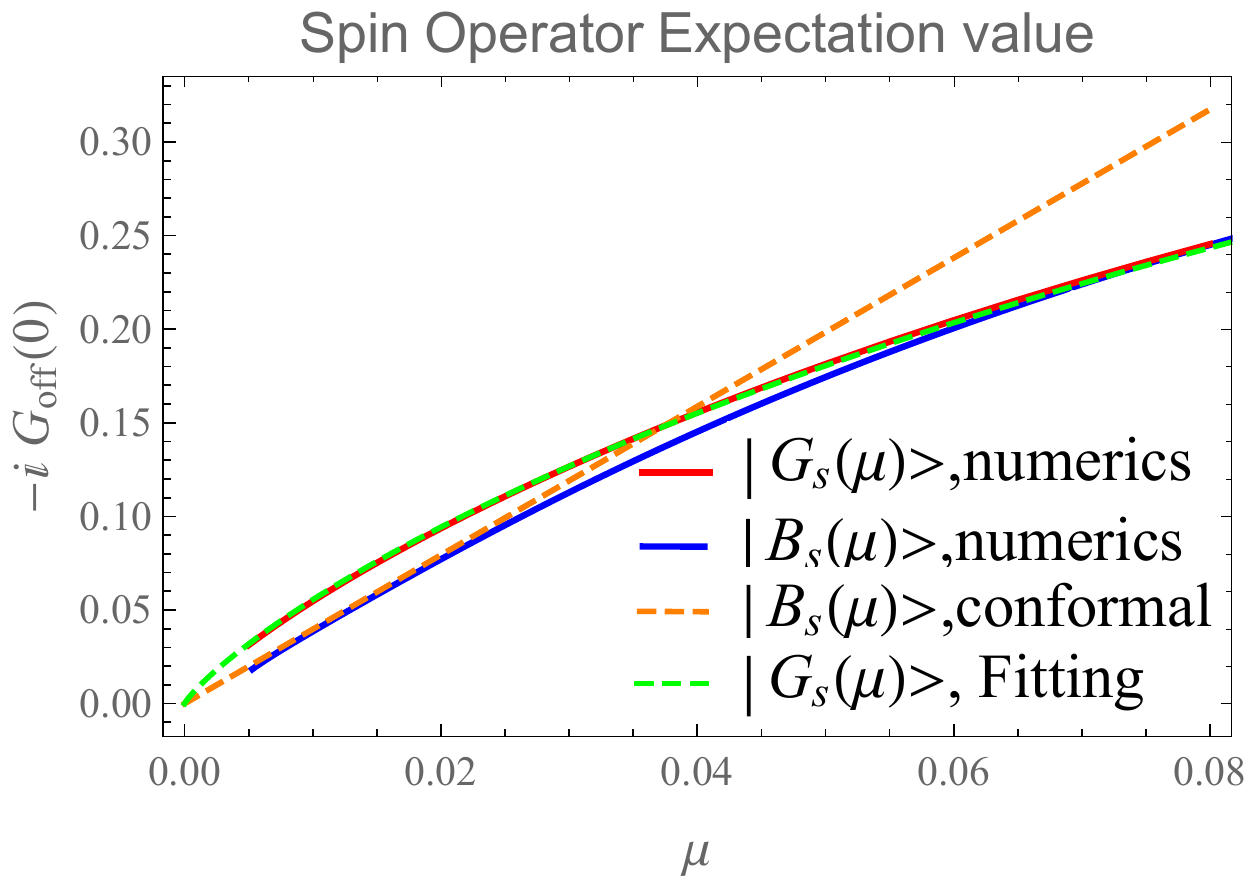}
\includegraphics[width=5.65cm]{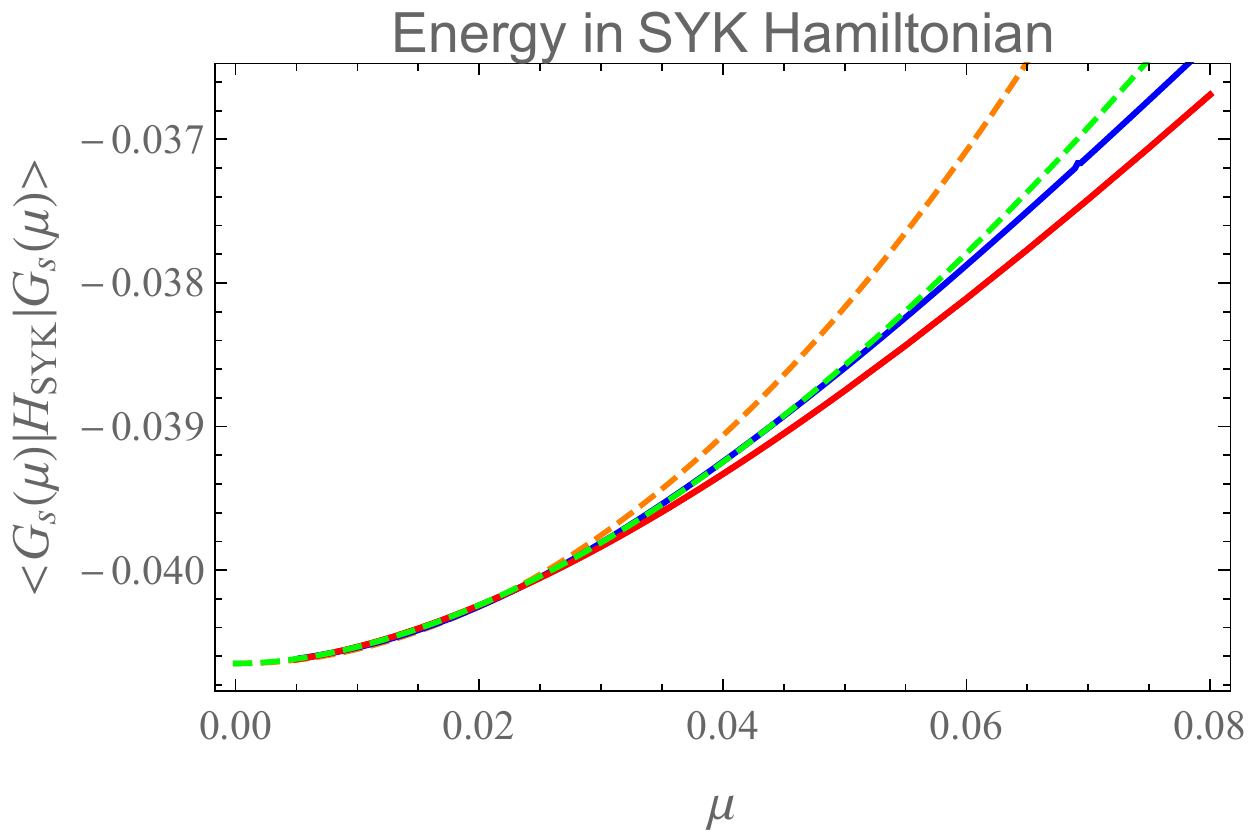}
\caption{The plot of observables both in the exact ground state $\ket{G_{\bm{s}}(\mu)}$ and the variational approximation $\ket{B_{\bm{s}}(\beta(\mu))}$.
Here we choose the parameter to be $q=4$ and $J = 1$.
As written in the central picture, the solid lines represent the numerics and the dashed lines represent the conformal limit answer.
{\bf Left:} The plot of the ground state $E_0$ as a function of $\mu$.
{\bf Middle:} The plot of the half of the absolute value of the spin operator expectation value $|\braket{S_k}|$, which is equal to the $\tau = 0$ off diagonal correlation function $-iG_{\text{off}}(0)$, as a function of $\mu$.
{\bf Right:} The plot of the energy in the SYK Hamiltonian $\braket{H_{SYK}}$ as a function of $\mu$.
 }  
\label{fig:ValVsExactq4}
\end{figure}

We have derived analytic formula for the conformal limit of spin operator expectation value (\ref{eq:SpinExConf}), the ground state energy (\ref{eq:GDEnConf}) and the energy in the SYK Hamiltonian (\ref{eq:SYKEnConf}).
These expression contains $\Gamma(1-4\Delta) = \Gamma(1-4/q)$, which diverges at  $q = 4$.
This does not mean that these observables are divergent but we should take into account the UV effects in the actual deformed SYK model.
Numerically we observe that the exact value is finite but the scaling behavior is violated by the UV effect.
We found that in the small $\mu$ regime the logarithmic behavior appears in the exact answer as 
\be
E_0(\mu) -E_0 \sim c_{E_0} \Big(\f{\mu}{J} \Big)^2 \log \f{\mu}{J}, \qquad -iG_{\text{off}}(0)  \sim  c_{G_{\text{off}}}\f{\mu}{J} \log \f{\mu}{J},\qquad \braket{H_{SYK}} \sim c_{H_{SYK}} \Big(\f{\mu}{J} \Big)^2 \log \f{\mu}{J}.
\ee
We use $0.05 < \mu/J < 0. 2$ region to fit the numerical data and the fitting gives  $c_{E_0} \approx 0.34$, $c_{G_{\text{off}}} \approx -1.2$ and $c_{H_{SYK}} \approx -0.26$.
We show the numerical plot of these observables and their comparison with variational approximation in Fig.~\ref{fig:ValVsExactq4}.
It may be possible  to derive these coefficient analytically and we leave these problems as future works.

\bibliography{KMNNbunken1_191226Nosaka.bib}

\end{document}